\documentclass[envcountsect]{llncs}
\usepackage{url}

\usepackage{ifpdf}
 \ifpdf
\usepackage[pdftex]{graphicx}
 \else
\usepackage[dvips]{graphicx}
 \fi

\spnewtheorem{defi}{Definition}[section]{\bfseries}{\rmfamily}
\spnewtheorem{prop}{Proposition}[section]{\bfseries}{\rmfamily}
\spnewtheorem{fact}{Fact}[section]{\bfseries}{\rmfamily}
\spnewtheorem{coro}{Corollary}[prop]{\bfseries}{\rmfamily}

\begin{document}

\def\mytitle{
On The Longest Chain Rule 
and
Programmed Self-Destruction
of Crypto 
Currencies
}
%
%


\title{\mytitle\\
\vskip-8pt
}

\author{
Nicolas T. Courtois
}

\institute{
$^{1}$
University College London, UK
}

\maketitle

%
%

\vskip-8pt
\vskip-8pt
\vskip-8pt
\begin{abstract}
In this paper we revisit some major orthodoxies
which lie at the heart of
the bitcoin crypto currency 
and its numerous clones. 
In particular we look at 
{\em The Longest Chain Rule},
the monetary supply policies 
and the exact mechanisms which implement them. 
We claim that these built-in properties are not as brilliant 
as they are sometimes claimed.
A closer examination reveals that they are
closer to being... engineering mistakes
which other crypto currencies have copied rather blindly. 
More precisely we show that 
the capacity of current crypto currencies to
resist double spending attacks
is poor 
and most current crypto currencies are highly vulnerable.
Satoshi did not implement a timestamp for bitcoin transactions
and the bitcoin software does not attempt to monitor double spending events.
As a result major attacks involving hundreds of millions of dollars
can occur and would not even be recorded, cf. \cite{MtGoxDoubleSpendingDecker}.
Hundreds of millions 
have been invested to pay
for ASIC hashing infrastructure
yet insufficient attention was paid to
ensure network neutrality and
that the protection layer it promises 
is effective 
and cannot be abused.

In this paper we develop a theory of
{\em Programmed Self-Destruction} of crypto currencies.
We observe that most crypto currencies have mandated abrupt and sudden transitions.
These affect their hash rate and therefore their protection against
double spending attacks which we do not limit the
to the notion of 51$\%$ attacks which is highly misleading.
Moreover we show that smaller bitcoin competitors are substantially more vulnerable.
In addition to lower hash rates,
many bitcoin competitors
mandate 
incredibly
important adjustments in miner reward.
We exhibit examples of `alt-coins'
which validate our theory and
for which the process of programmed
decline and rapid self-destruction
has clearly already started.
%
%

\vskip4pt
{\bf Note:} The author's blog is \url{blog.bettercrypto.com}.


\end{abstract}

\newpage
\vskip-6pt
\vskip-6pt
\section{Bitcoin and Bitcoin Clones}
\vskip-6pt



Bitcoin is a collaborative 
virtual currency and payment system.
It was launched in 2009 \cite{SatoshiPaper}
based on earlier crypto currency ideas
\cite{HashCash,WeiDaiBMoney}.
Bitcoin implements a certain type of peer-to-peer financial
cooperative without trusted entities such as traditional financial institutions.
Initially bitcoin was a sort of social experiment,
however bitcoins have been traded for real money for several years now
and their price have known a spectacular growth
\cite{TwoGoxBotsBought650KBTCAndManipulatedBTCPrice1000}.

Bitcoin challenges our traditional ideas about money and payment.
Ever since Bitcoin was launched \cite{SatoshiPaper,BitcoinMainSoftwareDistribution} in 2009
it has been 
clear that it is an experimental
rather than mature electronic currency ecosystem .
A paper at the Financial Cryptography 2012 conference explains that
Bitcoin is a system which
{\em uses no fancy cryptography, and is by no means perfect}
\cite{BitcoinFC12SecurityOverview}.
In one sense it is still a play currency in early stages of development. 
The situation is even worse for bitcoin competitors.
Their creators and promoters typically
just copy features of bitcoin without
any deeper insight into their consequences.

In this paper we are going to see that
the exact same rules which
might after all work relatively well (at least for some time)
for a large dominating crypto currency such as bitcoin,
are rather disastrous for 
smaller crypto currencies.

\medskip
On the picture below we explain
the organization of this paper.

\begin{figure}[!here]
\centering
\begin{center}
\vskip1pt
\vskip1pt
\includegraphics*[width=5.0in,height=2.7in,bb=0pt 0pt 730pt 350pt]{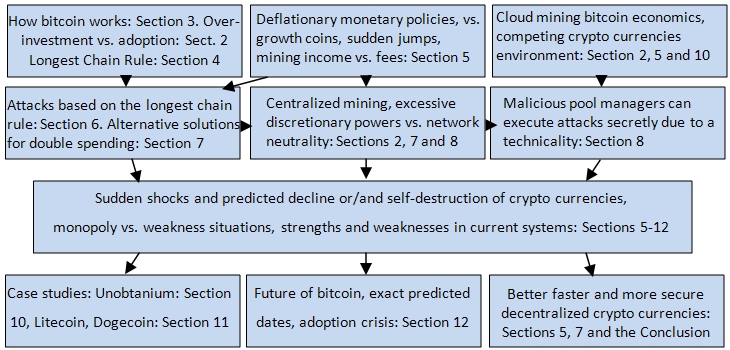}
\vskip-0pt
\vskip-0pt
\end{center}
\caption{
Our roadmap: risks and dangers of bitcoin and
other digital currencies.
}
\label{BitcoinDestrRoadmap}
\vskip-3pt
\vskip-3pt
\end{figure}

\newpage
\section{
Bitcoin As A Distributed Business:
Its Key Infrastructure and Investor Economics}
\label{SectionBitcoinInfrastructureHistory12MInvestorEconomocis}

Bitcoin digital currency \cite{SatoshiPaper}
is an electronic payment system based on cryptography
and a self-governing open-source financial co-operative.
Initially it was just a social experiment and
concerned only some enthusiasts.
However eventually a number of companies have started trading bitcoins for real money.
One year ago, in April 2013,
the leading
financial magazine
The Economist
recognized bitcoin as a major disruptive
technology for finance and famously called bitcoin ``digital gold''.
We can consider that the history of bitcoin
as a mainstream financial instrument
started at this moment.

\vskip-7pt
\vskip-7pt
\begin{figure}[!ht]
\centering
\begin{center}
\hskip-6pt
\hskip-6pt
\includegraphics*[width=4.9in,height=1.7in,bb=0pt 0pt 1170pt 450pt]{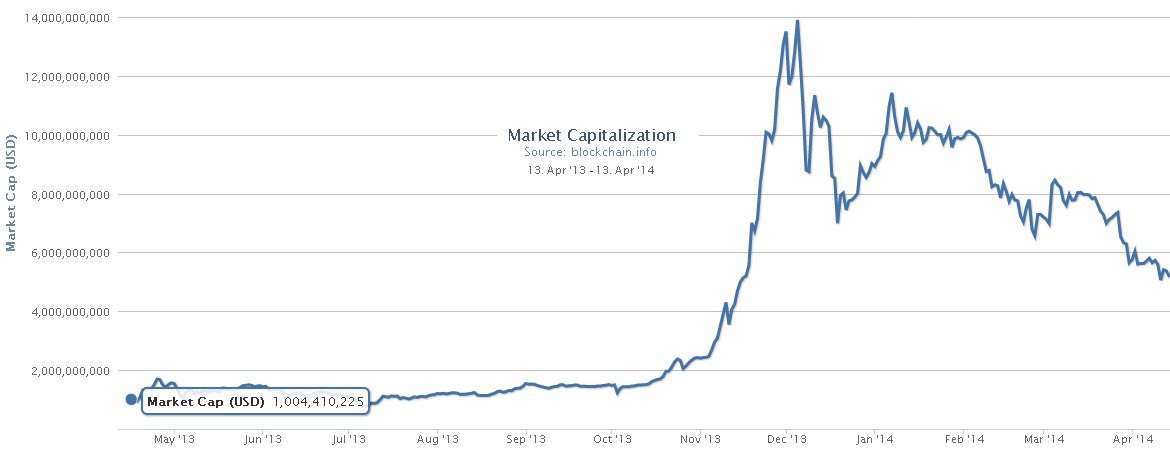}
\end{center}
\vskip-5pt
\vskip-5pt
\caption{
The bitcoin market capitalization
in the last 12 months.
}
\label{BitcoinGraph1YMarketCap}
\end{figure}
\vskip-7pt

Our starting point of April 2013 coincides more or less
with bitcoin achieving prices of 50 USD (and above),
the market capitalization exceeding 1 billion dollars,
and an important shift in the nature of the ownership of the bitcoin infrastructure.
In a great simplification, before April 2013,
one bitcoin was rarely worth more than 5-50 dollars,
and new bitcoins were produced
by amateurs 
on their PCs.
Then a new sort of high-tech industry emerged.
Specialized equipment (ASIC machines) 
whose only purpose is to produce new bitcoins.
Such machines are called miners and are increasingly sophisticated \cite{MiningUnreasonable}.
Bitcoin then rapidly switched to the phase where new bitcoins
are produced by a restricted\footnote{\label{foot1}
The inventor of bitcoin has postulated that each peer-to-peer network node should be
mining 
cf. Section 5 of \cite{SatoshiPaper}.
In practice a strange paradox is that
miners mine in very large pools
cf. 
\cite{RosenfeldPoolRewardMethods} and Table 2 in \cite{MiningSubversive}
and
the number of ordinary
peer-to-peer network nodes is in comparison incredibly low,
falling below 8,000 recently cf. \cite{NumberOfReachableNodesIsDroppingRecently}
}.
group of some 100,000
for-profit `bitcoin miners'
which people have invested money to purchase
specialized equipment.

These last 12 months of bitcoin history,
April 2013-April 2014,
have seen an uninterrupted
explosion of {\bf investment} in bitcoin infrastructure. 
Surprisingly 
large sums of money have been spent on
purchasing new mining equipment.
All this investment has been subject
to excessively rapidly decreasing returns.
Bitcoin mining
is a race against other miners
to earn a fairly limited fraction 
of newly created bitcoins.
We examine these questions in detail.

\newpage
\subsection{Investment in Hashing Power and Incredible 1000x Increase}

\begin{figure}[!ht]
\centering
\begin{center}
\hskip-6pt
\hskip-6pt
\includegraphics*[width=4.9in,height=1.7in,bb=0pt 0pt 1170pt 450pt]{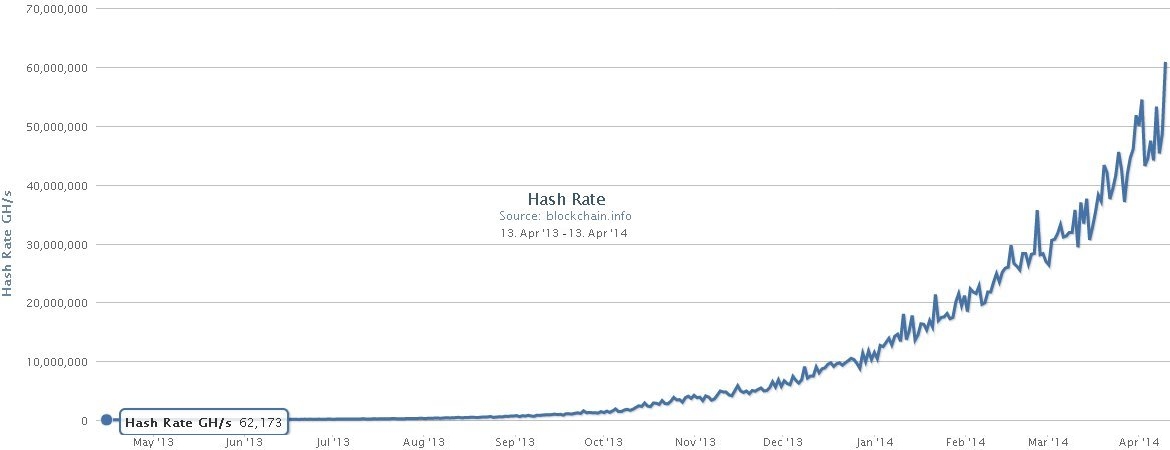}
\end{center}
\vskip-5pt
\vskip-5pt
\caption{
The combined computing power in the
collectively owned bitcoin `hashing infrastructure'
has nearly {\bf doubled each month}
and overall it has increased 1000 times in the last 12 months
while the monetary supply has increased only by 1 $\%$ each month.
The mining profitability has also been eroded accordingly.
The income from any existing miner
was divided by half nearly every month, cf Section \ref{WickedErosionOfIncomeDividedByTwoEachMonth}.
}
\label{BitcoinGraph1YHashRate}
\end{figure}

The combined power of bitcoin mining machines
has 
been multiplied by 1000 in the last 12 months
cf. Fig. \ref{BitcoinGraph1YHashRate}.
However due to built-in excessively conservative
monetary policy cf. \cite{MiningUnreasonable},
during the last 12 months,
miners have been competing for a modest fraction
of bitcoins yet to be generated.
The number of bitcoins in circulation
has increased only by 15 $\%$, from 11 million to 12.6 million.

\vskip-7pt
\vskip-7pt
\vskip-7pt


A 1000-fold increase in hash power is a very disturbing fact.
We lack precise data in order to investigate
how much of this increase
was due to improved technology
(important increase in the speed
of bitcoin mining machines, cf. \cite{MiningUnreasonable}),
and how much was due to a surge in investment:
more people bought bitcoin miner machines.
%
%
However it is certain that
{\bf a monumental amount of money}
has been invested
in these ASIC miner machines.
It is not easy to estimate it accurately.
If we consider that the current hash rate
is composed primarily of KNC Neptune 28 nm miners
shipped in December 2013
which for the unit price of 6000 USD
can deliver some 0.5 TH/s,
we obtain that miners have spent in the last four months
maybe 600 millions of dollars on approximately 120,000 ASIC machines
which are already in operation\footnote{\label{footnote600M}
Similar estimations can be found in \cite{SimilarFigures600M1Billion}.
If we consider that more recent miners with capacities
between 1-3TH/s were already available
for the same price to some privileged buyers
many months before officially
sold on the retail market,
the total cost could be less than our 600M USD estimation.}.
In addition knowing that more miners
were ordered and not yet delivered,
it is quite plausible to assume
that miners have spent already more than 1 billion dollars
on ASIC miners.

As we have already explained,
we don't know exactly how this investment has evolved with time.
However the near-doubling of the hash rate every month
does certainly mean one thing:
{\bf excessively rapid decline in mining revenue} for every existing ASIC machine.

\subsection{Investors Facing Incredibly Fast Erosion of Profitability}
\label{WickedErosionOfIncomeDividedByTwoEachMonth}

This is due to the fact all miners are in competition for
a fixed number of bitcoins which can be mined in one month.
The rule of thumb is that exactly 25 bitcoins are produced every 10 minutes.
Doubling the aggregate hash rate for all bitcoin miners
means dividing each individual miner's income by 2 each month\footnote{
Assuming that the cost of electricity is low compared to the income generated
and that the price of bitcoin is relatively stable.
}.
It means that investors can only hope for fast short-term gains,
and that their income tends to zero very quickly.

Let us develop this argument further.
Imagine that a miner invests 5,000 USD and that the income from mining
in the first month was 2,000 USD.
Is this investment going to be profitable?
Most investors will instinctively believe it will be.
However in actual bitcoin it isn't.
In the recent 12 months
the hash power has been decreasing approximately twice each month.
We need to look at the following sum: 

\vskip-4pt
\vskip-4pt
$$
1
+\frac{1}{2}
+\frac{1}{4}
+\frac{1}{8}
+\ldots
=2
$$
\vskip-1pt

We see that the total income is only
{\bf twice the income for the first month}.
This is not a lot.
In our example the investor will earn only 4,000 USD
and has spent 5,000 USD.
The investor does not make money, he makes a loss.


\subsection{Dividend From Hashing}
\label{DividendHashing}

It is easy to calculate exactly how much money
has already been earned by miners in freshly minted bitcoins
multiplied by their present market price.

\vskip-7pt
\vskip-7pt
\begin{figure}[!ht]
\centering
\begin{center}
\hskip-6pt
\hskip-6pt
\includegraphics*[width=4.9in,height=1.7in,bb=0pt 0pt 1170pt 450pt]{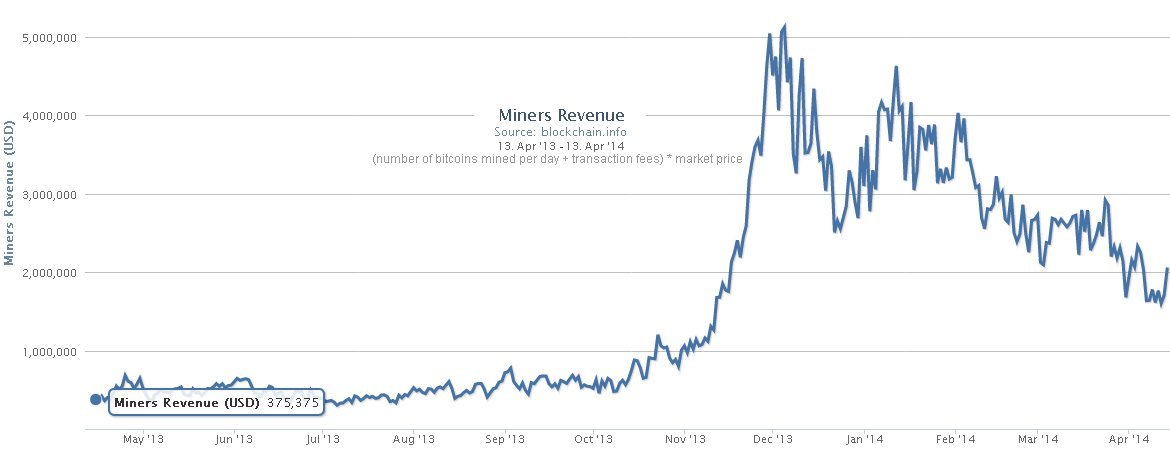}
\end{center}
\vskip-5pt
\vskip-5pt
\caption{
The daily market price of freshly created bitcoins in the last 12 months.
}
\label{BitcoinGraph1YMinersDailyRevenue}
\end{figure}
\vskip-7pt
%

If we estimate the area under Fig. \ref{BitcoinGraph1YMinersDailyRevenue}
we see that currently all miners combined
make some 60 millions of dollars only per month
and have been paid roughly some 400 million dollars in
mining dividend 
most of which was earned in the last 4 months.
In this paper we neglect the cost of the electricity.
Contrary to what was suggested in some press reports
\cite{MiningMegaWattsEnvirDisaster},
this cost has so far remained
relatively low for bitcoin mining
in comparison to the high cost of ASIC miners
which cost needs to be amortized over surprisingly
short periods of time of no more than a few months
as shown in Section \ref{WickedErosionOfIncomeDividedByTwoEachMonth}.


\vskip-7pt
\vskip-7pt
\subsection{Investors' Nightmare}
\label{InvestorsNightmare}
\label{ASICDeliveryNightmare}
\vskip-5pt

The market for ASIC miner machines is very far from being fair and transparent.
There is only a handful of ASIC companies \cite{MiningHardwareOrdersWhatHappensNext}
and from their web pages it seems that they
might have manufactured and sold only a few thousands units each.
In fact 
most manufacturers have omitted to tell their customers
the actual size of their production.
It has been much higher than expected, as shown by
the hash rate, cf. Fig. \ref{BitcoinGraph1YHashRate}.
Most manufacturers worked with pre-orders.
Customers were never able to know when machines were going to be delivered
and how much the hash rate would increase in the meantime.
Many manufacturers had important delays in delivery,
frequently 3, 6, 8, 12 months \cite{MiningHardwareOrdersWhatHappensNext}.
Such delays decrease the expected income from mining by an {\bf incredibly large factor}.
We give some realistic examples which
based on personal experiences of ourselves and our friends:

\vskip-5pt
\vskip-5pt
\begin{enumerate}
\item
If for example a miner have ordered his device
from ButterflyLabs and the device is delivered 12 months later.
He earns roughly 1000 times less than expected
(
cf. Fig. \ref{BitcoinGraph1YHashRate}),
and even if the price of bitcoin rises 10 times during this period,
cf. \cite{TwoGoxBotsBought650KBTCAndManipulatedBTCPrice1000},
he still earns maybe 100 times less than expected (!).

\item
ButterflyLabs are not the worst.
Many miners ordered devices
from suppliers which do NOT even exist,
and were pure criminal scams,
even though they advertise on the Internet and
their machines are frequently compared to legitimate ASIC manufacturers
on web sites such as \url{https://en.bitcoin.it/wiki/Mining_hardware_comparison}
which have NOT attempted to distinguish
between criminal scams and genuine manufacturers.
See Appendix of \cite{MiningSubversive} and \url{http://bitcoinscammers.com}
for specific examples.

\item
A San Francisco-based startup HashFast
currently embroiled in many federal fraud lawsuits
related to production delays (3 months or longer)
and the fact that they promised to refund their customers in bitcoins.
However the market price of bitcoins went up significantly.
In May 2014 they denied bankruptcy rumors
and announced that they will lay off 50 $\%$ of its staff \cite{HashFastTroubles,MiningHardwareOrdersWhatHappensNext}.

\item
Another miner ordered his device from BITMINE.CH
(also near bankruptcy)
and the device was delivered with 6 months delay,
he earns roughly 64 times less than expected.
Even if the price of bitcoin rises 4 times during this period,
and even if BITMINE.CH compensates customers
by increasing hash rate by 50 $\%$,  
he still earns maybe 10 times less than expected (!).

\item
In another example a miner ordered his device from KNC miner
or Cointerra, and the device was delivered with just a one month delay
compared to the predicted delivery date.
Here the miner earns just half of what was expected,
which is already problematic but might be OK.
\end{enumerate}
\vskip-7pt

Overall it is possible to see that most miners were mislead
when they ordered the ASIC machines.
Miners were probably confused and expected mining profitability
to be much higher 
than what they actually experienced when machines were finally delivered.
%
Accordingly many people {\bf lost money} in the bitcoin mining business
(see also Section \ref{DividendHashing}).
In addition, many of those who made profits
have seen their bitcoins disappear in
large-scale thefts, cf. \cite{MtGoxDoubleSpendingDecker}.

\subsection{Bitcoin Popularity and Bitcoin as Medium of Exchange}
\label{IsBitcoinUsedEconomyMOE}

\begin{figure}[!ht]
\centering
\begin{center}
\hskip-6pt
\hskip-6pt
\includegraphics*[width=4.9in,height=1.2in,bb=0pt 0pt 800pt 220pt]{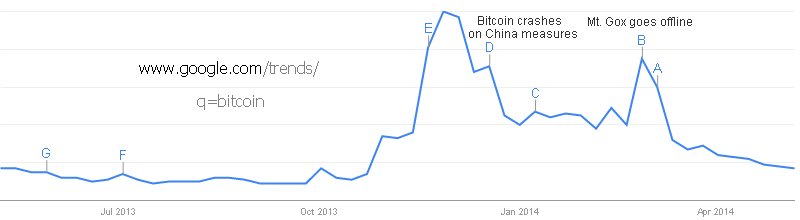}
\end{center}
\vskip-5pt
\vskip-5pt
\caption{
Bitcoin popularity as a keyword in Google web search queries.
}
\label{BitcoinGraph1YGoogleSearchPopularity}
\end{figure}

Bitcoin has certainly been very popular among investors in the last 12 months.
Has it been popular among the general public?
Are they adopting bitcoin as a currency in order to carry ordinary transactions?
In Fig. \ref{BitcoinGraph1YGoogleSearchPopularity}
we show the Google trends for the keyword bitcoin.
We see that the interest in bitcoin\footnote{It has been observed for a very long time
that the bitcoin market price
cf. Fig. \ref{BitcoinGraph1YMarketCap}
and the popularity of bitcoin in Google search
cf. Fig. \ref{BitcoinGraph1YGoogleSearchPopularity}
are strongly correlated.}
 is not growing. 
In May 2014 there were alarming reports about the
total number of full bitcoin network nodes dropping to
dangerously low levels of less than 8,000 nodes,
cf. \cite{NumberOfReachableNodesIsDroppingRecently}.

%
\vskip-7pt
\vskip-7pt
\vskip-6pt
\vskip-6pt

\begin{figure}[!ht]
\centering
\begin{center}
\hskip-6pt
\hskip-6pt
\includegraphics*[width=4.9in,height=1.7in,bb=0pt 0pt 1170pt 450pt]{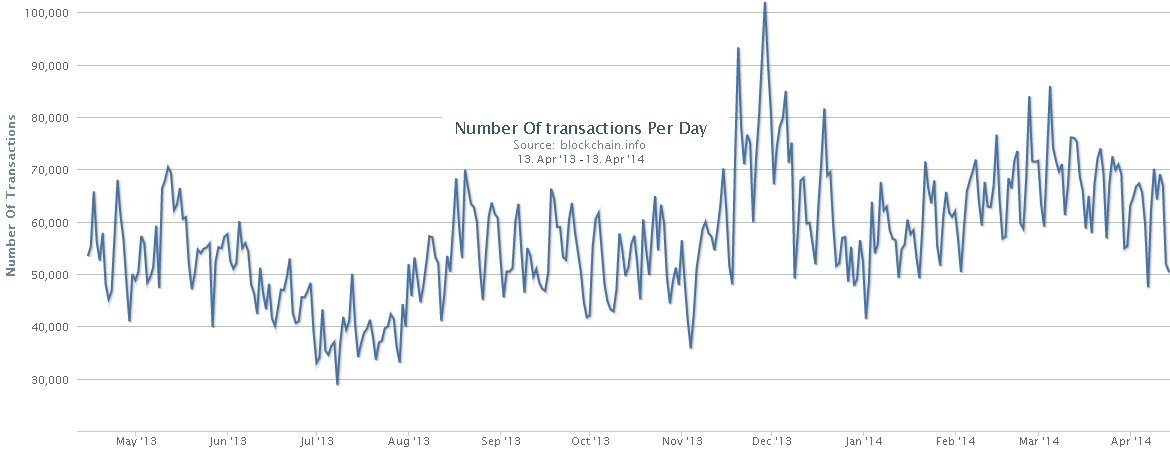}
\end{center}
\vskip-5pt
\vskip-5pt
\caption{
The average number of transactions per day
has remained relatively stable in the last 12 months.
It remains between 40,000 and 80,000
and it can decline rather than increase
during certain months of activity.
}
\label{BitcoinGraph1YTransactionsPerDay}
\end{figure}

It appears that bitcoin is {\bf not used a lot}
as a currency or payment instrument. 
The number of transactions in the bitcoin network is NOT growing,
cf. Fig. \ref{BitcoinGraph1YTransactionsPerDay}
and it can sometimes decrease.
%
The number of merchants accepting bitcoin has been growing
recently cf. \cite{BlockChainTransactionsAnalysisSpikes}
however the number of transactions wasn't.

\vskip-7pt
\vskip-7pt
\vskip-6pt
\vskip-6pt

\begin{figure}[!ht]
\centering
\begin{center}
\hskip-6pt
\hskip-6pt
\includegraphics*[width=4.9in,height=2.0in,bb=0pt 0pt 762pt 350pt]{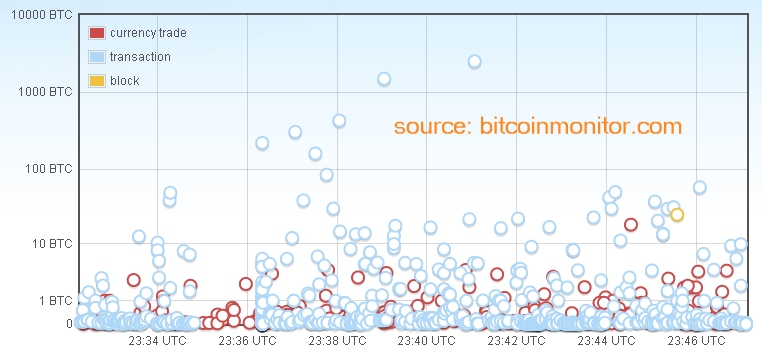}
\end{center}
\vskip-7pt
\vskip-7pt
\caption{
Bitcoin transactions and the amounts involved
displayed in real time
over a period of 15 minutes.
Each circle represents a single transaction,
a yellow circle is the initial 25 BTC mining event,
blue circles are bitcoin transactions on the blockchain,
and red transactions are currency exchange transactions
(not necessarily recorded in the bitcoin blockchain).
}
\label{BitcoinA5MVariousTransactions}
\end{figure}

Things get more complicated if we want to look at the transactions in volume.
An interesting tool which allows
to distinguish between small and large transactions
and to visualise their distribution 
are the real-time graphs
produced by
\url{http://www.bitcoinmonitor.com/}
cf. Fig.
\ref{BitcoinA5MVariousTransactions},
cf. also \cite{BlockChainTransactionsAnalysisSpikes}.
However these graphs and much of the other data
on transaction volume remain very seriously biased
by the amounts which bitcoin users return to themselves.
This is mandatory in all bitcoin transactions
and makes analysis difficult$^{\ref{ImpossibleTaskDailyVolume}}$.

\vskip-7pt
\vskip-7pt
\vskip-7pt
\vskip-7pt

\begin{figure}[!ht]
\centering
\begin{center}
\hskip-6pt
\hskip-6pt
\includegraphics*[width=4.9in,height=1.7in,bb=0pt 0pt 1200pt 600pt]{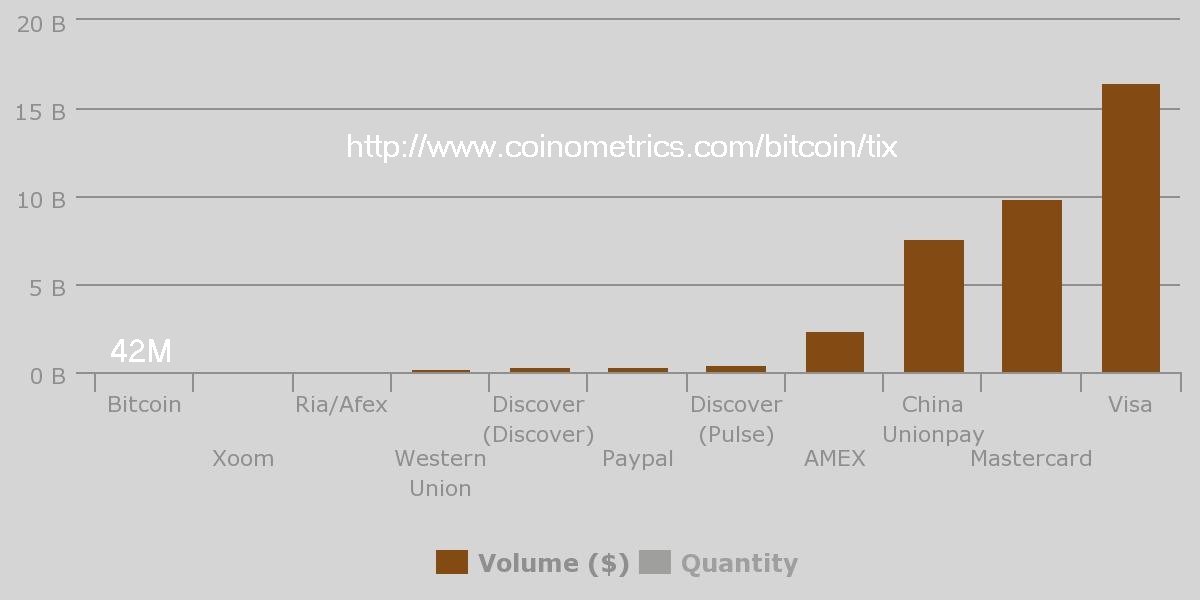}
\end{center}
\vskip-5pt
\vskip-5pt
\caption{
The daily volume in USD comparison for Bitcoin, Paypal VISA and other major payment processors,
September 2014.
}
\label{BitcoinTransactionsVolumeVsVISA}
\end{figure}

Several press reports have WRONGLY
claimed that bitcoin has surpassed Western Union
and is catching up with PayPal \cite{FakeBitcoinVolumeSurpassesWesternUnion,Fake300MBitcoinDaily}.
These reports are based on
bitcoin transaction volume figures
which are artificially inflated.
They do NOT reflect the actual bitcoin economy.
It is easy to see that there is NO easy to way to
reliably estimate the transaction volume
from the blockchain data\footnote{\label{ImpossibleTaskDailyVolume}
It is very difficult to reliably estimate the transaction volume
from the blockchain data alone.
Truly accurate estimations 
are impossible to obtain.
A particular problem are the actions
of some bitcoin addresses which hold very large balances
and return change to themselves at new freshly created addresses. 
Another problem are outliers cf. \cite{BlockChainTransactionsAnalysisSpikes}.
}.
%
The Fitch rating agency
has attempted to obtain more accurate data
\cite{MoreAccurateVolumeReutersFitch}.
We learn that
bitcoin transaction volume is 68 M$\$$ per day [2 April 2014]
and it remains
``small relative to [...] 
traditional payment processors
''.
A recent press report claims that the 
transaction volume was 
at the lowest level in 2 years \cite{BitcoinTransactionVolume}
based on one imperfect method\footnote{
Blockhain.info provides both the misleading artificially inflated figures
at \url{http://blockchain.info/charts/output-volume} and
their estimation of the actual transaction volume by their own
(imperfect) proprietary method
cf. \url{http://blockchain.info/charts/estimated-transaction-volume},
cf. also
\cite{BitcoinTransactionVolume,MoreAccurateVolumeReutersFitch,BlockChainTransactionsAnalysisSpikes}.
}
to eliminate the amounts people
return to themselves$^{\ref{ImpossibleTaskDailyVolume}}$.
%
The nature of bitcoin makes that we do NOT have a truly reliable
source of data on actual bitcoin transactions.
However it is possible to see that bitcoin is still about 400 times smaller than VISA,
cf. Fig. \ref{BitcoinTransactionsVolumeVsVISA}.
\vskip-7pt
\vskip-7pt
\vskip-7pt

Another method to measure the success of bitcoin
is to count the unique users of bitcoin wallet applications.
Their number has reached 1 million in January 2014
1.5 million was attained in April 2014,
and the were 1.6 million in May 2014
\footnote{
Cf. \url{http://www.coindesk.com/blockchain-info-reaches-one-million-wallets/}
then
\url{https://blog.blockchain.com/2014/04/14/blockchain-15m-users/}
and
\url{https://coinreport.net/blockchain-passes-1-6-million-users-mark/}
}.
This growth is quite positive even though
the number of bitcoin transactions is not increasing,
as seen in Fig. \ref{BitcoinGraph1YTransactionsPerDay}.


We propose an alternative measure
of the success of bitcoin as a currency:
it will be the transaction fees.
The more people are willing to pay in order to
transfer money from one person to another using the bitcoin technology,
the more successful it is.
However we should NOT report fees in bitcoins
(as in earlier version of this paper),
but in US dollars.

\vskip-7pt
\vskip-7pt
\begin{figure}[!ht]
\centering
\begin{center}
\hskip-6pt
\hskip-6pt
\includegraphics*[width=4.9in,height=1.5in,bb=0pt 0pt 1041pt 400pt]{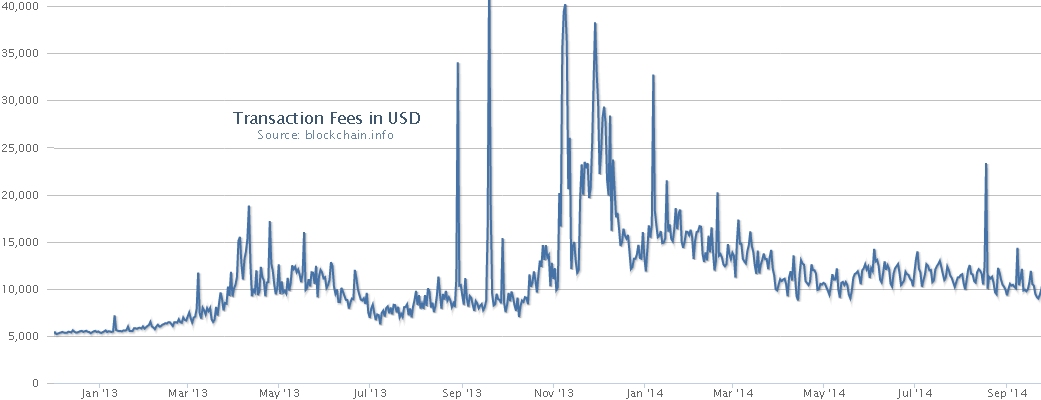}
\end{center}
\vskip-5pt
\vskip-5pt
\caption{
The daily bitcoin transaction {\bf fees} in USD, Jan. 2013-Sept. 2014.}
\label{BitcoinGraph1YTxFeesPerDay}
\end{figure}
\vskip-7pt
\vskip-7pt

We don't have great news in this space, 
the income from fees has been stable or declining.
cf. Fig. \ref{BitcoinGraph1YTxFeesPerDay}.


\newpage

\subsection{Analysis of Bitcoin From The Point of View of Investors}
\label{InvestorEconomicsAB2Billion}

In the previous sections we have seen that
the bitcoin `investment economy'
(mining or holding bitcoins for profit)
has been thriving in the recent 12 months,
while it is very hard to claim
that we have seen any growth
in adoption of bitcoin in ordinary e-commerce
cf. Fig. \ref{BitcoinGraph1YTxFeesPerDay}.
Moreover we were surprised to discover that the number of active miners seems
was much larger than the number of ordinary bitcoin users
see see Section \ref{SectionBitcoinInfrastructureHistory12MInvestorEconomocis}
and \cite{NumberOfReachableNodesIsDroppingRecently}.

Consequently we consider that
until now the bitcoin business
was primarily about some investors (A) spending some 1000
million dollars on mining hardware,
and other investors (B) which preferred
to buy or use 
these newly created bitcoins for 400 million dollars
and holding them. 

We can now argue that
the second group (B)
has potentially spent MUCH more than 400 million dollars.
This is due to the fact that
only a small fraction of bitcoins was manufactured
in the last 12 months.
Investors who in the last 12 months
have purchased newly created bitcoins for 400 million dollars
(due to Fig. \ref{BitcoinGraph1YMinersDailyRevenue})
have also purchased a lot more bitcoins
from previous owner of bitcoins who are free riders:
people who have paid/invested very little
mining or purchasing some bitcoins earlier.
We lack any precise data
but in order to
be able to pay some 400 M in to miners (A)
\footnote{which has paid for some of their 600+ millions of dollars in hardware expenses},
investors (B) must have injected into the bitcoin economy
a possibly much larger sum of cash money (dollars).
Let us assume that this was 2 billion dollars.
This amount is hard to estimate from available data but
it is probably a small multiple of 600 M
and it cannot be higher than
5 billion dollars,
the peak value at Fig. \ref{BitcoinGraph1YMarketCap}.

We can observe that
the reason why so much money was made by owners of older coins
was the monopoly rent:
miners (A) were convinced
to mine for this particular crypto currency
which has influenced further investors (B)
to provide additional funds also for this market.
It is probably correct to assume that this is
substantially more than the total amount of money
invested in mining Litecoin and other crypto currencies,
based on the fact that the total Market capitalization of
all alternative currencies combined remains small compared to bitcoin,
cf. \url{http://www.cryptocoincharts.info/v2/coins/info}.

\medskip
Both investment decisions (A,B)
have been made
on expectation that
the bitcoin market price will rise.
In fact during the last 12 months
the price has been increasing\footnote{This spectacular increase
is now suspected to be an effect of a monumental market manipulation.
An anonymously published report claims that
up to 650,000 bitcoins were bought by two algorithms
with money which is suspected to be paid
from the customer money held as outstanding balances
at the infamous MtGox exchange,
cf. \cite{TwoGoxBotsBought650KBTCAndManipulatedBTCPrice1000}.
%
}
(a lot) just during just one month at the end of 2013
after which we have seen a long painful correction
cf. Fig. \ref{BitcoinGraph1YMarketCap}.

\medskip
The idea that bitcoin market price in dollars
will appreciate in the future
is based on several premises
which in our opinion
are more irrational than rational:

\vskip-5pt
\vskip-5pt
\begin{enumerate}
\item
Bitcoin is expected to imitate the scarcity
of rare natural resources such as Gold
\cite{TheEconomistDigitalGold}
and for this purpose
bitcoin has a fixed monetary supply.
\item
However the scarcity of bitcoins is not natural.
It is NOT a hard reality.
It is really totally
{\bf artificial}.
It is mandated by the bitcoin specification and software
\cite{SatoshiPaper,BitcoinMainSoftwareDistribution}.
This property is not written in stone.
It is frequently criticized
\cite{MiningUnreasonable,WiredKrollLMustKeepCreating}
and it CAN be changed if a majority of miners agree,
cf. \cite{MiningUnreasonable,FeltenMinersProbablyCanChangeMonPol}.
\item
Investors might be overestimating
the importance of bitcoin in the economy in the future:
the adoption of bitcoin
as a currency or payment instrument
cf. Section \ref{IsBitcoinUsedEconomyMOE}.
\item
This expectation does not take into account the `alt-coins'
(competitors to bitcoin).
Alt-coins clearly break the rule of fixed monetary supply
of coins and can be created at will,
cf. Section \ref{sec:CompetingCryptoCurrencies}.
It cannot be guaranteed that the current monopoly situation of bitcoin is going to last.
\end{enumerate}
\vskip-5pt

\noindent
Various surveys 
show that about 50 $\%$ of people involved with bitcoin
do very naively believe
that bitcoin will be worth 10,000 USD at the end of 2014,
see \cite{50PcBelieve10000USD}.
Extremely few people have predicted that bitcoin would collapse:
one university professor have claimed that bitcoin will
go down to 10 USD by June 2014
\cite{BitcoinToCrash10USDInJune2014}.
This prediction was already largely proven wrong.

\newpage
\vskip-8pt
\vskip-8pt
\subsection{What Does This Monumental Investment Pay For?}
\vskip-6pt

We have estimated that
for-profit bitcoin miners (A)
have invested some 1,000 M dollars
in bitcoin infrastructure,
while at the same time other investors (B)
have invested a yet larger sum of cash money,
maybe 2,000 M
on buying bitcoins probably driven by a naive\footnote{
The bitcoin market price
is rather going down ever since December 2013
cf. Fig. \ref{BitcoinGraph1YMarketCap}
and \cite{TwoGoxBotsBought650KBTCAndManipulatedBTCPrice1000}.
}
expectation that they will rise
in the future.

Now the interesting question is,
what these {\bf monumental investments} pay for?
Knowing that the bitcoin adoption
as a medium of exchange is not expanding
as suggested by Fig. \ref{BitcoinGraph1YHashRate}
these investments went {\bf mostly into
building an excessive quantity of
hashing power} (1000x increase).
In \cite{SamsMarginal} Sams writes:

\begin{quote}
"The amount of capital collectively burned hashing
fixes the capital outlay required of an attacker
to obtain enough hashing power to have a meaningful chance
of orchestrating a successful double-spend attack 
on the system [...]
The mitigation of this risk is valuable, [...]"
\end{quote}

We have this expensive and powerful
hashing infrastructure.
We could call it (ironically) {\bf the Great Wall of Bitcoin}
which name is justified by the fact
that bitcoin miners have invested roughly about 1 billion dollars to build it
and it is expected to protect bitcoin against attacks.
This leads to the following working hypothesis
which is really about
economics of information security
and which we will later dispute. 
{\bf Maybe} one must spend a lot of money
on the bitcoin hashing infrastructure
in order to achieve good security.
Maybe there is a large cost
associated with
building a global distributed financial infrastructure
totally independent from governments,
large banks, the NSA, etc.
Maybe one can hardly hope to spend less
and security against double spending attacks
has some inherent price which needs to be paid.

We claim that this sort of conclusion is {\bf MISTAKEN} and the devil is in the details.
In this paper we are going to show
that the amount of money needed
to commit for-profit double spending attacks 
remains {\bf moderate},
it has nothing to do with the 600 M dollars spent on ASIC miners in activity.
It is a fallacy to consider that money burnt in hashing
could or should serve as effective protection against attacks.
This is because money at risk,
for example in large transactions,
can be substantially larger
than the cost of producing a fork in the block chain. 
We claim that nearly anybody can commit
double spending attacks,
or it will become so in the 
future.
We claim that the current 1 billion dollar
investment in 
bitcoin infrastructure is
{\bf neither necessary nor sufficient} to build a secure digital currency.
It simply
{\bf does NOT serve as effective protection} 
and does not deliver the security benefits claimed.
This is due to misplaced ideology
such as the so called
{\em The Longest Chain Rule}, 
important technicalities,
dangerous centralization and insufficient network neutrality,
and lack of the most basic features
in Satoshi bitcoin specification.
We intend to show that it is possible to fix
the double spending problem in bitcoin
with cryptography and timestamping,
and the cost of doing so is in general
much lower than expected.


\newpage
\section{Short Description of How Bitcoin Works}

We have essentially
one dominant form of bitcoin software \cite{BitcoinMainSoftwareDistribution} and
the primary ``official'' bitcoin protocol specification
is available at \cite{BitcoinTechSpec}.
However bitcoin belongs to no one and the specification is subject to change.
As soon as a majority of people run a different version of it,
and it is compatible with the older software,
it becomes the main (dominating) version.

Bitcoin is a sort of distributed electronic notary system
which works by consensus. We have a decentralized network of nodes with peer-to-peer connections.
The main functionality of bitcoin is to allow transfer of money from one account to another.
At the same time network participants create new coins and
perform necessary checks on previous transactions which are meant to enforce ``honest'' behavior.
Integrity of bitcoin transactions is guaranteed by cryptographic hash functions,
digital signatures and
a consensus about what is the official history of bitcoin.
Below we provide a short, concise 
description of how bitcoin works.

\vskip-6pt
\vskip-6pt
\begin{enumerate}
\item
We have a decentralized network of full bitcoin nodes which resembles a random graph.
Network nodes can join and leave the network at any moment.
\item
Initially, when bitcoins are created,
they are attributed to any network node
willing and able to spend sufficient computing power
on solving a difficult cryptographic puzzle.
We call these people ``miners''.
\item
It is a sort of lottery in which
currently 25 bitcoins are attributed to one and unique
``winner'' every 10 minutes.
\item
With time this quantity decreases which has been decided by the creator(s)
of bitcoin in order to limit the monetary supply of bitcoins in the future.
\item
The legitimate owner of these 25 bitcoins is simply identified by a certain public key
(or several public keys).
\item
A public ledger of all transactions is maintained and
it is used to record all transfers of bitcoins from one account (one public key)
to another.
\item
Bitcoins are divisible and what is stored on the computers
of the network participants are just the private keys.
\item
The amount of bitcoins which belongs to a given key at a given moment is stored
in the public ledger, a copy of which is stored
at every full network node application and constantly kept up to date.
\item
Miners repeatedly compute a double SHA-256 hash H2 of
a certain data structure called a block header
which is a combination of
events in the recent bitcoin history
and which process is described in more detail in \cite{MiningUnreasonable,MiningUnreasonable2,BitcoinTechSpec}.
\item
This H2 must be such that when written as an integer in binary
it will have some 64 leading zeros
which corresponds to the difficulty level in the bitcoin network
at a given moment (cf. \cite{MiningUnreasonable}).
\item
The difficulty level can go up and down depending on how many people participate in mining
at a given moment. It tends increase and it does rarely decrease
\footnote{In bitcoin it
has increased at truly unbelievable speed, cf. Fig. \ref{BitcoinGraph1YHashRate}.
In other crypto currencies it is more likely to decrease in a substantial way
as we will see in this paper}.
\item
More precisely,
in order to produce a winning block,
the miner has to generate a block header such that
its double SHA-256 hash H2
is smaller than a certain number called \mbox{target}.
\item
This can be seen as essentially a repeated experiment where H2 is chosen at random.
The chances of winning in the lottery are very small
and proportional to one's computing power multiplied by $2^{-64}$.
This probability decreases with time as more miners join the network.
The bitcoin network combined hash rate
increases rapidly, see
Fig. \ref{BitcoinGraph1YHashRate}.
\item
If several miners complete the winning computation only one of them
will be a winner which is decided later by a consensus.
\item
Existing portions of the currency are defined {\bf either} as outputs of a block
mining event (creation)
{\bf or} as outputs of past transactions (redistribution of bitcoins).
\item
The ownership of any portion of the currency is achieved through
chains of digital signatures.
\item
Each existing quantity of bitcoin identifies its owner by specifying his public key or its hash.
\item
Only the owner of the corresponding private key
has the power to transfer
this given quantity of bitcoins to other participants.
\item
Coins are divisible and transactions are multi-input and multi-output.
\item
Each transaction mixes several existing quantities of bitcoins
and re-distributes the sum of these quantities of bitcoin
to several recipients in an arbitrary way.
\item
The difference between the sum of inputs and the sum of all
outputs is the transaction fee.
\item
Each transaction is approved by all the owners of each
input quantity of bitcoins
with a separate digital signature approving the transfer
of these moneys to the new owners.
\item
The correctness of these digital signatures is checked by miners.
\item
Exactly one miner approves each transaction which is included in one block.
However blocks form a chain and other miners will later approve this block.
At this moment they should also check the past signatures,
in order to prevent the miner of the current block from cheating.
With time transactions are confirmed many times and it becomes
increasingly hard to reverse them.
\item
All this is effective only for blocks which are in the
dominating branch of bitcoin history (a.k.a. the Main Chain).
Until now great majority of events in the bitcoin history made it to become the part of this official history.
\item
In theory every bitcoin transaction could later be invalidated.
A common solution to this problem is to wait for a small multiple of 10 minutes
and hope that nobody will spend additional effort just in order to invalidate one transaction.
These questions are studied in more detail
in Section
\ref{LongestChainRuleDoubleSpendingAttacks}.
\item
Overall the network is expected to police itself.
Miners not following the protocol risk that their blocks
will be later rejected by the majority of other miners. 
Such miners would simply not get the reward for which they work.
\item
There is no mechanism to ensure that all transactions
would be included by miners other than the
financial incentive in the form of transaction fees.
\item
There is no mechanism to store a complete history of events in the
network other than the official (dominating) branch of the block chain.
Memory about past transactions and other events in the network may be lost,
cf. \cite{MtGoxDoubleSpendingDecker}.
\end{enumerate}
\vskip-6pt


\vskip-5pt
\vskip-5pt
\section{Asynchronous Operation And The Longest Chain Rule}
\label{LongestChainaAynchronous}
\vskip-5pt

According to the initial design by Satoshi Nakamoto \cite{SatoshiPaper}
the initial bitcoin system is truly decentralized
and can be to a large extent asynchronous.
Messages are broadcast on the basis of {\em best effort}.
Interestingly the system can support important network latency
and imperfect diffusion of information.
Information does not have to reach all nodes
in the network in the real time
and they could be synchronized later
and can agree on a common history
at any later moment.

The key underlying principle which allows to achieve this objective
is {\bf the Longest Chain Rule} of Satoshi Nakamoto \cite{SatoshiPaper}.
It can be stated as follows:

\begin{enumerate}
\item
Sometimes we can have what is called {\em a fork}:
there are two equivalent solutions to the cryptographic puzzle.

\item
Currently a fork happens less than 1 $\%$ of the time,
see Table 1 in \cite{MiningSubversive}. 
However it clearly could and would be more frequent in poor network conditions
or due to certain attacks, cf.
\cite{BitcoinBrokenGrowingSelfishPoolStrategy,MiningSubversive}.

\item
Different nodes in the network have received one of the versions first and different miners
are trying to extend one or the other branch.
Both branches are legitimate and the winning branch
will be decided later by a certain type of consensus mechanism,
%
automatically without human intervention.

\item
The {\bf Longest Chain Rule} of \cite{SatoshiPaper}
says that if at any later moment in history one chain becomes longer,
all participants should switch to it automatically.
\end{enumerate}

With this rule,
it is possible to argue that due to the probabilistic nature of the mining process,
sooner or later one branch will automatically win over the other.
For example we expect that a fork of depth 2 happens with the frequency which is the square
of previous frequency, i.e. about 0.01 $\%$ of the time.
This is what was predicted and claimed 
by Satoshi Nakamoto \cite{SatoshiPaper}.
This is precisely what makes bitcoin quite stable in practice.
Forks are quite rare,
and wasted branches of depth greater than one 
are even much less frequent,
see Table 1 in \cite{MiningSubversive}.
All this is however theory or how the things
have worked so far in recent bitcoin history.
In practice it is more complicated as we will see in this paper.

\vskip-5pt
\vskip-5pt
\subsection{Why Do We Have This Rule?}
\label{LongestChainWhyRuleExists}
\vskip-5pt

This Satoshi rule can be seen as
an early and imperfect
attempt to solve the problem of double spending.
More generally in some way it also is
a yet another attempt to solve some version of
the long-standing so called "Byzantine Generals" problem \cite{Byzantine},
which is also solved by voting and has been studied by computer scientists since 1982.
This sort of problems are known to be very difficult to solve in practice.
In contrast in current bitcoin literature
{\bf the Longest Chain Rule}
is somewhat taken for granted without any criticism.
For example in the very highly cited recent paper
\cite{BitcoinBrokenGrowingSelfishPoolStrategy} we read:
"To resolve forks, the protocol
prescribes miners to adopt and mine on the longest chain.".
In this paper we are going to show that this rule is
highly problematic and it leads to very serious hazards. 

\vskip-5pt
\vskip-5pt
\subsection{Genius or Engineering Mistake?}
\label{LongestChainWhyRuleExistsGeniusMistake}
\vskip-5pt

It is possible to see that this consensus mechanism in bitcoin
has two distinct purposes:

\vskip-5pt
\vskip-5pt
\begin{enumerate}
\item It is needed in order to decide {\bf which blocks}
obtain a monetary reward.
It allows to resolve potentially arbitrarily
complex fork situations in a simple, elegant and convincing way.
\item It is also used to decide
{\bf which transactions} are accepted
and are part of official history,
while some other transactions are rejected
(and will not even be recorded,
some attacks could go on without being noticed,
cf. \cite{MtGoxDoubleSpendingDecker}).
\end{enumerate}
\vskip-5pt

Here is the crux of the problem.
The creator of bitcoin software Satoshi Nakamoto
has opted for a solution of extreme elegance and simplicity,
one single (longest chain) rule which regulates both things.
This is neat.

However in fact it is possible to see that this is rather a mistake.
In principle there is NO REASON why the same mechanism
should be used to solve both problems.
On the contrary.
This violates
one of the most fundamental principles of security engineering: 
the principle of {\em Least Common Mechanism} [Saltzer and Schroeder 1975],
cf. also \cite{ComputerSecurityPrinciples}.
One single solution rarely serves well two distinct problems
equally well without any problems.

We need to observe that the transactions are generated
at every second.
Blocks are generated every 10 minutes.
In bitcoin the
{\bf
receiver of money is
kept in the state of incertitude\footnote{
\label{FootIncertitideWait10Hours}
This period of incertitude is even much longer for large transactions:
for example we wish to withdraw some
1 million dollars which is currently about 2200 bitcoins,
we should probably wait for some 100 blocks or 10 hours.
Otherwise
it may be 
profitable to run 
the double spending attack
which we study later in Fig.
\ref{AltChainProfitDoubleSpendAttack}, page \pageref{AltChainProfitDoubleSpendAttack}.}
for far too long}
 and this
{\bf with no apparent reason}.
The current bitcoin currency produces
a situation of discomfort and dependency
or peculiar sort.
Miners who represent some wealthy people in the bitcoin network,
are in a privileged position.
Their business of making new bitcoins
has negative consequences on the smooth processing of transactions.
It is a source of instability
which makes people wait
for their transactions to be approved
for far too long time$^{\ref{FootIncertitideWait10Hours}}$.
This violates also another very widely accepted principle
of security engineering: 
the principle of {\em Network Neutrality}.
We claim that it should be possible
to design a better mechanism in bitcoin,
which question we will study later in Section \ref{DoubleSpendingSolutionCharacteristics}.


\vskip-5pt
\vskip-5pt
\subsection{Consensus Building}
\label{LongestChainaConsensus}
\vskip-5pt

The common history in bitcoin is agreed by
a certain type of democratic consensus.
In the initial period of bitcoin history people mined with CPUs
and the consensus was essentially of type {\bf one CPU one vote}.
However nowadays people mine bitcoins with ASICs
which are roughly ten thousand times more powerful than CPUs
(more precisely they consume ten thousand times less energy, cf. \cite{MiningUnreasonable}).
Bitcoin miners need now to invest
thousands of dollars to buy specialized devices
and be at the mercy of the
very few suppliers of such devices which tend
NOT to deliver them to customers
who paid them for extended periods of time,
see Appendix of \cite{MiningSubversive}.
It appears that the democratic base of bitcoin has shrunk
and the number of active miners has decreased.

Nevertheless in spite of these entry barriers
the income from mining remains
essentially proportional to the hashing
power contributed to the network
(in fact not always, see \cite{MiningSubversive,BitcoinBrokenGrowingSelfishPoolStrategy}).
This is good news:
malicious network participants which do not represent
a majority of the hash power are expected to
have difficult time trying to influence
the decisions of the whole bitcoin network.

In a first approximation
it appears that the Longest Chain Rule works well
and solves the problem of producing consensus
in a very elegant way.
Moreover it allows asynchronous operation:
the consensus can propagate slowly in the network.
In practice it is a bit different.
In this paper we are going to challenge this traditional
wisdom of bitcoin.
In Section
\ref{LongestChainRuleDoubleSpendingAttacks}
and in later Sections
\ref{UNODestructionSection}
and
\ref{DogeCoinandLitecoin}
we are going to argument that more or less anyone
can manipulate virtual currencies for profit.

In fact 
we are not even sure if
the Longest Chain Rule is likely to be applied by miners
as claimed.
This is what we are going to examine first.

\vskip-5pt
\vskip-5pt
\subsection{The Longest Chain Rule - Reality or Fiction}
\label{LongestChainMajorityIsItTrue}
\vskip-5pt

This rule is taken for granted and it seems to work. However.
We can easily imagine that it will be otherwise.
There are several reasons why the reality
could be different:

\begin{enumerate}
\item
We already have a heterogenous base of software which runs bitcoin
and the protocols are on occasions updated or refined with new rules.
On occasions there will be some bugs or ambiguities.
This has already happened in March 2013.
There were two major versions of the block chain.
For 6 hours nobody was quite sure which version should be considered as correct, both were correct.
The problem was solved because the majority of miners could
be convinced to support one version.
Apparently the only thing which could solve this crisis
was human intervention and influence of a number of key people
in the community, see \cite{blockchainfork}.

\item
Open communities tend to aggregate into clusters.
These clusters could produce distinct
major software distributions of bitcoin,
similar to major distributions of Linux
which will make some conflicting choices
and will not necessarily agree on how decisions can be made.
For example because they promote their brand name and some additional business interests.
We already observe a tendency to set up authoritative bitcoin
authorities on the Internet such as \url{blockchain.info}.
Software developers are tempted to rely on these web services
rather than work in a more ``chaotic'' fully distributed asynchronous way.
People can decide to trust a well-established web service rather than
network broadcasts which could be manipulated by an attacker.

\item
This is facilitated by the fact that bitcoin
community produces a lot of open source software
and free community web services.

\item
It is also facilitated by the fact that the great majority of miners
mine in pools. Moreover they tend to ``flock to the biggest pools''
\cite{MiningSubversive,RosenfeldPoolRewardMethods}.
Just one pool reportedly based in Ukraine was recently controlling
some 45 $\%$ of the whole bitcoin network,
see Table 2 in \cite{MiningSubversive}.

The pool managers and not individual miners are those who can decide
which blocks are mined and which transactions will be accepted.
The software run by pools is not open source and not the same as run by
ordinary bitcoin users.
In particular they can adopt various
versions or exceptions from The Longest Chain Rule.
In Section \ref{HiddenAttacks}
we will propose further new ways for pool managers
to attack the bitcoin network.

\item
More importantly participants could suspect or resist
an attack by a powerful entity
(which thing allows effectively to cancel past transactions and double spend)
and they will prefer to stick to what their trusted authority says.

\item
Even more importantly these sub-communities of bitcoin enthusiasts
will also contain professional for-profit bitcoin miners
who can be very influential because for example they will
be sponsoring the community.
Their interest will be that their chain wins
because they simply need to pay the electricity bill for it.
If another chain wins, they have lost some money.
\end{enumerate}

We see that sooner or later we could have a situation
in the bitcoin community such that people
could agree to disagree.
If one group have spent some money on electricity on one version of the chain,
their interest will be to over-invest now in order to win the race.
Over-investment is possible because there is always spare capacity in bitcoin
mining which has been switched off because it is no longer very profitable.
However the possibility to earn money also for previous blocks
which money would otherwise been lost can make
some operations profitable again.
%
Such mechanisms could also be used
to cancel large volumes of transactions
and commit large scale financial fraud,
possibly in combination with cyber attacks.
This can be done in such a way that nobody is to blame
and everything seems normal following the Longest Chain Rule.
{\bf Losses will be blamed on users} not being careful enough
or patient enough
to confirm their transactions.

\vskip-5pt
\vskip-5pt
\subsection{Summary: Operation in Normal Networks}
\label{LongestChainVsNormalFastNetworks}
\vskip-5pt

We have seen that bitcoin has been designed to operate
in {\bf extreme} network conditions.
Most probably bitcoin could operate in North Korea or in Syria torn by
war operations, or in countries in which the government is trying to ban bitcoin
or is very heavily limiting the access of the citizens
to fast computer networks such as the Internet.

In contrast in the real life,
the propagation in the global network of bitcoin client applications
is quite fast:
the median time until a node receives a block is 6.5 seconds
whereas the average time is 12.6 seconds,
see \cite{ForksPropagationDecker,ForksPropagationDecker2}.
The main claim in this paper is that in normal (fast) networks
the Longest Chain Rule is not only not very useful,
but in fact it is sort of {\bf toxic}.
It leads to increased risks of attacks
or just unnecessary instability and
overall slower financial transactions
\cite{FTACourtoisVideoHorseCarriage,FasterBitcoin}.

Before we consider how to reform or replace
the Longest Chain Rule, we look at the questions of monetary
policy in bitcoin. Later we will discover that both questions are related,
because deflationary policies erode the income of honest miners
which in turn increases the risk of for-profit block chain manipulation attacks,
cf. Sections
\ref{UNODestructionSection}, \ref{DogeCoinandLitecoin} and \ref{FutureBitcoin}.

\newpage

\section{Deflationary Coins vs. Growth Coins}
\label{DeflationaryCoinsGrowthCoinsTheorySketch}


It is possible to classify crypto currencies in two families:
\vskip-6pt
\vskip-6pt
\begin{enumerate}
\item
{\bf Deflationary Currencies} in which the monetary supply is fixed\footnote{These are also called Log Coins in \cite{WiredKrollLMustKeepCreating}
which is not quite correct because the monetary supply in bitcoin does not grow logarithmically.}.
For example in bitcoin and Litecoin.
\item
{\bf Growth Currencies} in which the monetary supply is allowed to grow at a steady pace,
for example in Dogecoin.
\end{enumerate}
\vskip-6pt

\noindent
Bitcoin belongs to the first family.
This is quite unfortunate.
In \cite{WiredKrollLMustKeepCreating} we read:

\vskip-6pt
\vskip-6pt
\begin{quote}
"This limited-supply issue is the most common argument against the viability of the new currency.
You read it so often on the web. It comes up time and again".
\end{quote}

In the following three subsections we look at the main arguments
why a fixed monetary supply in bitcoin is heavily criticized.
We need to examine the following four questions:

\vskip-5pt
\vskip-5pt
\begin{enumerate}
\item
comparison to gold, other currencies and commodities
\item
volatility
\item
miner reward vs. fees
\item
competition with other cryptocurrencies.
\end{enumerate}
\vskip-5pt

\subsection{Comparison to Gold Other Currencies and Commodities}

Bitcoin is frequently compared to gold
and The Economist called it ``Digital Gold''
in April 2013, cf. \cite{TheEconomistDigitalGold}.
However actually gold belongs to the second category:
the worldwide supply of gold grows every year due to gold mining
and other factors,
with a yearly increase of the quantity of gold by some 0.5 - 1 $\%$.
In fact when bitcoin mandates a fixed monetary supply,
ignoring the growth of the bitcoin economy,
arguably we enter
an area of misplaced ideology and monetary non-sense.
If the economy grows substantially, the monetary supply should probably
follow or the currency is not going to be able
to make a correct connection between the past and the future.
It is widely believed that business does not like instability.
It is well known in traditional economics that deflation discourages
spending, creates an expectation that prices would further decrease with no apparent limit.

To the best of our knowledge, no currency and no commodity has ever had
in the human history a totally fixed quantity in circulation.
This is clearly an artificial property which makes that bitcoin
is like no other currency and like no other commodity.
This is expected to have very serious consequences and could be potentially
fatal to bitcoin in the long run.

\subsection{The Question of Volatility}

Here the argument is that basically deflationary currencies are expected to have
{\bf higher volatility}
due to the existence of people holding large balances for speculation.
In \cite{SamsMarginal} Robert Sams claims that deflationary currencies lead to a
``toxic amount of exchange rate volatility''
providing yet another reason for users to
``run away'' from using these currencies as a medium of exchange.



This is actually not so obvious and requires some explanation.
We see one good reason for that.
In a recent report published by Bank of England \cite{BOECryptEconomicsReport2014}, we
read that one of the key problems of bitcoin
is that the supply of money does NOT respond to variations in demand.
As a consequence they predict
"welfare-destroying volatility in economic activity".
%
They point out that
"growth rate of the currency supply could be adjusted to
respond to transaction volumes in (close to) real time", cf. \cite{BOECryptEconomicsReport2014}.

\subsection{Miner Reward}

We need to recognize the role of miners in digital currencies.
In \cite{WiredKrollLMustKeepCreating} Sams writes:

"The amount of capital collectively burned hashing
fixes the capital outlay required of an attacker
to obtain enough hashing power to have a meaningful chance
of orchestrating a successful double-spend attack 
on the system [...]
The mitigation of this risk is valuable, [...]"

\label{KrollKeepHighIncentiveForMiners}
Now the deflationary currencies do with time decrease the reward for miners.
This is highly problematic. In \cite{WiredKrollLMustKeepCreating}
citing J. Kroll from Princeton university we read:
 "If you take this away, there will be no incentive for people
 to keep contributing processing power to the system [...] 
 "If the miner reward goes to zero, people will stop investing in miners,".
Then the hash rate is likely to decrease and bitcoin will no longer benefit from a
protection against double spending attacks,
cf. Section \ref{LongestChainRuleDoubleSpendingAttacks}.

Moreover Kroll explicitly says that the problem is NOT solved by transaction fees and says:
[...] 
You have to enforce some sort of standard payment to the miners, 
[...] change the system so that it keeps creating bitcoins.
In a paper presented at WEIS 2013 and co-authored by Kroll \cite{KrollEconomicsRulesCouldBeChangedAtAnyTime}.
this is presented as a clear dilemma, either break the monetary policy or increase the fees:

\begin{quote}
The only way to preserve the system's health will be to change the rules,
most likely either by maintaining mining rewards at a level higher than originally envisioned,
or making transaction fees mandatory.
\end{quote}

\subsection{Problems With Increasing The Fees}
\label{IncreasingFeesArgument}

The question of whether higher fees could be effectively mandated 
in the current bitcoin
is discussed by Kroll in Sections 4.2 and 6.2 of \cite{KrollEconomicsRulesCouldBeChangedAtAnyTime}.

Now it is possible to see that it would be a very bad idea to increase the fees.
This is brilliantly explained by Robert Sams in \cite{SamsMarginal}.
The argument is that basically sooner
or later ``deflationary currencies'' and ``growth currencies'' will be in competition.
Then all the other things being more or less in equilibrium,
in deflationary currencies most of the profit from appreciation will be received by holders of
current coins through their appreciation.
Therefore less profit will be made by miners in these currencies.
However miners control the network and they will impose higher fees.
In contrast in growth coins,
there will be comparatively more seignorage profit
and it will be spent on hashing.
Miners will make good profits and transaction fees will be lower.
Thus year after year people
will prefer growth currencies due to lower transaction fees.

Overall we see that this is crucial question of how the cost of
the infrastructure necessary for the maintain a digital currency
is split between new adopters (which pay for it through appreciation)
and users (which pay through transaction fees).
It is obvious that there exists an optimal equilibrium between these two
sources of income, and that there is no reason why the creator of bitcoin
would get it right, some adjustments will be necessary in the future.

\subsection{The Appreciation Argument}

There is yet another argument:
it is possible to believe that bitcoin will appreciate
so much that halving the reward every 4 years
will be absorbed by an increase in bitcoin price.
This means an extreme amount of deflation (double every 4 years)
making it tempting to hoard bitcoins,
which further decreases the amount of bitcoins
in actual usage and makes people hoard bitcoins even more.


We claim that this is very unlikely.
This is mainly because the digital economy is not expected to double every 4 years
and even less it ie expected to grow by sudden jumps at the boundaries of the intervals
arbitrarily decided by the creator of bitcoin.
We refer to Part 3 of \cite{MiningUnreasonable},
Sections \ref{UNODestructionSection}, \ref{DogeCoinandLitecoin} and \ref{FutureBitcoin}
for further discussion and concrete examples of predicted and actual
devastating effects of sudden jumps in the miner reward.


\medskip
\medskip
\subsection{On Self-Defeating Monetary Policies and Alt-Coins}
\label{sec:CompetingCryptoCurrencies}

The bitcoin monetary policy is challenged
by the very existence of alternative crypto currencies.
In \cite{CompetingCryptoCurrencies} we read:

\begin{quote}
[...]
the constant volume of Bitcoins faces an unlimited number
of alternative crypto-currencies and,
therefore, an unlimited number of alternative coins.
[...]
Clearly, an investor may move his assets from Bitcoins to a competing currency,
thereby freely moving in a space with
an unlimited number of coins.
\end{quote}

It is easy to see that the bitcoin restricted monetary supply
is a {\bf self-defeating} property:
if bitcoin is limiting the monetary supply beyond what is `reasonable',
and if as a result of this bitcoin economy suffers from excessive deflation,
bitcoin adopters are likely to circumvent this limitation by using alternative coins.
This can erode the dominant position of bitcoin.

\medskip
\medskip
\subsection{The Future}

Can Bitcoin change its reward rules and the monetary policy given
that fixed monetary supply
is problematic as shown above?
User DeathAndTaxes, a highly respected frequent
contributor in \url{bitcointalk.org} forum wrote on 10 May 2014:

\vskip-0pt
\vskip-0pt
\begin{quote}
"The bitcoin protocol reward is not going to be changed.  Period."
\end{quote}
\vskip-0pt

Source: \\
\url{https://bitcointalk.org/index.php?topic=600436.msg6657579#msg6657579}

\subsection{Who Can Change The Bitcoin Monetary Policy?}

There is an interesting additional question who has the power to change the bitcoin monetary policy,
is it the majority of miners, ordinary bitcoin users, bitcoin developers,
or is it that all must agree?
This is a very complex and highly controversial question on which opinions differ rally a lot,
see Sections
\ref{PowerControversyFelterSirer} through \ref{MinersHavePowerYes2LaterCombined}
and \cite{CornellVotingPowerAmazingAntiEthicalClaims,FeltenMinersProbablyCanChangeMonPol}.

\newpage

\newpage

\section{Is The Longest Chain Rule Helping The Criminals?}
\label{LongestChainRuleDoubleSpendingAttacks}

This section is the central section in this paper.
We are going to show a simple attack which allows double spending.
The attack is not very complicated and we do not claim it is entirely new.

Our attack could be called a 51 $\%$ attack
however we avoid this name because it is very highly misleading.
There are many different things which can be done with 51 $\%$ of computing power,
(for example to run a mining cartel \cite{MiningSubversive}
or/and cancel/undo any chosen subset of past transactions)
and many very different attacks
have historically been called a 51 $\%$ attack.

We are in general under the impression that a 51 $\%$ attack
is about holding more than 50 $\%$ of the hash power
kind of permanently or for a longer period of time,
while our attacks are rapid short-term attacks 
cf. Fig. \ref{AltChainProfitDoubleSpendAttack} page \pageref{AltChainProfitDoubleSpendAttack}.

\subsection{Common Misconceptions About 51$\%$ Attacks}
\label{LongestChainRuleDoubleSpendingAttacksMisconceptions}
\label{LongestChainRuleDoubleSpendingAttacksReasonWhySoBadlyUnderstood}

There many reasons why such attacks
has not been properly
understood and studied before
in bitcoin community and
in the bitcoin literature.

\begin{enumerate}
\item
There is a large variety of attacks which could be or have been called a 51 $\%$ attack.
Opinions or statement
which might be true for some of these attacks are simply not true for other attacks.
This creates a lot of confusion in the bitcoin community.

\item
Great majority of people who discuss bitcoin
make an implicit {\bf wrong} assumption
about a static nature of threats and attacks about bitcoin.

\item
We hear about 51 $\%$ attack etc,
entities who own or control
51 $\%$  of hash power
and it seems that only
incredibly powerful or very wealthy
entities \cite{CoinDeskIgnorantPaper51,AntonopoulosIgnorant51PcLAJan2014}
could execute such attacks
and that they are
"so amazingly cost-prohibitive to perform that we're basically talking about a government focusing the full power of every top-secret ridiculously expensive supercomputer", cf. \cite{PerryPlainenglishBitcoinAttacksVeryNaiveCostProhibitive}
%

\item
Many commentators stress that
51 $\%$ attack are only {\em theoretical} attacks,
cf. \cite{LiteCoinPoolNear51Percent,AntonopoulosIgnorant51PcLAJan2014},
try to convince us to ``stop worrying''
e.g. \cite{PerryPlainenglishBitcoinAttacksVeryNaiveCostProhibitive}.
The official bitcoin wiki, does even consider that there are any real problems in bitcoin.
The section about 51$\%$ attacks does NOT even get into the part entitled "Might be a problem".
It appears in the following part entitled "Probably not a problem", cf. \cite{BitcoinWiki}
which many people would maybe not read, why bother if it probably is not a problem?

\item
In the original paper Satoshi
have portrayed
"a greedy attacker" being
"able to assemble more CPU power than all the honest nodes",
see Section 6 of Satoshi paper \cite{SatoshiPaper}.
The attacker is also portrayed as having considerable "wealth"
which he would endanger by engaging in the attack.
It is clearly suggested that the attack would have little to gain
and a lot to lose from 
being dishonest.

\item
Satoshi has invented a term "CPU power" and always explicitly
states the principle of "one-CPU-one-vote".
In reality
nowadays it is rather "one-ASIC-one-vote"
and in the future it could be something yet different.
A reasonable term is {\bf "hash power"}\footnote{
It can be measured in GH/s
(Giga Hashes per second)
which notion is almost never 
properly defined 
in a non-ambiguous way:
one hash per second is capacity to hash one block header,
which is two applications of SHA256
and which in turn is three applications
of the underlying block cipher.
In repeated hashing some of these computations
do not have to be done, this is why we speak about
"capacity to hash" rather than hashing,
see \cite{MiningUnreasonable}
for a detailed analysis of this problem.
}

\item
In general a very common but also one of the
{\bf most serious mistakes}
is to claim that
51$\%$ attacks occur when the attacker
{\bf owns} or is in {\bf possession} of
51$\%$ of all the hash power.

This mistake is committed again and again by major Bitcoin experts and evangelists,
cf. for example \cite{SatoshiPaper,CoinDeskIgnorantPaper51,PerryPlainenglishBitcoinAttacksVeryNaiveCostProhibitive} 
to cite just a few.
%
%
The official bitcoin wiki \cite{BitcoinWiki}
has a subsection with this super highly misleading title:
"Attacker has a lot of computing power".
Quite happily just below they correct it and say it is rather about
temporary control not ownership\footnote{
They explain that the exact scenario
is when he "controls more than $50\%$ of the network's computing power"
and they make it clear it can be temporary: "for the time that he is in control".
However almost to make things worse again,
this official wiki at numerous places refers to another article about Bitcoin attacks
written for more general audience
\cite{PerryPlainenglishBitcoinAttacksVeryNaiveCostProhibitive}
in which we see the repetition of the basic mistake to consider that
$51\%$ attacks are "so amazingly cost-prohibitive to perform".
}

Nevertheless, the same confusion was made more recently
by Cornell researchers in \cite{CornellTimeForHardForkAvoid51GHashWithholdingEtc}
which clearly very badly confuse
between A) having 51 $\%$ of the mining power
and B) launching a 51 $\%$ attack
trying to convince the reader that A does not have to imply B
while the real problem is that B can be executed without A.

Again attacks are presented as being exclusively about
powerful entities who "can turn dishonest" all of the sudden,
\cite{CornellTimeForHardForkAvoid51GHashWithholdingEtc}.
They fail to see that the key problem is
the control (not ownership) of hash power
for the purpose of mining blocks,
and this can be {\bf a lot} easier and cheaper.

\item
Less people admit
that the attacker could indeed
be one single malicious pool
which gathers more than
51 $\%$  of hash power under his sole control
(controlling but not owning hash power).

It is worth noting that this
has already happened at least once
in both Bitcoin
\cite{CornellTimeForHardForkAvoid51GHashWithholdingEtc}
and Litecoin
\cite{LiteCoinPoolNear51Percent}.
However then it was claimed that pools
reaching more than 51 $\%$
would have no reason to execute
any sort of attack. 

\item
Another serious mistake is to consider that "control" is exclusive.
For example in the Abstract of his paper Satoshi writes:
"As long as a majority of CPU power is controlled by nodes that are not cooperating to
attack the network they'll
[...]
outpace attackers".
This is not correct in general.
The key point is that {\bf control is NOT exclusive},
both the miners and the attacker can have some control on the mining process.
So "a majority of CPU power is controlled by nodes" as Satoshi says and
also at the same time it could be controlled by the attacker
in a more or less subtle and more or less invasive ways,
cf. Section \ref{TechnicalityFurtherManipulation}.

\item
Many people stress that that 51$\%$ attacks, and for example double spending events
would be visible to anyone to see
on the public blockchain \cite{CoinDeskIgnorantPaper51}.
This is simply not true, the blockchains does NOT record double spending events,
it rather hides them and would show only on transaction our of two,
cf. also \cite{MtGoxDoubleSpendingDecker}.


\item
In reality the notion of a 51 $\%$ attack
takes a very different meaning in a cloud computing world:
the attacker does not need to own a lot of computing power,
he can rent it for a short time,
and then 51 $\%$ attack can have a surprisingly low cost.

\item
Alternatively an attacker could also trick miners to help
him to execute the attack without their knowledge and consent
(man in the middle attacks).

This is particularly easy with mining pools: the attacker
just needs to compromise extremely few web servers used by
tens of thousands of individual miners
and he can command very substantial hash power
without owning any of it.
At this moment
less than 10 pools control over two-thirds 
of all the hash power, cf.
\cite{ToddBitcoinEcosystemWillMaybeBreakDown,MiningSubversive}.

\item
It is important to remember that not only Satoshi did not predict ASIC mining
and mining pools, but also he did {\bf NOT specify} bitcoin
fully in the sense that the mining pools typically
use the Stratum protocol \cite{StratumProtocol}.
which was specified in 2012 
and which at some moment took
{\bf an important strategic decision}
which is clearly stated in documented in \cite{StratumProtocol}
in order to move the choice and the control
of which transactions are included
in a block from miners to the pool managers,
see \cite{StratumProtocol}.

This decision {\bf broke the bitcoin peer network}
because miners do no longer have any incentive whatsoever
\footnote{This decision also has definitely infringed
on the initial intentions of Satoshi explicitly
stated in Section 6 of his paper \cite{SatoshiPaper}
where he explains that the
fact that a block provides a monetary reward for the "creator of the block"
is something which "adds an incentive for nodes to support the network".
This incentive is now broken.}
to support
this network by running peer nodes,
and the bitcoin network is now very seriously declining cf.
\cite{NumberOfReachableNodesIsDroppingRecently}.

\item
In fact,
even if large pools had only 10 $\%$ of hash power each,
we should see reasons to worry:
it would be sufficient to hack just
5 pool manager servers in order to be able to
execute double spending attacks.

\item
Nobody has yet stated under which
exact assumption bitcoin is expected to be secure
and there is a lot of ambiguity in this space.
Knowing the assumption is crucial because
if we have stated our assumption
and bitcoin is later shown to be broken insecure,
we can blame {\bf either} the real world which does not
satisfy our assumption,
or the designers and engineers of bitcoin
which have not been able to design
a secure system based on this assumption.
In other worlds we could 
determine without ambiguity who is to blame.
In this respect Satoshi shows a bad 
example
of not being clear about what his assumption is
and yet explicitly several times claiming
that his system is secure:

\vskip-1pt
\vskip-1pt
\begin{enumerate}
\item[A.]
For example in the abstract of his paper \cite{SatoshiPaper}
Satoshi
says that he assumes that
"majority [...] are not cooperating to attack the network".
Here Satoshi claims the system is secure under this assumption,
which security claim is {\bf not true}
as people can easily be part of
an attack without cooperating (as already explained above).

\item[B.]
Now in the conclusion of his paper Satoshi again claims that the system
is secure if
"honest nodes control a majority of CPU power".
which is a very different and STRONGER assumption than A. above:
nodes could be not honest and deviate from the protocol for fun or for profit
in a variety of creative ways without "cooperating" with any attacker.

Does this stronger assumption make that bitcoin becomes secure?
Of course not, the security result claimed by Satoshi is wrong again
if you take it literally:
even if honest nodes control a majority of hash power,
because the control is not exclusive, bitcoin can still be attacked.

\end{enumerate}
\vskip-3pt

\item
It is {\bf nonsensical} to claim that the attacker would
prefer to behave honestly,
and that it is "more profitable to play by the rules" \cite{SatoshiPaper}.

This is claimed by Satoshi on the grounds that the attacker
should be able to "generate new coins"
which would be an honest way to use his hash power,
see 
Section 6 of \cite{SatoshiPaper}.
Many other authors repeat this mistake,
for example in \cite{CornellTimeForHardForkAvoid51GHashWithholdingEtc}
we read about "miners which may "hold 49 $\%$ of the [mining] revenue".
%


\item\label{PointFakeChoice}
In reality, in almost all\footnote{
With exception of attacks described in \cite{RedirectedMiningStealingCoinsAugust2014}.
}
 bitcoin mining scenarios known to us,
the attacker does NOT control the money from mining:
he does NOT have the private keys used for mining.
This is because
{\bf the whole process of mining requires exclusively the public keys}.

It would simply be an unnecessary mistake for any miner or for any mining pool
to have the private keys around to be stolen by the attacker which targets the mining process.
Therefore the attacker typically
does NOT have an honest option at all\footnote{
In contrast Satoshi have claimed that he always has such an option,
in Section 6 of \cite{SatoshiPaper} we read:
"he would have to choose between using it to defraud people
by stealing back his payments, or using it to generate new coins."}.

\item
The notion of  51 $\%$ attacks is also very highly misleading because presenting
the hash power as a percentage figure does NOT make sense
because the hash rate is measured at two different moments.
Therefore the proportion of hash power used in attack is NOT
a number between 0 and 100 $\%$.
It can easily be larger than 100 $\%$.

In fact the relative hash power
at one moment can be easily of the order of 500$\%$
and many times bigger than a few minutes later,
see Fig. \ref{DOGEIncredible5xIncreaseDogecoin} on page \pageref{DOGEIncredible5xIncreaseDogecoin}
for an actual historical example.

\item
It was also wrongly assumed that the bitcoin adopters are more or less the same as miners,
they own the devices and the computing power cannot change hands very quickly.

\item
It is in general not sufficient to trust 
the pools not to be malicious.

Attacks could be executed without
the knowledge and consent of these companies by a single rogue developer.

\item
Many bitcoin adopters did not anticipate that in the future bitcoin will have to compete
with other crypto currencies and that hash power could instantly be moved from
one crypto currency to another.

\item
Attacks could also operate
through re-direction of hash power in bulk to another pool,
see for example later Sections
\ref{HiddenAttackAcrossCurrencies}
and
\ref{TechnicalityFurtherManipulation}.

\item
People have wrongly assumed that bitcoin
achieves very substantial computing power which no one can match,
which is still the case today
however it is highly problematic to see if this will hold in the future.

\item
Many people
did not predict that an increasing fraction of all available computing power
is going to exist in the form of rented cloud miners which further facilitates the attacks.

This is due to several factors.
Investing in wholly owned mining equipment has been excessively risky.
this is both due to the impossibility to know if and when miners will effectively be delivered
(cf. Appendix of \cite{MiningSubversive} and Section \ref{ASICDeliveryNightmare})
and due to the price volatility.
In contrast investing in rented capacity could be nearly risk-free.

Another reason is that some large investors may have over-invested in
large bitcoin mining farms consuming many Megawatts of electricity
(we know from the press that such facilities have been built in Sweden, Hong Kong, USA, etc..) 
and now they want to rent some parts of it in order to get immediate cashflow and return on their investment.

\item
Furthermore rented cloud miners can be seen as a method to
absolve owners of hash power from any legal responsibility. 

This does in addition lead to the possibility
of running for-profit attacks with cooperating peers
who may or not be aware of participating in an attack,
see Section \ref{UnthinkableDoubleSpendingService}.

\item
There is some sort of intuitive understanding in the bitcoin community that
the Longest Chain Rule solves all problems in this space,
and there is simply no problem of this sort,
and if there is,
people naively believe that it is not very serious.
In other terms nobody wants to admit that
the brilliant creator(s) of bitcoin
could have created a system which has
serious security problems.

\item
For example many authors claim that the problem has already been fixed:
and that the fix is to wait for 6 confirmations,
cf. \cite{PerryPlainenglishBitcoinAttacksVeryNaiveCostProhibitive}.
More generally it is frequently claimed that the probability of reverting a transaction in a block
decreases exponentially with the number of blocks $t$ mined on the top of the current block
%
cf. \cite{SatoshiPaper}.
In fact if a lot of money is at stake in a large transaction
(or in many small transactions) it is possible to see that
a larger attack could be mounted.
According to \cite{DelegatedProofStakeFastTxVs120}
core developers require 120 blocks (about 1 day\footnote{
So it is in fact faster to take a plane to Switzerland,
withdraw money from a bank, and travel back,
than to use bitcoins to withdraw larger sums of money,
cf. \cite{FTACourtoisVideoHorseCarriage,FasterBitcoin}.})
before they consider the network sufficiently protected from the potential of a longer attack-chain.
In general as the money at stake involved in each block is likely to grow in the future,
the risk will also increase\footnote{
Later we are going to see that 51 $\%$ attacks
will get worse with time
due to the build-in monetary policy in bitcoin
(money at risk grows in comparison to the cost of attack)
and moreover there will be sudden transitions
because the monetary policy
mandates sudden jumps in the miner reward
(cf. also Part 3 in \cite{MiningUnreasonable}).
}
 and we believe that
"no amount of confirmations" can fix such problems,
citing \cite{BitcoinWiki2}.
See also \cite{BlockDiscardingNetworkSuperiority}
and Section \ref{LongestChainRuleDoubleSpendingAttacks}.
\end{enumerate}

Overall we see that 51 $\%$ attacks are
a huge problem and cannot be easily dismissed.


\subsection{The Basic Attack}
\label{LongestChainRuleDoubleSpendingAttacksBasicAttack}

Our basic attack is self-explanatory,
some attacker produces a fork in order to
cancel some transaction[s] by producing a longer chain
in a fixed interval of time,
see Fig. \ref{AltChainProfitDoubleSpendAttack} below. 

The attack clearly can be profitable.
The question of actual feasibility of this attack
is a complex one,
it depends on many factors and we will
amply study this and related
questions later throughout in this paper.

\begin{figure}[!h]
\centering
\vskip-4pt
\vskip-4pt
\begin{center}
\hskip-7pt
\hskip-7pt
\includegraphics*[width=5in,height=1.6in,bb=0pt 0pt 748pt 240pt]{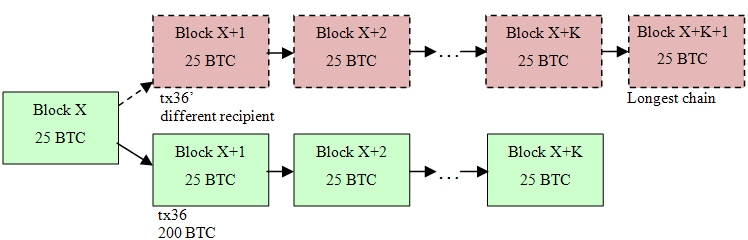}
\end{center}
\vskip-2pt
\caption{
A simple method to commit double spending.
The attacker tries
to produce the second chain of blocks
in order to modify the recipient
of some large transaction(s) he has generated himself.
Arguably
and under the right conditions,
this can be quite easy to achieve.
The attack is clearly profitable
and the only problem is {\bf the timing}.
To produce these blocks on time
requires one to temporarily ``command'' 
very substantial computing power
such for example 51 $\%$ of the current capacity
or higher.
It is totally incorrect to believe that this requires the attacker
to be very powerful such as owning 51 $\%$ of the bitcoin hash power
\cite{SatoshiPaper}.
This needs only to be done only for a very short time,
like less than 1 hour,
for example through redirection (man-in-the-middle attack)
of hash power which is in the physical possession
of other miners but under ``logical''
control of extremely few
pool manager servers.
}
\label{AltChainProfitDoubleSpendAttack}
\end{figure}
\vskip-2pt
\vskip-2pt

In the following sections we are going to analyse the risks
which result form this and similar attacks.

\vskip-7pt
\vskip-7pt
\subsection{Large vs. Smaller Transactions}
\label{LongestChainRuleDoubleSpendingAttacksDiscussion1}
\vskip-6pt

Our attack does NOT limit to defraud people who would accept a single large payment in exchange of goods
or another quantity of a virtual currency (mixing services, exchanges, some sorts of shares).
The attacker (for example a bitcoin exchange or a bitcoin lottery)
can in the same way issue a large number of smaller transactions
and cancel all of them simultaneously in the same way
and at exactly the same cost.


\vskip-7pt
\vskip-7pt
\subsection{Feasibility Discussion}
\label{LongestChainRuleDoubleSpendingAttacksDiscussion}
\label{LongestChainRuleDoubleSpendingAttacksDiscussion2}
\vskip-6pt


The attacker does NOT need to be very powerful, on the contrary.
The most shocking discovery is that 
{\bf anyone} can commit such fraud and steal money.
They just need to rent some hashing power
from a cloud hashing provider.
Bitcoin software does not know a notion of a double spending attack
and if it occurs possibly nobody would notice:
only transactions in the official dominating branch of the blockchain
are recorded in the current bitcoin network, cf.
\cite{MtGoxDoubleSpendingDecker}.
It may also be difficult to claim that something wrong happened:
one may consider that this is how bitcoin works
and the attacker has not done anything wrong.

In a competitive market they do not need to pay a lot for this.
Not much more than 25 BTC per block
(this is because miners do not mine at a loss,
the inherent cost of mining per block should be less than 25 BTC).
The attacker just needs to temporarily displace the hashing power from other crypto currencies
for a very short period of time which is easy to achieve
by paying a small premium over the market price.

There is another very serious possibility,
that the spare hash power could also be obtained from older miner devices
which have been switched off because they are no longer profitable
(or a combination of old and new devices).
However they may be profitable for criminals
able to generate an additional income from attacks.
Given the fact that the hash rate
increases steadily, cf. Fig. \ref{BitcoinGraph1YHashRate},
it is quite possible to imagine that the hash power
which has been switched off is very substantial
and comparable in size to the active hash power.


\vskip-7pt
\vskip-7pt
\subsection*{How to Achieve 500 $\%$ or More}
\label{LongestChainRuleDoubleSpendingAttacksDiscussion3}
\vskip-6pt

There is yet another way to execute such attacks:
to offer a large number of miners a small incentive
(as a premium over the market price)
to go mine for another crypto currency,
{\bf before} the attack  begins.
This can lead to massive displacement of hash power before the attack starts.
Then at the moment when block X+1 is mined
following the notations of Fig. \ref{AltChainProfitDoubleSpendAttack},
the double spending attack costs less.
The hash rate goes down dramatically at the very beginning of the attack, and raises back again.
In this way it is possible also to achieve
500 $\%$ hash power or more.
More precisely the attacker can for example
re-do this block $X+1$, and potentially few more blocks
with  hash power which could be literally 500 $\%$
compared to the (reduced) hash power
with which first block $X+1$
was initially mined.
Now the attacker is going to modify the recipients of one or many transactions
included in this block to cancel his own transactions\footnote{
He can also cancel transactions of many other people
with  double spending as a service \url{bitundo.com},
see Section
\ref{UnthinkableDoubleSpendingService}}.

Further advanced attacks scenarios
with malicious pool managers and
which can easily be combined
with this
preliminary displacement of hash power are
proposed and studied in Section
\ref{HiddenAttackAcrossCurrencies}.

\vskip-7pt
\vskip-7pt
\subsection{The Question of Dominance}
\label{LongestChainRuleDoubleSpendingAttacksDiscussion4}
\vskip-6pt

It is important to understand
that what we present in Fig. \ref{AltChainProfitDoubleSpendAttack}
is already feasible to execute today for nearly anyone,
not only for rich and powerful attackers.
Then as we advance in time, such attacks are expected to become easier.

At this moment bitcoin is a dominating crypto currency:
its hash power is substantially larger than for other crypto currencies combined.
It appears that bitcoin could claim to be a sort of natural monopoly:
it is able to monopolize the market and its competitors find it hard to compete.

Now the attack will become particularly easy when bitcoin
ceases to be a dominant crypto currency.
At this moment the attacker needs for example to hack some (very few) pool manager servers
in order to execute the attack.
But when there is plenty of hash power available to rent outside of bitcoin,
the attacker will be able to execute the attack without doing anything illegal
(except possible legal consequences of canceling some bitcoin transactions).
At this moment it is quite easy to execute double spending
attacks on many existing crypto currencies
cf. for example Section
\ref{UNODestructionSection}
and
\ref{DogeCoinandLitecoin}.
For example in April 2014 one single miner owned 51 $\%$
of the hash rate of Dogecoin.

In this respect things are expected to considerably change in the future for bitcoin.
We do not expect bitcoin to remain dominant forever.
Here is why!
Unhappily due to the cost of adopting bitcoin as a currency
(the necessity to purchase bitcoins which have already been mined
at a high price)
one cannot prevent users from creating their own crypto currency
(cf. Section \ref{sec:CompetingCryptoCurrencies} and \cite{CompetingCryptoCurrencies}).
Gold does not give people and major countries any choice:
some countries have gold mines or gold reserves, others don't.
Digital currencies put all the countries and all the people at an equal footing.
There will be always a large percentage of the population which will
not be happy about the distribution of wealth and will try to promote a new
crypto currency which gives (new) investors a better chance than having
to buy coins already mined by other people.

The fact that bitcoin is expected to lose its dominant position
is also due to another factor, built-in decreasing returns for miners
and the predicted consequences of this fact, see
Section \ref{DeflationaryCoinsGrowthCoinsTheorySketch}.
At the same as miner rewards decreases substantially with time,
the money at risk increases (compared to the cost of mining a new block).

{\bf Phase Transitions.}
All these factor combined,
we expect that most crypto currencies
will undergo ``destructive'' transitions
from a secure state to an insecure state.
For many crypto currencies all these things are already happening, see
Section \ref{UNODestructionSection} and \ref{DogeCoinandLitecoin}.
The question whether it can also happen to bitcoin
and what might be further consequences of it is
further studied in Section \ref{FutureBitcoin}.

\newpage

\vskip-5pt
\vskip-5pt
\section{Alternative Solutions For Double Spending}
\label{FixesSoultionsDesirableCharsAndTimeStampingSolutionsWholeSection}
\vskip-5pt

{\small
{\bf Note:} this section is work in progress.
Not everything can be covered inside this paper and
many questions are really not obvious.
We thank all the authors of very valuable comments on early versions
of this paper posted at \url{bitcointalk.org}.
We plan to develop these questions further and publish another paper on this topic.
}
\medskip

In this paper we heavily criticize the longest chain rule of
Satoshi Nakamoto.
A single rule which offers apparent elegance and simplicity
and regulates two things at one time.  
It is responsible for deciding which freshly mined blocks are ``accepted'' and obtain
monetary reward and at the same for deciding which transactions are finally accepted
and are part of the official common history of bitcoin.
However as we have already explained in Section \ref{LongestChainWhyRuleExistsGeniusMistake},
it is problematic to solve both problems with one single ``blunt'' rule,
{\bf there is NO REASON why the same mechanism should govern both areas}.
It should be possible to design better mechanisms
in bitcoin and other digital currencies,
this NOT in order to replace the blockchain by another solution,
but as a complement,
in order to improve the security and the speed of transactions.

\vskip-5pt
\vskip-5pt
\subsection{Our Objectives}
\label{DoubleSpendingSolutionMainObjective}
\vskip-5pt

Our primary goal is to design and build
{\bf Fast Consensus Mechanisms} for bitcoin transactions.
We approach the problem from a conservative angle:
we do not think it is realistic un bitcoin to
try to change the speed at which blocks are mined.
We want to improve bitcoin in such a way that
payments can be accepted much faster than the speed
of mining the next block.

\vskip-5pt
\vskip-5pt
\subsection*{Desired Characteristics}
\label{DoubleSpendingSolutionCharacteristics}
\vskip-5pt

Let us examine what kind of solutions would be desirable.

\vskip-4pt
\vskip-4pt
\begin{enumerate}
\item
Order and timing of transactions should matter
and should be hard to modify
(protection against malicious manipulation
in the timing and network propagation of transactions).

\item
The solutions should be incremental and should NOT destroy the existing order
in the bitcoin network.
They should offer some benefits even if not
every network participant adopts them initially.
They should not require a permission of everybody in the bitcoin network.

\item
Earlier transactions should be
preferred 
and as time goes by
it should be increasingly difficult to emit a second (double spending) transaction.
\item
Instead of instability and all or nothing behavior where
large number of transactions could be put into question,
we should get stability and convergence.
\item
Relying parties should get increasing probabilistic certitude
that the transaction is final as times goes by, second after second.

They should also be able to get obtain some tangible evidence
in form of network events which are difficult to forge,
which allows them to evaluate their risks.

\item
Unique transactions which spend some quantity[ies]
of money in bitcoin should be always accepted
with very large probability.

\item
Double spending transactions should simply be resolved on the (objective)
basis of earlier transaction, if one transaction is much earlier than the other.

\item
Only in rare cases where competing transactions are emitted within a certain time frame
there could be an ambiguity about which transaction will be accepted.
We should also ask the question that maybe no transaction
should be accepted in this case, as it would show that either the payer is trying
to cheat or his private key has been compromised.

\item
In particular though it is possible and does not cost a lot
to rewrite bitcoin history in terms of which blocks get the reward,
it should be somewhat STRICTLY HARDER
and/or cost more
(the exact criteria to be determined)
to rewrite bitcoin
history in terms of who is the recipient of moneys.

\item
{\bf Network neutrality:} the criteria to decide which blocks are approved
should be as objective as possible.
Even though miners can produce competing blocks
and no one can decide which block obtains the reward later,
incentives in place
should be such that
all blocks are likely to include
the same transactions.

\item
Ordinary peer-to-peer network nodes
and ordinary people who use bitcoin for payments
and peers
should be empowered by the new solutions.
We need a self-defence mechanism against
potentially abusive behavior of miners.

\item
A decentralized solution should mean
more than one solution could be used and running concurrently.
Solutions should be designed in such a way as to cooperate
and not conflict with other similar solutions.

\item
Fast zero confirmation transactions should be encouraged
cf. \cite{AcceptingZeroConfirmationFalselyClaimedSecure}
and risks of accepting them should be reduced.

In order to achieve this we propose that some
small cash premiums would be offered by volunteers
who want their transactions
to be certified or re-confirmed by others and accepted faster.
These mechanisms should be decentralized and
several methods for doing this could be tried.
These certification and re-confirmation events
can and should be chained.

\item
The solution should incentivize ordinary network nodes\footnote{\label{foot8}
The current bitcoin community
has let down very badly ordinary people who
support the network at considerable expense in terms of CPU,
network and energy usage,
online availability, excessive hard drive space usage, etc.
A network 
which benefits primarily a restricted cartel of miners
was probably not exactly the intention of Satoshi Nakamoto
who has clearly postulated that each network node should be
mining 
cf. Section 5 of \cite{SatoshiPaper}. See also next footnote below$^{\ref{foot9}}$.
}
and miners$^{\ref{foot9}}$
to be active network nodes and help
improving the security of the network\footnote{\label{foot9}
Moreover pooled mining makes that miners do NOT even need
to be there support the network as full network nodes.
Recently there were alarming reports about the number
of full bitcoin network nodes dropping to dangerously low levels,
cf. \cite{NumberOfReachableNodesIsDroppingRecently}. }.
The cash premiums discussed above could be used precisely here.

\item
(Optional)
In addition the solution could incentivize ordinary network nodes to
spend money (use bitcoins and pay transaction fees)
through cash premiums.
This is in order to promote the adoption of bitcoin as a currency
which is not doing well, cf. Fig. \ref{BitcoinGraph1YTransactionsPerDay}
in a very similar way in which
some credit card companies offer bounties and rewards.

\item
Holders of balances in bitcoins
and especially those 
who do some efforts to manage their security (private keys) correctly
should also be encouraged to participate in supporting the network:
they should be able to generate
some additional income. 

\item
Confirmations should be chained and the
mechanisms should be designed in such a way that
the attacker in order to commit double spending
needs to corrupt several entities not known in advance.

\item
(Optional) Miners could be asked to apply certain rules regarding on how exactly
they order their transactions in their Merkle trees.
This in order to provide evidence
about the timing of transactions received by
given network nodes. 

\item
(Optional) There could also be some protection against spam or DOS attacks:
it should be difficult to jam the P2P network
with too many transactions.

\item
Double spending attacks should be visible and monitored.

\item
People who deliberately execute attacks
on the bitcoin network
or help others execute such attacks
should
NOT 
be rewarded
cf. Section
\ref{UnthinkableDoubleSpendingService}.
\end{enumerate}
\vskip-5pt

How exactly this can be done
is not totally obvious,
however it appears that bitcoin does not
really provide an optimal solution
and we need to propose something better.
We are not going to claim to provide
the ultimate solution.
This is expected to be a solution slightly better
than status quo, subject to further improvement
and detailed tuning to adapt it to the realities of bitcoin.

\subsection*{Decentralized Consensus: Historical Background and Related Research}
\label{DoubleSpendingSolutionsConsensusBackgroundByzantyne}

It is clear that our problem has potentially many solutions.
However, do these solutions work well? Are they secure?
This is closely related to the well-known 1982
Byzantine Generals problem\footnote{
In theory this problem is already partly solved in bitcoin
by Satoshi bitcoin mining process and Longest Chain Rule,
however in practice this is very slow and unstable.
Therefore the problem needs to be solved again
on a more practical level.
}
 in computer science \cite{Byzantine}.

\subsection{Proposed Solutions}
\label{DoubleSpendingSolutionProposed}

It surprising to discover that {\bf Satoshi did NOT introduce a transaction timestamp}
in bitcoin software. It is NOT known WHY neither the original creator of bitcoin
nor later bitcoin developers did not mandate one.
This could be seen as an expression of
{\bf misplaced ideology}.
Giving an 
impression showing that maybe the Longest Chain Rule
does solve all the problems in an appropriate way.
Unhappily it doesn't\footnote{Or at least in the current bitcoin and many other current crypto currencies
it doesn't, they are permanently vulnerable to double spending attacks and transactions are slow.}.

Currently only an approximate timing of transactions
is known in the bitcoin network,
it comes from the number of block in which a given transaction is included:
this gives a precision of approx. 10 minutes.
Transactions without a fee could be much older than the block.
However all blocks are broadcast on the network and it is very easy for
the bitcoin software to obtain
more precise timing
of transactions with a precision of 1 second, maybe better.
A  number of web sites such as \url{blockchain.info}
are already doing this: they publish timestamps
for all bitcoin transactions which correspond to the earliest moment at which these transactions
have been seen.

A preliminary remark is that in the current bitcoin system,
each quantity of bitcoins such as created or attributed to a certain public key
by some previous transaction, can be used only once.
There should be at most one digital signature
which transfers
this quantity to another set of public keys
(there can be multiple recipients for each transaction).
Two distinct signatures indicate double spending\footnote{
Things get more complicated with transactions
which contain multiple signatures.
Moreover there are transaction malleability attacks
and signatures themselves can also be easily modified
to appear as another distinct signature,
cf. \cite{MtGoxDoubleSpendingDecker}}.

\medskip
{\bf The 20 Second Solution}.
We sketch a solution to our problem: 

\vskip-4pt
\vskip-4pt
\begin{enumerate}
\item
First of all, all signatures should be
converted to some sort of normal form to avoid
identical signatures which look different,
for example in ECDSA if $r,s$
is a valid signature
$r,-s$ is also valid.
Also all large integers should be converted to a standard 256-bit integer format
in the interval $0,\ldots, q-1$ where
$q$ is the order of the elliptic curve group used in bitcoin\footnote{
In contrast current bitcoin network data is full
of incorrectly formatted signatures,
for example due to the presence of unnecessary leading zeros.}.

\item
In case of double spending if the second
event is older than say 20 seconds after the first transaction,
the first transaction will simply be considered as valid and the second as invalid.

It should be based on the earliest timestamp in existence
which proves that one transaction
was in existence earlier.
This seems reasonable knowing that the median time until a node receives a block
is 6.5 seconds 
cf. \cite{ForksPropagationDecker,ForksPropagationDecker2}.
The exact implementation of such a mechanism 
will be studied later$^{\ref{TheOlder20Sec2011Foot}}$.

This type of solutions have been studied for some time,
cf. \cite{FasterBitcoin} from 2011
which is not identical than our proposal\footnote{\label{TheOlder20Sec2011Foot}See Section \ref{DoubleSpendingSolutionProposedEnhancementsandLimitations}
sub-point \ref{joeEnhancement} for a further discussion and
additional variants/enhancements.
}
and more recently in Ripple\footnote{\label{RippleVs20SecFoot}
In September Ripple has presented a specific detailed solution called
Ripple Protocol consensus algorithm (RPCA) in which
the first transaction will be confirmed during
a 
voting process which takes a few rounds
and is claimed  to reach consensus in a matter of seconds
\cite{RippleConsensusAlgoCoindesk}
and such that everybody is expected to reach the same decision,
cf. \cite{RippleConsensusWhitePaper}.
As in our 20 second solution,
in this process a second (later) transaction
will simply be rejected
cf. \cite{RippleConsensusWhitePaper}.}.
However these (older and more recent) solutions are rather expected to work
on the basis of order in which transactions
are received rather than some timestamps
(privileged in this paper).

\item
In case of double spending if both events
come within at most 20 seconds
of each other,
we propose that miners should NOT include
any of these transactions in block they mine\footnote{
If we mandate this we would also need rules to handle additional
third fourth etc. spending transactions issued later.
One way to solve would to forget older attempts after some very long time
such as 1 month and then eventually
accept only proper single spends.}.
It would also be possible to accept
just any one of these two transactions
as proposed in \cite{BitcoinForumFastZeroConfTransactions2011}.
It remains an open problem what is the best decision in this case,
cf. also \cite{FasterBitcoin}.
%

\item
We propose the following 
mechanism to facilitate zero confirmation transactions.
Transactions should pay a small donation to a public key of volunteers in the bitcoin network
which should be ordinary full network nodes which accept connections and
advertise this additional software capability.
This in order to incentivize more people to run network nodes,
see \cite{NumberOfReachableNodesIsDroppingRecently}
which people should work on very low latency
and immediately spend their attributions few seconds later (or faster!).
We call these network nodes {\bf transaction confirmer} peers.
They are going to confirm the transactions by spending their input immediately,
and at the same time facilitate the diffusion of information
about these transactions in the network.


These confirmations should and will be {\bf chained}:
confirmation transactions will be themselves confirmed in other transactions for a fee paid from the initial fee.
Confirmation transaction should spend simultaneously
several incoming fees from previous transactions in order to link them together.

\item
(Optional)
We also propose to re-use ``shares'' which are already computed by miners
in vast quantities or select only certain shares with a sufficient number of zeros.
These can also be used to confirm that transactions are already in existence
at a certain moment.
For these shares 
we can in addition mandate that
if transactions are hashed in a certain order in a Merkle hash tree,
it means that this miner have seen certain transactions earlier.

\end{enumerate}
\vskip-5pt

In other terms a mined block could be considered as invalid
if it only includes one transaction while two
were already in existence say 20 seconds before it was produced
AND if these transactions were close in time.
If one was much earlier, it could be included.
Again this decision on whether to include or not
a given transaction could be decentralized and
requires some form of [secure or not] timestamping
and should be complemented by various forms of attestation by peers
which allows for better security against manipulation of these timestamps.

A big question is whether timestamps are needed at all,
see Section \ref{DoubleSpendingSolutionTimestampsDenial}.
An alternative to timestamps could be various pure consensus mechanisms
without timestamps by which numerous network nodes would certify that
they have seen one transaction earlier than another transaction.
In this paper we take the view that timestamps should be present by default
and further confirmed by (the same) sorts of additional mechanisms.

{\bf Remark:}
This solution is not an urgent need for larger crypto currencies
which enjoy a dominant position and command a lot of hash power.
They can probably survive for years without it.
It is however
{\bf vital }
for all small crypto currencies which are more vulnerable
and subject to risk of very rapid self-destruction
if it is not applied, as shown in this paper.

\subsection{More Details On TimeStamps}
\label{DoubleSpendingSolutionProposedMoreDetailsTimeStamps}

The exact implementation of timestamping is not obvious.
Initially it could be left to the free market, some timestamping
is better than no timestamping at all
which is the current situation.
Several mechanisms could function simultaneously.
For example one can immediately use timestamps published by \url{blockchain.info}
and later (simultaneously) use more secure timestamp solutions from other sources.

For solutions which would prevent for-profit manipulation of timestamps
we need to propose additional mechanisms, such as secure bitstamps
or additional distributed consensus mechanisms.
We have already proposed two solutions to this problem
in points 4) and 5) in the previous section.
Below we discuss these`peer/miner confirmation solutions in more detail.
We plan to develop all these questions in another paper.

\subsection{More Details On The
[Multiple Chained] Peer Transaction Re-Confirmation Mechanism}
\label{DoubleSpendingSolutionProposedMoreDetailsConfirmationTransaction}

We recall that some network nodes
are going to become {\bf transaction confirmer peers}.
They are going to confirm the transactions by spending their input immediately.
These confirmations will be chained as already explained.
It is important to remark that current bitcoin DOES allow
transaction outputs to be spent immediately without any delay
(0 seconds delay) in the next transaction
which can be included directly in the same block, cf.
\cite{ImmediateSpending}.

The main idea is not that the {\bf transaction confirmer peers}
do NOT have to be entities working for profit which would advertise and sell their services.
It could rather be ordinary network nodes.
They should just run the right version of ordinary
Satoshi full network node software
which implements the additional mechanisms
and should ensure a high level availability
and reactivity to the network events.
All bitcoin users which have decent PCs or other devices which are always
connected and always on should be invited to participate.
We do NOT need a reputation mechanism, it will be easy to check in the blockchain
and evaluate their past reliability, speed and capacity to reach many other network nodes
for these services.

For these transactions we postulate that there should be a standard fee $C_f$ per confirmation
fixed by a certain market mechanism (like a majority vote).
Nodes could compete in terms of the confirmation speed however it is more important that
we have large list of peers which work reasonably well.
We postulate that in order to increase randomness in the choice of peers
the fee should be fixed in most cases (so that many peers will appear as exactly equivalent choices).
A standard practice should be to send a multiple
$K\cdot C_f$ of this fee to TWO transaction confirmer peers,
these peers are expected to immediately send
the amount of $(K-1)\cdot C_f$
to two other transaction confirmer peers,
which two peers should be chosen
in a deterministic pseudo-random way using a hash function
from a public list of confirmer peers active in the recently mined 2016 blocks.

{\bf Remark.}
It is illusory to make the fees depend on the amounts of money hold
by the private keys participating in confirmations
and claim that nodes which hold larger amounts can be trusted. 
This is because network nodes could agree to participate in the attacks as a service without
revealing their private keys and without putting their money at risk.
A standard fee is the way to go.


\subsection{More Details On How To Use Shares Generated By Miners}
\label{DoubleSpendingSolutionProposedMoreDetailsReuseShares}

In typical pooled mining miners produce shares in which
H2 starts with say 42 zeros
\cite{MiningSubversive,MiningUnreasonable}
and
send them to the pool managers in vast quantities
in order to prove that they have done the work for which
they should be paid.
For example if the current difficulty is such that
H2 must starts with 66 zeros,
which is very close to what we had recently,
a staggering number of $2^{24}$ shares are generated
every 10 minutes.

We propose that only shares with at least 48 zeros
should be used as evidence in the bitcoin network.
This gives roughly $2^{16}$ events every 10 minutes,
or one event every 10 miliseconds on average.
This probably gives sufficient precision
for certifying the timing of transactions in bitcoin network
(even though one cannot force miners to disclose these events,
and a large percentage of these events might be lost).
We say at least 48 zeros, as it is more or less clear that miner pools
will in the future increase the difficulty of shares
and they will have more than 48 zeros, giving
less that $2^{16}$ events every 10 minutes\footnote{
If this happens it will be a sign of further very dangerous centralization
of mining, and unhappily the robustness of our solutions against attacks will decrease,
however some security will remain.}.

Our key proposal is that network nodes which are
{\bf transaction confirmer} peers can publish these
data in order to make more peers use them
which will increase their expected income.
This should encourage miners to also participate in the peer
network and to publish these shares,
which just by the fact
of becoming public will improve the security of
zero confirmation transactions in the bitcoin network.

{\bf Disclaimer.}
It is however important to understand that individually
such events are relatively inexpensive to produce.
The idea is that many such events produced by different miners
will be used,
combining the concepts of proof of stake and proof of work.
These events will be chained for extra security.
Moreover some of these events will achieve substantially smaller values of H2,
with 49, 50 and more zeros.
Such events will be more valuable.

%
%
%
%

\subsection{Enhancements and Limitations}
\label{DoubleSpendingSolutionProposedEnhancementsandLimitations}

What we describe above is NOT yet a full solution.
It requires further work to specify
and analyse if it does the job reasonably well
and if it does not lead to new attacks.
We also need to consider a number of enhancements and improvements.
Below we list some ideas.

\vskip-6pt
\vskip-6pt
\begin{enumerate}
\item
Probably we need to require more than a timestamp for all bitcoin transactions.
We could also require timestamps for all individual signatures.
A digital signature gives security guarantees which answer two questions:
Who? (signs) and What? (is signed).
A digital signature which includes a timestamp also answers the question When?
(the transaction was authorized).
\item
It is NOT correct to believe that miners have no other choice than to
rely on the current bitcoin network
where the median time until a node receives a block is 6.5 seconds
whereas the average time is 12.6 seconds, etc.
cf. \cite{ForksPropagationDecker,ForksPropagationDecker2}.
This is like the "zero-fee propagation", it costs nothing.

Miners could actually - because they work for profit - PAY a tiny little bit of money to have access to a much faster
and more accurate data about all transactions,
super-fast latency data based on a set of some 1000 randomly chosen full network nodes
which are connected to a faster `backbone' network.
Then it is easy to imagine and easy that miners have access to all transactions within miliseconds rather than seconds. 
Such additional network could be run by business providers
or as a cooperative belonging to miners themselves
and could also provide double-spending alerts automatically.

\item
\label{joeEnhancement}
The following enhancement to our solution
was proposed by user joe, see
\url{https://bitcointalk.org/index.php?topic=3441.msg48484#msg48484}, \\
which post goes back to February 2011,
part of a discussion thread on fast transaction acceptance
\cite{BitcoinForumFastZeroConfTransactions2011}, cf. also \cite{FasterBitcoin}.
The author proposes exactly the same solution as our points
2) and 3)
in Section \ref{DoubleSpendingSolutionProposed}
with the same 20 s threshold,
with additional rules
regarding rejecting blocks which include (later) transaction2 out of two,
a transaction which should normally not be accepted
if there was another earlier transaction1 which was in existence more than 20 s before.

Normally blocks which contain transaction2 should be rejected
by miners, except if they already have 6 confirmations or more,
in which case they should be nevertheless accepted.
This is claimed to avoid permanent block forks\footnote{\label{ExtraRuleAttackerWins6}
The author does not explain what exactly is the threat,
so we have invented our own scenario to illustrate this point.
Imagine that the Chinese government firewall is abused in a very subtle way by a disgruntled government employee, so that nobody notices but certain bitcoin transactions never make it to China for 1 hour,
some packets are dropped by the IP network.
Chinese miners might already have 51 $\%$ (because it is a big country),
and simply do never receive earlier transaction1,
so in good faith they include later transaction2 is a block, it becomes official history of bitcoin in China.
Now these blocks mined in China are not recognized by people outside Chine because they contain an invalid transaction2,
so people outside Chine mine only on their chain, and maybe always have a shorter chain because they have 49 $\%$.
Thus 49 $\%$ of hash power is permanently wasted.
This is why after 6 confirmations miners could join the non-orthodox branch.
However this means that the attacker has succeeded.}
which rule however is somewhat problematic, the attack can eventually succeed$^{\ref{ExtraRuleAttackerWins6}}$.
\end{enumerate}
\vskip-8pt

{\bf Limitations:}
A major factor which is expected to affect the development and adoption of solutions
to our problem is the size of the blockchain in bitcoin
which is stored at every full network node
and takes about 20 gigabytes,
which is one of the reasons why the
number of people who support the current
bitcoin network has been falling dramatically
cf. \cite{NumberOfReachableNodesIsDroppingRecently}.
This is however also an opportunity to recruit additional people
to work for the bitcoin P2P network
which is already a part of our proposed solution.


\vskip-5pt
\vskip-5pt
\subsection{Timestamps - Controversy And Discussion}
\label{DoubleSpendingSolutionTimestampsDenial}
\vskip-5pt

In this paper we sketch one possible solution, not every possible solution
to instability of bitcoin and its poor ability to defend users against double spending attacks.
Timestamping is one of the key elements in this solution.
This comes as a shock to many people
who get used to consider that bitcoin is a neat system
and nothing could possibly be wrong with it.
However the role of academic research is
not to assume that by default bitcoin is perfect,
and to challenge even what is considered as
an obvious and well-established truth.
Using timestamps is a disturbing 
proposition,
it is somewhat contrary to a certain idea
of what bitcoin should be\footnote{
A self-governing 
asynchronous system in which ``the invisible hand'' of brute hash power
makes all the important decisions.
Unhappily this is very slow in current bitcoin network
and leads to further instability
with blockchain forking attacks.
}
\footnote{
For example timing information certainly
provides some additional information which
goes against improving the anonymity of transactions.
However anonymity is not really a strong point of bitcoin.},
and not everybody agrees.

Few days after this paper was released, on 08 May 2014,
the following comment was posted on a bitcoin forum,
cf. \cite{BitcoinForumOnMyPaperDavis}.
It was written by Gerald Davis, 
also known as the user {\it DeathAndTaxes},
a highly respected and very 
frequent contributor
to this bitcoin forum, responsible for some 14,000 posts.
Here is what he writes about this paper.
This is really his first reaction (which was later expanded). 

\vskip-0pt
\vskip-0pt
\begin{quote}
"Utter nonsense. \\
It is sad that they wrote a paper based on the premise that timestamps
can be used to solve the double spend problem (they can't)" 
\end{quote}
\vskip-0pt

In other terms the author claims that maybe timestamps
do not help to solve this problem, on the contrary. 
It seems that this paper has created a genuine controversy.
Are timestamps really needed? Are they actually useful?

\vskip-6pt
\vskip-6pt
\subsubsection{Timestamps as a Quick Fix.}
We do not have a strong opinion whether timestamps are absolutely necessary
in the case of an ideal crypto currency.
Potentially additional built-in consensus mechanisms
which depend on the network propagation of different transactions,
could achieve a similar effect as already explained earlier.
However we claim that:
\vskip-5pt
\vskip-5pt
\begin{enumerate}
\item
current bitcoin has very {\bf slow} confirmation, which is bad for its adoption,
\item
we need to {\bf add some low latency mechanisms} to the current bitcoin,
\item
{\bf the order and timing of transactions SHOULD matter}
and it should somewhat be used
in order to decide which transactions are accepted,
\item
fast zero-confirmation transactions should be {\bf encouraged}, not discouraged,
\item
double spending attacks should be made {\bf increasingly difficult with time},
after the initial transaction was broadcast in the peer-to-peer network.
%
\end{enumerate}
\vskip-5pt

One of the methods to achieve this (but probably not the only one)
is to use timestamps.
It is difficult to redesign the whole of bitcoin,
make it substantially faster and more secure,
produce more than 1 block every 10 minutes,
and convince everybody to upgrade.
The current method is proposed mostly as a quick-fix
for a crypto currency such as bitcoin
which is as it is:
slow at approving transactions, especially for large transactions
\cite{FTACourtoisVideoHorseCarriage,FasterBitcoin}.
We are trying to develop
some proposals for the future of bitcoin
digital currency which would improve it (even slightly)
which would fix some of the current problems
and such that 
they would not be too complicated to adopt.

\vskip-4pt
\vskip-4pt
\subsubsection{Are Timestamps Really Necessary?}
In the same Internet forum \cite{BitcoinForumOnMyPaperDavis}
Davis writes:
``Satoshi did not include tx timestamps
because proving timestamps in a decentralized environment
is an incredibly difficult (some would say impossible) task''
and later
``Satoshi understood that timestamps are very difficult to authenticate [and]
as such limited them to areas where there is no solution which doesn't involve timestamps.''

In the same Internet forum user {\it telepatheic} writes:
``Satoshi didn't put much thought into the problem of time stamping, although he realised timekeeping was important''.
We learn that Satoshi has written the following comment inside the code:
``Never go to sea with two chronometers; take one or three.''
Another user jonald$\underline{~}$fyookball explains that:
``we need time stamps for the difficulty change''.

This is very interesting.
Satoshi DID mandate timestamps in blocks,
even though knowing their exact values are of secondary importance\footnote{
They are apparently needed in bitcoin in order to prevent miners from manipulating the difficulty level in bitcoin,
see \url{https://bitcointalk.org/index.php?topic=600436.msg6622244#msg6622244}}
and they do not play yet a very important role in bitcoin.
These timestamps are already ``certified'' by the blockchain which unhappily
is a very slow process. 
%

It probably is a difficult task to obtain additional (higher resolution)
timestamps which could be trusted.
However it is {\bf needed}.
We hypothesize that additional timestamps
with a certain level of security
will always be better than no stamps at all.
We observe that without additional timestamps,
there is no way to distinguish between:

\vskip-6pt
\vskip-6pt
\begin{enumerate}
\item
double spending events which could be rejected on a purely conventional basis
in any reasonably fast network:
if one transaction is broadcast many minutes later,
it could just be rejected without any justification,
and the first transaction could still be accepted;
\item
double spending events which occur quasi-simultaneously,
in which case both transactions could in turn be rejected
on a purely conventional basis,
if miners accept to reject\footnote{
Alternatively in this second case, also by convention,
one transaction could be accepted,
we do not recommend this variant.}
such transactions,
based on the idea that
the signing key has been misused
AND this fact is already known at an early stage.
\end{enumerate}
\vskip-6pt

This distinction is crucial in order to:
\vskip-8pt
\vskip-8pt
\begin{enumerate}
\item
substantially {\bf improve the transaction speed} in the bitcoin network,
\item
achieve {\bf better network neutrality} in bitcoin.
Being able to decide
in a short time which transactions
should be approved by miners in a short time
and in a more objective and transparent way. 
\item
We want to have
a method less prone to discretionary 
decisions taken by miners regarding which of the two transactions is accepted
\footnote{
Which is currently the case and leads to
greater risk of for-profit blockchain manipulation.}.

\item
On the contrary, we want to increase the role
played by ordinary 
peers which post transactions in the network.
Nodes need to be encouraged to stay connected and active,
or bitcoin is going to disappear,
cf. \cite{NumberOfReachableNodesIsDroppingRecently}.
\end{enumerate}
\vskip-3pt
\vskip-3pt

To summarize, timestamps should be highly recommended.
In the same bitcoin forum another senior member Cryddit writes:

%

\vskip-4pt
\vskip-4pt
\begin{quote}
"we don't really have a practical distributed-timestamp scheme.
But there may be a simpler one
[...] (not requiring a distributed timestamp)
that works.  [...] 
it's certainly in the best interests of honest miners and honest transaction makers
to provide accurate timestamps if it improves security against dishonest ones"
\end{quote}
\vskip-3pt



\vskip-6pt
\vskip-6pt
\subsubsection{Could Timestamps Be Hacked?}
This is a serious and valid question which requires more work.
In the same Internet forum \cite{BitcoinForumOnMyPaperDavis}
Davis writes:
``So the decentralized currency is based on the timestamps as decided by some centralized "super peers".
If I bribe the timestamp servers to say my tx is older
then I can double spend without even using hashing power.''.
And then he goes into an argument to the effect that
the only way to actually solve it would be to...
reinvent bitcoin and the blockchain.
This is very interesting:
\vskip-7pt
\vskip-7pt
\begin{enumerate}
\item
We must reinvent bitcoin every day. It is not perfect.
\item
We really need to avoid this situation:
where the proposed modification would {\bf help} to double spent
without using hash power (or at a lower cost).
It is should be at least as hard as previously to double spend, and it should be rather strictly harder.

Even if some timestamping authority certifies that my freshly created transaction tx2 is very old
and it should be accepted as older, other network participants are NOT forced to believe this authority.
They will observe that they have received tx1 much earlier and they will either certify tx1 by various existing means
and include it in Merkle trees.

This is a serious question which requires further research.
The risk is then that they might decide to exclude both transactions
(given the evidence of key misuse).
Then tx2 will work as a denial of service attack on tx1
but then this is NOT double spending.
In principle tx2 should not make tx1 rejected by miners.

\item
Double spending without using hash power is prevented
because the attacker needs to know
many private keys used in previous history of bitcoin and the sum of the balances
hold at these keys provides evidence that the attacker does not know the keys
(or he could steal the money) however he could corrupt some people without them reveling
their keys, just ask them to generate confirmations.
So it really is about inability of the attacker to corrupt many peers NOT known in advance,
within minutes.

Peers to be corrupted are not known in advance due to the chaining of these confirmations.

\item
{\bf Bribing} timestamp authorities increases the cost of double spending attacks
which will be executed only if they are profitable.
Therefore it will already be a valuable improvement to bitcoin and other crypto currencies. 

\item
The time is a reality which is far bigger and far {\bf more objective} than the bitcoin blockchain.
It should therefore be easier and less costly to develop reasonable and effective solutions for this problem.
It should be possible to use any crypto currency other than bitcoin,
and re-use many existing Internet services and/or digital notary services to certify events.
It is also possible to use shares generated by bitcoin miners as already suggested.
We believe that there is plenty of solutions to this
problem. We intend to further develop this question in future papers.
\item
We {\bf do need} some additional "peers" to help the network.
Miners mine in pools and currently miners
are not interested in supporting the bitcoin network:
the number of active network nodes is falling very badly,
cf. \cite{NumberOfReachableNodesIsDroppingRecently},
and it is 
much smaller
than the number of active miners,
cf. Section \ref{SectionBitcoinInfrastructureHistory12MInvestorEconomocis}.
Bitcoin popularity is in decline
cf. Fig. \ref{BitcoinGraph1YGoogleSearchPopularity}
page \pageref{BitcoinGraph1YGoogleSearchPopularity}
and transaction activity is in decline
cf. Fig. \ref{BitcoinGraph1YTxFeesPerDay}
page \pageref{BitcoinGraph1YTxFeesPerDay}.

\item
However we do not need ``super peers'' or new ``privileged'' entities.
More precisely we do not want new mechanisms
to be {\bf centralized}. 
We have already indicated our preference to decentralized solutions,
which however need to be developed and deployed progressively.
The real centralization is the current situation
where the number of network nodes involved in checking
the bitcoin transactions is declining below reasonable levels
cf \cite{NumberOfReachableNodesIsDroppingRecently}.
In addition pooled mining is {\bf super-centralized},
cf. 
\cite{RosenfeldPoolRewardMethods} and Table 2 in \cite{MiningSubversive}
and we cannot even trust the miners to be honest
and not manipulated by others,
cf. Section \ref{HiddenAttacks}.
\end{enumerate}
\vskip-4pt
\vskip-4pt

{\bf Remark.}
Even if timestamps do not solve our problems very well,
they still should probably be recommended,
as a mesure of transparency, accountability,
promoting trust and better reputation of bitcoin.
They are needed
because they give better visibility to
various forms of problematic events,
as explained above
and allow to
better distinguish between different situations
and better understand the spectrum of actual double-spending events and attacks.
Current bitcoin network is basically
somewhat tolerant to fraud
and it is not trying to make it more visible. 
%

%

\newpage

\subsection{Peer Voting Solutions and Ripple RPCA}
\label{DoubleSpendingSolutionProposedMoreDetailsConfirmationTransactionvsRippleRPCA}

Solutions to our problem of peer confirmation
can also be achieved by peer voting.
Below we discuss how more generally the double spending
is solved by peer voting in the Ripple network.

In September 2014,
Ripple have published a white paper
in which they explain how this is done.
The solution has apparently been already
operational for some time.
The main objective is to
achieve fast consensus in a peer network
without expensive (and slow) proof of work.
They describe a mechanism called
Ripple Protocol Consensus Algorithm
(RPCA) such that:

\vskip-7pt
\vskip-7pt
\begin{quote}
"Each server in the Ripple network is tasked with voting on a new batch of candidate transactions
during rounds that take place every few seconds."
\end{quote}
\vskip-7pt


%

\noindent
As a result, transactions are expected to be approved "in a matter of seconds",
cf. \cite{RippleConsensusAlgoCoindesk,RippleConsensusWhitePaper}.

\vskip7pt
\vskip7pt
More precisely they specify a precise solution
called {\bf Ripple Protocol consensus algorithm (RPCA)}
which is such that:

\vskip-7pt
\begin{enumerate}
\item
Each server maintains a Unique Node List (UNL)
for which group he trusts the majority vote of this group
(which does NOT require all these nodes to be honest).

\item
Each server "takes all valid transactions it has seen" and
"makes them public" 
This happens at each round, which take place every few seconds.

\item
Then servers vote in many rounds,
and only transactions which receive a certain
minim percentage of YES are approved.

\item
In the last round, 80 $\%$ threshold is required.

\item
These transactions form a new "closed-ledger"
and it is claimed that under certain conditions
"the last-closed ledger maintained by all nodes in the network will be identical",
cf. \cite{RippleConsensusWhitePaper}.

\item
Double-spending is prevented because
when the first transaction confirmed during
this DETERMINISTIC process,
"the second will fail",
and everybody is expected to reach the same decision,
cf. \cite{RippleConsensusWhitePaper}.
%
\end{enumerate}
\vskip-7pt

Both the Ripple solution and the spectrum of solutions proposed in this paper
ar trying to avoid "relying on proof-of-work infrastructure"
and achieve consensus at a lower cost than currently
(with exception of solutions we discuss in Section \ref{DoubleSpendingSolutionProposedMoreDetailsReuseShares}).


\newpage
\subsection{The Unthinkable Double Spending as a Service}
\label{UnthinkableDoubleSpendingService}

In the bitcoin community
there is already a service 
\url{bitundo.com}
which is trying to
convince miners to help to cancel
other people's
bitcoin transactions on demand.
This is done
by including a transaction
which is a genuine double spend transaction
(sending the same money to a different address).
It incentivizes miners to help to undo
bitcoin transactions for a certain fee
which can improve their mining income.
It appears that currently it focuses only 
on undoing
transactions within minutes,
before they are included in any block.
It is not (or not yet) trying to undo transactions later on:
when they are already approved.

This service is highly problematic from the ethical perspective:
it can be seen as a method to bribe miners in order to help one
to commit
a double spending attack.
It appears that his service is not illegal
and has some legitimate applications.
There are certainly people in bitcoin community
who think it should be legal and allowed.

An interesting feature of our solution sketched in Section
\ref{DoubleSpendingSolutionProposed}
is that it automatically makes such attacks very hard, close to impossible.
If the second transaction comes later,
chances to double spend should be very low.
If the two transactions occur quasi-simultaneously,
chances are that both would be rejected by the network.
Thus it is not necessary to make new laws to defend bitcoin
against \url{bitundo.com}.
This is very much in the spirit of bitcoin
as a 
public space
which does not require legal protection
because it is able to self-regulate
and mandate the right sort of protection
against threats.

\newpage

\section{Hidden Attacks: How To Abuse Miners}
\label{HiddenAttacks}

\subsection{A Small But Important Technicality}
\label{HiddenAttack}
\label{Technicality}


We examine the process of double hashing which is used
in bitcoin mining according to \cite{MiningUnreasonable,MiningUnreasonable2}.

\begin{figure}[!here]
\centering
\begin{center}
\vskip1pt
\vskip1pt
\includegraphics*[width=5.0in,height=3.2in,bb=0pt 0pt 1158pt 666pt]{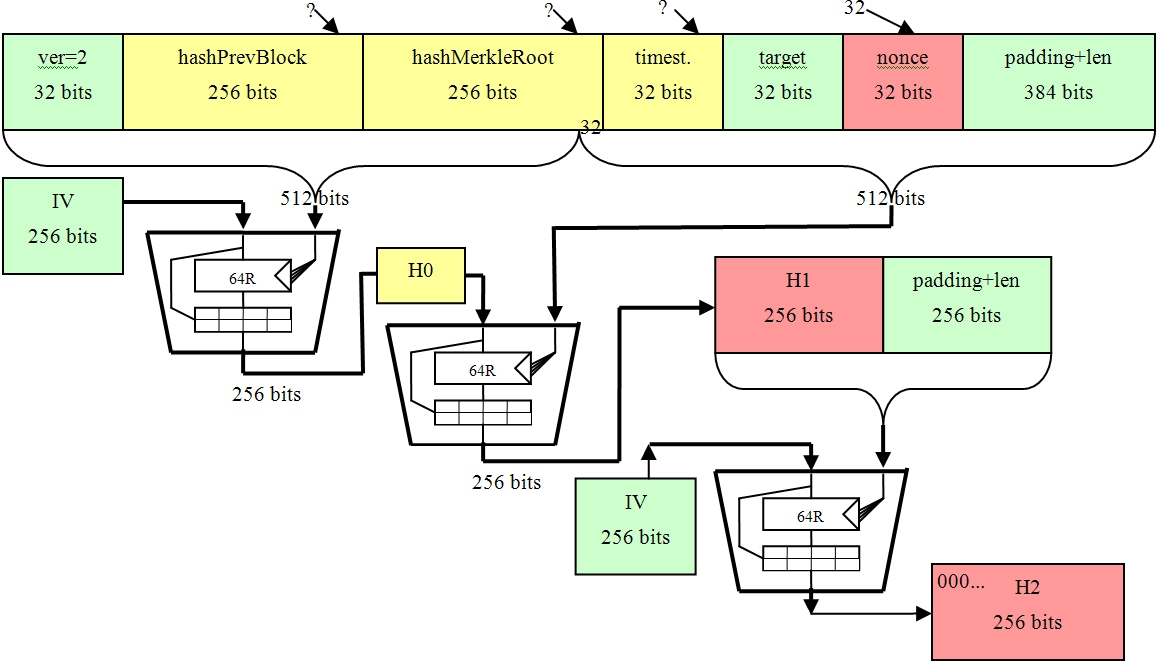}
\vskip-0pt
\vskip-0pt
\end{center}
\caption{
The process of bitcoin mining according to \cite{MiningUnreasonable,MiningUnreasonable2}.
}
\label{BitcoinCISOProblem2}
\vskip-3pt
\vskip-3pt
\end{figure}

One thing jumps to our attention
[we thank Lear Bahack for observing this fact independently,
though we have observed that many months earlier].
For every H0, the miner needs to check many possible nonces.
The miners do NOT need to know on which block they are mining:
they do NOT need to know the value of hashPrevBlock
which computation is amortized over many hash operations and
the value of H0 changes very slowly.
They only need to know the value H0
which could be computed for them by the pool manager.
Miners can be made to
{\bf mine without any precise knowledge about
which block they are mining for} or whom they are mining for.

Only an excessively small number of miners,
will actually manage to find a winning block:
only a very small proportion of say $2^{-41}$ of all shares
found by miners are winning shares.
Only these miners might be able to know on which block they have mined
by examining the public data in the blockchain,
and this is not at all guaranteed.
In practice they can see it ONLY if they have also
recorded all hundreds of thousands of shares produced by their miner and sent
to the pool manager over the whole weeks and months.

We see that pool managers CAN implement arbitrary subversive strategies,
for example accept certain transactions only to overthrow them
within less than one hour and accept another transaction with another recipient.
{\bf Nobody will notice:}
miners will never know that they have been involved in some major
attacks against bitcoin such as producing two different versions
of the blockchain in order to double spend some large amount of money.

{\bf Remark 1.}
Moreover even those miners who have produced winning blocks
and therefore will be made aware of the previous block on which they have been mining,
still cannot claim they have participated in some sort of attack.
Fork events do happen in the bitcoin network.
Only overall higher frequency of fork events
mined by one large pool
could suggest that some attacks have been executed by that pool,
however the pools can execute such attacks
just within the limits of the standard deviation
\footnote{Standard deviation is excessively large as mining events are quite rare, 
cf. \cite{MiningSubversive,RosenfeldPoolRewardMethods}.}
and never attract any attention.

{\bf Remark 2.}
It is also possible to see that
even with the knowledge of all recent transactions from the network
and with the knowledge of H0,
it is not possible to guess how exactly the Merkle root hash is composed.
We are talking about preimage (inversion) attacks starting from H0
aiming at guessing which hashPrevBlock was used to produce this H0.
This is because the number of combinations is too large.
For example the number of ways
to permute the order of 100 transactions is
already more than $2^{500}$.

\subsection{Miner Hidden Abuse Attack Across Currencies}
\label{HiddenAttackAcrossCurrencies}

The same attack works {\bf across digital currencies}.
Some miners think that they mine bitcoin, while in fact
they are made to mine Unobtanium, and vice versa.
All this is the discretionary power of the pool manager, this is due to the fact
that one can mine only knowing H0 and most of the time no other information is disclosed to miners.
In rare cases miners could discover that they found a block for another crypto currency which they have never mined.
In practice miners do NOT store vast quantities of H0 values with which they have mined.
Miner devices do NOT have enough memory to store them.

\subsection{Further Manipulation Scenario With Deflected Responsibility}
\label{TechnicalityFurtherManipulation}

Our attack can also be made to work in the scenario in which it is not possible for the attacker to corrupt pool managers.
It can be run in a different way in which pool managers are going to corrupt themselves
and there will be no reason to accuse them of acting with any sort of malicious or criminal intention.

Basically it is possible for an attacker to manipulate
the price of a small crypto currency such
as Unobtanium to be 10 $\%$ MORE profitable than bitcoin mining
(typically such currencies are in  a sort of equilibrium situation in
which the profitability is similar as for bitcoin).
Then we can hope that the pool managers themselves
are going to implement code to switch to this crypto currency
for a short time
(real-time switching mechanism mining for the most profitable currency at the moment).
If not, the attackers can themselves release open source code of this sort in order to encourage
the adoption of this sort of gain optimization techniques among pool managers.
Pool manager can now re-direct 100 $\%$ of the hashing power they command
to another entity. 
They are NOT going to tell this to miners and simply pocket the difference,
and they will still pay miners in bitcoins.
Again, there is in principle no way in which miners
could see the difference. 

\subsection{Has It Already Happened?}
\label{HasItHappenedAndDetectionMITM}

In general it is possible to see that if miners use the Stratum protocol,
the miner cannot be cheated without being detected
and none of the subversive scenarios of this Section \ref{HiddenAttacks}
could be implemented.
Stratum is what the majority of ASIC miners and pools use at this moment:
GHash, DiscusFish, Eligius, Bitminter, etc. 
In the stratum protocol the hash of the previous block is always transmitted in cleartext to the miner.
If the miner sniffs the data transmitted (e.g. using Wireshark)
and checks against just a few recently mined blocks
he will detect if he is made to mine on a different block and if he is participating in an attacks.
We have verified in ourselves with the most popular mining pools and found that 
indeed 
the hash of the previous block is systematically transmitted\footnote{ 
The miner would not be able to mine if he doesn't know it, 
except if H0 is transmitted, 
cf. Section \ref{Technicality}}. 
Therefore one CAN detect the attack: 
one needs to record incoming packets with method being 
"mining.notify" and check 
if the second parameter after "params" 
is the hash of the last block in the blockchain, 
cf. our paper on this topic \cite{DetectionMITMBitcoin}. 
Unhappily most miners will not do these checks. 
It requires specialized hardware (a Network Tap) and software (e.g. Wireshark) 
to sniff network packets. 
Therefore in practice miners can still be abused. 

\subsection{Known Attacks which Have Happened}

Known attacks are not as sophisticated as what we describe in Section \ref{Technicality}.

In \cite{PoolServerHacksRedirectionWithholding2013} we find some reports
of suspected attacks on a mining pool 50BTC such as
"physical unauthorised access to [pool mining] servers"
and relay attacks
in which a miner formally connects to one pool which
communications are redirected to the victim pool 
with block withholding
\footnote{
See Sections IX, X.B, XI.A in \cite{MiningSubversive} for
a more detailed study and discussion of block withholding attacks. }
and apparently also other attacks, cf.
\cite{PoolServerHacksRedirectionWithholding2013}.

A major attack with redirection of hash power were reported in August 2014,
in this attack the hacker was more powerful than we generally assume in this paper
and was able to steal coins of users as a man in the middle
cf. \cite{RedirectedMiningStealingCoinsAugust2014}.
The attacker has hacked some major Internet service providers,
and the attack could be prevented by standard network security techniques such as TLS.

\subsection{Is It Possible to Fix It? - Reactions in the Bitcoin Community}

In the following bitcoin forum user Cryddit and senior member of this forum writes:

\begin{quote}
"The author is right about increasing the security of mining by requiring miners to know both the hash of the current block and the hash of the previous block - the hashing operation they need to do is essentially the same,
and making sure miners know what block they're building on
would make certain classes of attack
(diverting pool miners to another coin
using pool miners to build an unpublished blockchain, etc)
[...]
leave incontrovertible evidence.

That is a good idea and we should do it."
\end{quote}

Source: \\
\url{https://bitcointalk.org/index.php?topic=600436.msg6626004#msg6626004}.
Another post tries to absolve bitcoin developers from any responsibility
for the
current situation,
%
%
\url{https://bitcointalk.org/index.php?topic=600436.msg6657579#msg6657579}.
%

\subsection{Is It Possible to Fix It? - Solutions}

There are two main questions 
to be considered when considering possible solutions to this problem. 
The first question is detection. 
Maybe miner software/hardware interface should be modified to display at any moment the hash of the previous block,
in order to know on which they are mining and obtain the appropriate evidence. 
This has is always transmitted and must be known to the miner in order to mine correctly,  
cf. Fig. \ref{BitcoinCISOProblem2}, 
Section \ref{HasItHappenedAndDetectionMITM} and \cite{DetectionMITMBitcoin}. 
Until this is implemented, some miners can detect the attack 
using specialized hardware (a network Tap) and software (e.g. Wireshark) 
to sniff network packets and inspect packets which contain "mining.notify" 
and check the second parameter after "params", 
see Section \ref{HasItHappenedAndDetectionMITM} and \cite{DetectionMITMBitcoin}. 

In general the attack of Section \ref{Technicality} is a serious security flaw
in the Satoshi bitcoin specification. 
It is therefore impossible to claim that bitcoin cryptography is perfect, cf. \cite{SatoshiCryptoGenius}.
It seems to be an inherent problem due to double hashing, 
and maybe bitcoin needs to go back to solutions
using a single application of a (sufficiently robust) hash function. 
Our original solution to this problem is the concept 
of "plaintext aware hash functions" 
which is briefly described below. 

\vskip-4pt
\vskip-4pt
\subsection{Plaintext Aware Hashing}
\label{PlaintextAwareH}
\vskip-3pt

Before we try to define what is
a plaintext aware hashing,
we are going to explain what it isn't.
A double application of a hash function
commits to the plaintext yet
it is NOT plaintext aware:
from the first hash there is no way to
recover the message being hashed.

\vskip-3pt
\vskip-3pt
$$
H2=\mbox{SHA256}(\mbox{SHA256}(\mbox{block header}))
$$
\vskip-2pt
\vskip-2pt

In bitcoin we need the opposite to happen:
people who contribute hash power to the network should not be abused
in order to produce hashes for the attacker
which actually for example
is currently making them participate in a double spending
attack on another currency as explained in
Section \ref{HiddenAttacks}.
The miner needs to be certain that he mines
 honestly (Cf. Section \ref{HiddenAttacks})
on the top of the current block 
and also that the Longest Chain Rule is actually applied. 

The solution is quite simple.
We need to modify Fig. \ref{BitcoinCISOProblem2}
in such a way that
the last hash H2 has hashPrevBlock
as input (repeated) and that the key expansion function is a
combination of a solid pseudo-random function
with a dispersed repetition of bits from the input
so that we can be confident that in order to compute a valid H2
the miner must know hashPrevBlock.
%


It remains to develop and standardize a
concrete proposal of a plaintext-aware hash function.
One simple solution would be to compute

\vskip-3pt
\vskip-3pt
$$
H2=
\mbox{SHA256}(hashPrevBlock\oplus \mbox{SHA256}(\mbox{block header})).
$$
\vskip-2pt
\vskip-2pt

This is not backwards compatible.



\newpage

\vskip-6pt
\vskip-6pt
\section{Towards A Theory of Programmed Self-Destruction}
\vskip-4pt

In this section
we are going to try to combine all the elements
which we have studied so far
in order to see what is the overall landscape.
We can now formulate a certain theory
or set of claims about
the predicted future of crypto currencies.
based on what we learned.

Our main claim is that
the combination of four things:
\vskip-5pt
\vskip-5pt
\begin{enumerate}
\item
the longest chain rule,
\item
deflationary monetary policies which heavily limit the production of new coins
(with or without sudden jumps in miner reward),
\item
poor network neutrality, centralization and related moral hazards
\item
and a competitive environment
where hash power can shift rapidly from one coin to another,
\end{enumerate}
\vskip-5pt
is a {\bf fatal} combination.
It leads to predicted destruction of crypto coins.

\begin{figure}[!here]
\centering
\begin{center}
\vskip1pt
\vskip1pt
\includegraphics*[width=5.0in,height=3.0in,bb=0pt 0pt 730pt 350pt]{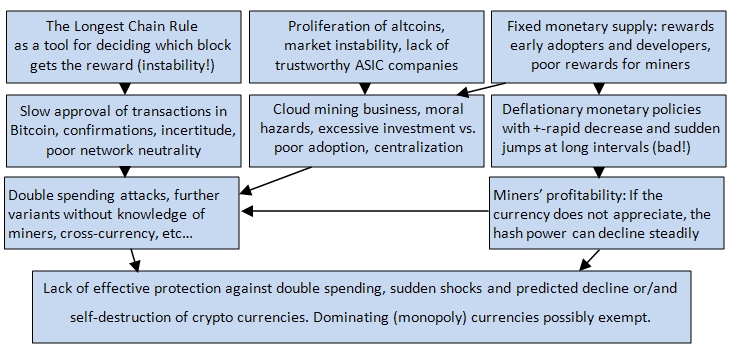}
\vskip-0pt
\vskip-0pt
\end{center}
\caption{
Theory of programmed self-destruction of crypto currencies: 
major factors and influencers which are also the main premises of our theory.
}
\label{BitcoinDestrTheorySummary3Things}
\vskip-3pt
\vskip-3pt
\end{figure}

On Fig. \ref{BitcoinDestrTheorySummary3Things} we summarize
again the main premises in our theory
and also try to show
some additional influencers.

\vskip-5pt
\vskip-5pt
\vskip-5pt

The remaining part of the paper will be a study of particular use cases.
Does our theory work? Does it allow us to
understand the past and 
and somewhat predict future of various crypto currencies?

\newpage

\vskip-6pt
\vskip-6pt
\section{Case Study: Unobtanium}
\label{UNODestructionSection}
\vskip-4pt

Unobtanium is a clone of bitcoin which is in operation since October 2013
(cf. \url{unobtanium.io}).
Unobtanium uses SHA256 and can reuse bitcoin ASICs for mining,
and it has a non-negligible value.
In March 2014 it was worth some 0.01 BTC which at the current hash speed
made Unobtanium mining roughly as profitable as standard bitcoin mining.
(note: later in April 2014 the profitability of UNO mining has declined).
It is traded at several exchanges.
Transactions are substantially faster than bitcoin:
blocks are generated and transactions
are confirmed 
once per 1.24 minutes instead of every 10 minutes for bitcoin
(it is 1.24 minutes and not 3 minutes as reported incorrectly by many sources).
At the first sight this currency seems therefore a quite promising clone of bitcoin
and the current market value of all Unobtanium in circulation is roughly about 0.5 million dollars.
On the official web page \url{unobtanium.io} we read
that Unobtanium is expected to be
``the cryptocurrency for serious traders''
and that ``Unobtanium is safe''.
At the first sight we see no problem with this currency whatsoever
apart from the fact that there are very few actual transactions in the blockchain.

Unobtanium is quite rare: only 250,000 will be ever made,
and the production of new currency
is halving every 2.88 months which is {\bf incredibly fast}.
There are only a few halving periods however,
and in September 2014 the miner reward settles forever
at a surprisingly 
small value. 

\vskip-7pt
\vskip-7pt
\begin{table}[h!]
  \caption{The Unobtanium Reward}
\label{BitcoinUNOReward}
\vskip-7pt
\vskip-7pt
$$
\begin{array}{|c|c|c|c|}
\hline
\mbox{blocks}
&
\mbox{approx. dates}
&
\mbox{UNO/block}  \\
\hline
1-102K    & \mbox{18 Oct 2013-}& 1 \\
102K-204K & \mbox{15 Dec 2013-}& 0.5 \\
204K-300K & \mbox{12 Feb 2014-}& 0.25 \\
300K-408K & \mbox{4 April 2014-}& 0.125 \\
408K-510K &\mbox{1 Jun 2014-}& 0.0625 \\
510K-612K &\mbox{1 Aug 2014-}& 0.03125 \\
612K-    & \mbox{after 29 Sep 2014}& \mathbf{0.0001} \\
\hline
\end{array}
$$
\vskip-1pt
\vskip-1pt
\end{table}
\vskip-5pt
\vskip-5pt
\vskip-5pt
\vskip-5pt

{\bf Remark added 01102014: These data are now INACCURATE
(ahead of time).
They will be updated later. On 1 Oct 2014 UNO has only mined
block 495900 and the UNO clock was ticking slower than expected. }

\medskip
In fact this crypto currency smells {\bf programmed self-destruction}.

\vskip-6pt
\vskip-6pt
\subsection{Double or Die}
\vskip-4pt
\label{UnoDoubleOrDie}

At the moment of writing some 2/3 of all coins were already made.
In March 2014 the current price of Unobtanium (UNO)
was about 6 USD
and we again Unobtanium mining was
roughly as profitable as standard bitcoin mining.
However because Unobtanium
uses {\bf the same SHA256 ASICs} as in bitcoin mining,
the computing power (hash power) can shift in both directions instantly.
In particular the computing power in Unobtanium currency is NOT
growing, it is rather declining.  

When the next rewards block halving comes in April,
the price of UNO needs to be at 12 USD
in order to keep mining equally profitable
(cf. later Theorem
\ref{LawDecreasing2xHashRate}
page \pageref{LawDecreasing2xHashRate}).
Then in June it would need to become 24 USD,
then in August it would need to become 48 USD.
Such rapid appreciation at an exponential rate is unlikely to happen
and the hash rate must decline accordingly,
until mining becomes profitable.

\vskip-6pt
\vskip-6pt
\subsection{The Self-Destruction of Unobtanium}
\vskip-4pt

\begin{figure}[!here]
\centering
\begin{center}
\vskip1pt
\vskip1pt
\includegraphics*[width=5.0in,height=2.2in,bb=0pt 0pt 759pt 296pt]{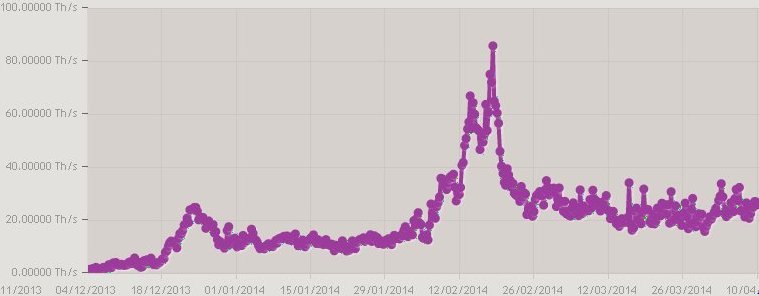}
\vskip-0pt
\vskip-0pt
\end{center}
\caption{
The growth and decline of UNOBTANIUM hash power in the last few monts.
we observe sudden (speculative?) jumps and periods
of intensive mining followed by steady decline
in days following each block halving
date (15 Dec and 12 Feb)
in the hash power
}
\label{BitcoinUNOHashGraph}
\vskip-3pt
\vskip-3pt
\end{figure}


\begin{figure}[!here]
\centering
\begin{center}
\vskip-1pt
\vskip-1pt
\hskip17pt
\hskip17pt
\includegraphics*[width=4.3in,height=1.7in,bb=0pt 0pt 500pt 393pt]{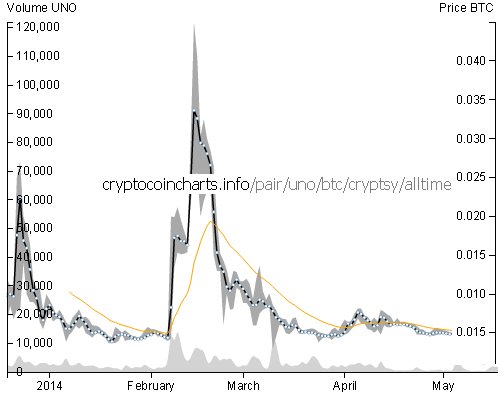}
\vskip-0pt
\vskip-0pt
\end{center}
\caption{
The UNOBTANIUM market price in the same period of time (grey curve) and volume
(yellow) have seen very similar perturbations.
}
\label{BitcoinUNOPriceGraph}
\vskip-3pt
\vskip-3pt
\end{figure}

On Fig. \ref{BitcoinUNOHashGraph}
we see that miners are already running away from this crypto currency.
This happens in sudden slumps as predicted.
There is important decline in the hash rate which occurs some a few days
after block halving dates
after some sort of short period of instability.
We see that the process of rapid self-destruction
has already started for this crypto currency\footnote{
We claim that similar periods of decline are hash power
are also likely to happen for bitcoin, though not before 2015/2016
see Section \ref{FutureBitcoin},
and more quickly for Dogecoin,
at several moments during 2014, see Section \ref{DogeCoinandLitecoin}.}.
The market price of UNO has suffered very similar speculative increases
followed by a periods of collapse as shown in Fig. \ref{BitcoinUNOPriceGraph}.

Unobtanium is a crypto currency which is already destroying itself.
It is bound to always have very small market cap, which implies small anonymity and small adoption.
In bitcoin the decline in mining profitability could be compensated by massive adoption and fees,
and miners do not have a better crypto currency to escape to.
Here the adoption as a payment instrument is close to zero,
fees are zero 
and miners have very good alternatives to switch to.

\newpage
\vskip-6pt
\vskip-6pt
\subsection{A Kill Switch}
\vskip-4pt
\label{UnoDoubleOrDieKillSwitch}

There is much worse than that.
After 29 September 2014
({\bf predicted date, it is now expected rather to happen in November 2014} )
the miner reward is going to
be {\bf divided by 312.5 overnight}.
Then if we want the mining profitability to be the same as today
and the hash rate not to decline,
the price of UNO would need to be 15,000 USD each to compensate for that again
(or mining will not be profitable and hash power protection will go elsewhere).
This would make UNO achieve
a market capitalization of about 4 billion dollars from 0.5 million today.
Unbelievable 8000x growth in a few months.

Of course it obvious that this is not going to happen.
We expect rather that there will be a very fast outflow of hash power
at each reward halving
(cf. Fig. \ref{BitcoinUNOHashGraph})
until we reach again
an equilibrium situation where again mining Unobtanium
will be as profitable as mining bitcoin. 
%
Overall on and before September/November 2014 (exact date is not yet clear, see above)
we predict very rapid spectacular collapse in Unobtanium hash power.

At the same time there can be some appreciation of Unobtanium
due to their increasing rarity and increased popularity.
However this appreciation is unlikely to happen by sudden jumps,
and it is obvious that it cannot achieve 100$\%$ appreciation
every 3 months and 30,000 $\%$ appreciation (300 times increase)
on one single day in September/November 2014.

\vskip-6pt
\vskip-6pt
\subsection{Further Decline?}
\label{UNOCriminalBusiness}
\vskip-4pt

Our prediction is that the hash power in Unobtanium will
{\bf decline to a ridiculously small value} (for example 1000x smaller than today).
If we assume (being VERY conservative and optimistic) 
that Unobtanium miners
mine at the same profitability
threshold as bitcoin miners,
and if UNO pays less miners would switch to bitcoin,
following Table \ref{BitcoinUNOReward}
in September/November 2014
the hash rate is going
to be at most 1250 times lower
than the peak of 80,000 TH/s of February 2014.
This is at most 70 TH/s.
In September/November 2014 anybody should
be able to execute a 51 $\%$
attack on Unobtanium.
For example we can estimate that in order to execute the attack of Section
\ref{DogeCriminalBusiness} based essentially in Fig. \ref{AltChainProfitDoubleSpendAttack}
which is expected to last only about 5 minutes,
the attacker needs to rent 35 TH/s of SHA-256 for about 5 minutes.
It is easy to see that this  would cost only a few dollars.

\medskip
A decline in hash power will inevitably lead to several major problems:
\vskip-6pt
\vskip-6pt
\begin{itemize}
\item
It will become easy to double spend older coins,
there will be permanent for-profit criminal activity
(cf. also Section \ref{DogeCriminalBusiness}).

Yes {\bf in September/November 2014 it will cost only a few dollars to execute a 51 $\%$ attack on Unobtanium}.

\item
It will become easy to run a ``mining cartel attack''
only accept blocks mined by members of a certain group, cf. \cite{MiningSubversive}.
\item
A sudden collapse of this crypto currency will probably
occur much earlier,
as soon as any of these two starts happening,
totally destroying confidence of investors and users in this crypto currency.
\end{itemize}
\vskip-6pt


{\bf Remark.}
It is clear that Unobtanium is in trouble,
and later in April 2014
we observed that
the profitability of UNO mining has declined
and apparently some miners are artificially sustaining it
and accept to mine with lower profitability,
probably in a bid to avoid total collapse of this currency.
We also observed on 28 April that the official web site for Unobtanium is
not even displaying the current hash rate anymore for the second half of April.

\newpage

\section{Another Case Study: Dogecoin vs. Litecoin}
\label{DogeCoin}
\label{DogeCoinandLitecoin}

In this section we we look at two currencies
Litecoin (long time established)
and Dogecoin (started end of 2013)
which are quite comparable
\footnote{
There was a very strong asymmetry between
bitcoin and Unobtanium, bitcoin was always many thousands of times larger
and it was never able to challenge bitcoin in any way.}.
Both currencies use the same hash function (SCRYPT)
and they have historically known comparable hashrates.
The hash power can move freely and
it is possible to see that throughout
most of the recent history of Dogecoin
{\bf EACH currency could be used to attack
each other with a 51 $\%$ attack}.
We are going now to show that
this ``symmetric'' situation is changing very rapidly,
and we will attempt to predict the future of these currencies.

\begin{figure}[!here]
\centering
\begin{center}
\hskip-11pt
\hskip-11pt
\includegraphics*[width=5.1in,height=1.75in,bb=0pt 0pt 1160pt 310pt]{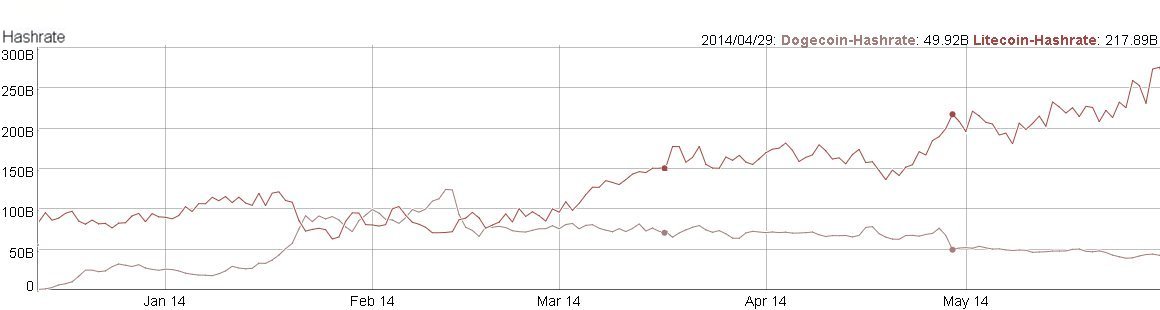}
\vskip-0pt
\vskip-0pt
\end{center}
\caption{
DOGE hashrate compared to LTC hashrate in the last 6 months}
\label{DOGELTCHashRate}
\vskip-3pt
\vskip-3pt
\end{figure}

Dogecoin is a newcomer which has challenged
the incumbent Litecoin very seriously
in terms of achieving a higher hash rate at moments.
However the market capitalization of Litecoin remains at least 8 times bigger
(300 M USD vs. 37 M USD at the moment of writing).
This is because Litecoin has been mined for longer
and more people hold some balances in Litecoins. 


\subsection{Block Halving and Programmed Self-Destruction of Dogecoin}
\label{DogeDoubleOrDie}

In Litecoin no block halving is planned until 30 August 2015, then the reward is halved,
and then the reward remains stable until 2019.
Then it has countless block halving events programmed over a period of some 100 years.

In Dogecoin block reward halving events are only very few but they are all planned to occur very soon
at the very early stage of existence of Dogecoin in the coming months of 2014.
Important events are unfolding before our eyes.

In excessively short time after its creation,
Dogecoin has been able to achieve a comparable and even higher hash rate than Litecoin.
This has lasted until March 2014 cf. Fig. \ref{DOGELTCHashRate}.
On this figure we also observe very strong negative correlation between the two hash rates.
When one goes up, the other goes down, the sum is nearly constant at times.
We take it as a strong 
evidence that the hash power has already been shifting
in both directions between these two currencies.

\begin{figure}[!here]
\centering
\begin{center}
\vskip-5pt
\vskip-5pt
\hskip-14pt
\hskip-14pt
\includegraphics*[width=5.20in,height=2.5in,bb=0pt 0pt 840pt 460pt]{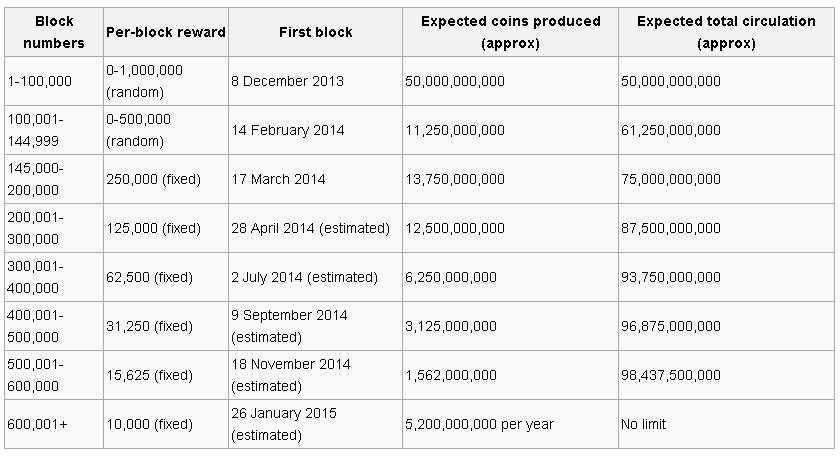}
\vskip-0pt
\vskip-0pt
\end{center}
\vskip-9pt
\vskip-9pt
\caption{
Programmed sudden jumps in DOGE block reward
}
\label{DOGELTCRewardJumps}
\vskip-3pt
\vskip-3pt
\end{figure}

Then on 17 March 2014 the reward was halved cf. Fig. \ref{DOGELTCRewardJumps}.
At this moment the hashrate in Litecoin has immediately adjusted
and switched to another curve,
very precisely in days following 17 March 2014,
cf. Fig. \ref{DOGELTCHashRate}.
This ratio has then been quite stable 
with the hash rate of Dogecoin remaining at or below half of the hash rate of Litecoin.

In this paper we claim that
this is strict mathematics.
When the reward halves,
miners will either see the value of Dogecoin double
or a fraction of miners will
switch and mine for a competing crypto currency.
More precisely miners will be leaving this crypto currency until a new equilibrium is reached:
less miners will be there to share the new (decreased) reward
and therefore the profitability
of their mining operations will be restored.
We have the following result:

\begin{theorem}[Law Of Decreasing Hash Rates]
\label{LawDecreasing2xHashRate}
If the miner reward of crypto currency is decreased 2 times
and the market price remains the same
and if the price of electricity is relatively low compared to the miner income,
the hash rate will be divided by 2 approximately.
\end{theorem}
%

\begin{figure}[!here]
\centering
\begin{center}
\hskip-14pt
\hskip-14pt
\includegraphics*[width=5.20in,height=2.1in,bb=0pt 0pt 883pt 338pt]{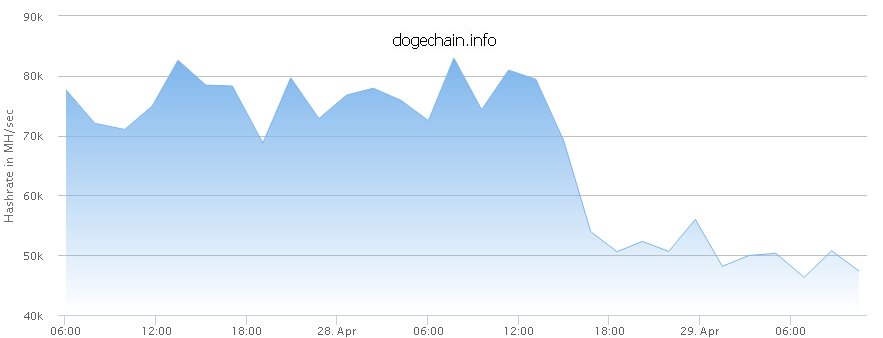}
\vskip-0pt
\vskip-0pt
\end{center}
\caption{
Rapid decline in DOGE hash rate in hours after block halving.
}
\label{DOGELTCRewardHalving28042014}
\vskip-3pt
\vskip-3pt
\end{figure}

Dogecoin has failed to appreciate 2x in value,
therefore the hash rate must decrease 2x.\footnote{
The same phenomenon of rapid decline in hash rate
at moments of block halving,
was also observed with Unobtanium currency,
cf. Fig. \ref{BitcoinUNOHashGraph} in Section \ref{UnoDoubleOrDie}.}
We will see this happen again in Fig. \ref{DOGELTCRewardHalving28042014}.

%

A few more successive block halving events in Dogecoin
are programmed every 69 days
leading to rapid decline in hashing power.
This is again {\bf unbelievably fast speed} for a financial asset,
not less crazy than with Unobtanium cf. Section \ref{UnoDoubleOrDie}.

\subsection{How Vulnerable Is DogeCoin?}
\label{DogeVulnerable}

In this paper we show that Dogecoin is threatened
by the 51 $\%$ attack in more than one way.
For example in April 2014
it was reported that one single pool in DogeCoin was controlling
50.3 $\%$ 
of the network hashrate 
\url{http://www.reddit.com/r/dogecoin/comments/22j0rq/}
\url{wafflepool_currently_controls_503_of_the_network/}
.
Moreover the pool managers can execute attacks without the knowledge of miners,
see Section \ref{HiddenAttacks}.
However bigger threats come from the
fact that the hash power in Dogecoin is declining and
the hash power available outside Dogecoin is becoming
many times larger than the whole of Dogecoin,
knowing that the hash power
used to mine for one currency can be reused
(with our without the knowledge of the miner)
to mine for another currency,
cf. Section \ref{HiddenAttackAcrossCurrencies}.

\subsection{Latest News: Decline Under Our Eyes}
\label{DogeDoubleOrDieActuallyHappening}

The latest Dogecoin halving event
has occurred on 28 April 2014 at 14:32. 
Our theory predicts that at this moment either Dogecoin market price goes up abruptly (not very likely) or
the hash power should be then divided by 2 in a short time. 
At this moment Dogecoin
capability to be protected against double spending
attacks will be seriously affected.

In order to verify if our theory is exact,
we have observed the hash rate of Dogecoin
at \url{dogechain.info}
in the hours following the block halving on 28 April 2014.
We have observed exactly what we expect:
a decline to achieve roughly half of the previous hash rate.
We were in fact surprised by the rapidity of this decline.

In a few hours the Dogecoin hash rate has declined below 50 Gh/s
while AT THE SAME time one single miner had 21.70 GH/s
\url{
http://wafflepool.com/miner/14t8yB3PDGfZT3VppxMY4J9xiBaXUcZvKp},
which data are updated every 15 minutes.

\subsection{Is Dogecoin Under Attack?}
\label{DogeUnderAttack51PcAnd10xEphemeralIncrease}

At one moment at 15h44 we have actually observed
that the hash rate went down to 40 GH/s
for a short moment and conditions for a 51$\%$ attack have been met.
{\bf One single miner had 51 $\%$}
for a short while.

\begin{figure}[!here]
\centering
\begin{center}
\hskip-14pt
\hskip-14pt
\includegraphics*[width=5.00in,height=1.7in,bb=0pt 0pt 883pt 338pt]{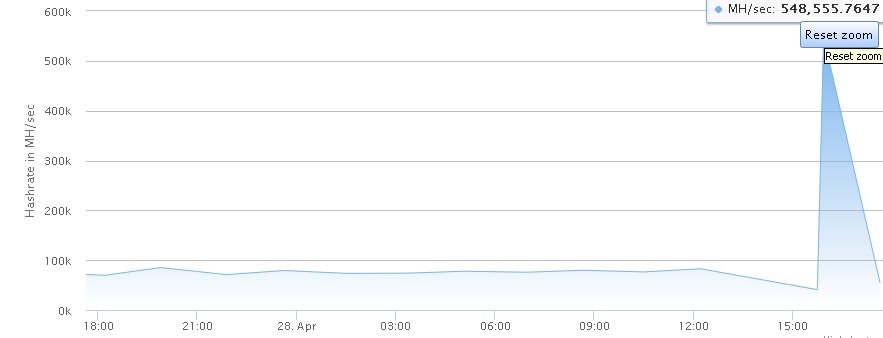}
\vskip-0pt
\vskip-0pt
\end{center}
\caption{
A rapid increase in DOGE hash rate observed in hours after block halving.
}
\label{DOGEIncredible5xIncreaseDogecoin}
\vskip-3pt
\vskip-3pt
\end{figure}

At another moment we have observed that the hash rate
has increased 10 times in a very short time,
see Fig. \ref{DOGEIncredible5xIncreaseDogecoin},
and went back to normal few minutes later.
We do not know if this was an attack on Dogecoin
of the precise sort we study in this paper,
and we do not know how much the
data reported by \url{dogecoin.info} are reliable.
The peak hash rate of 548 TH/s shown at this moment seems too large to be true
and would exceed the hash rate of Litecoin.

\subsection{Near Future - Is There A Criminal Business Case?}
\label{DogeNearFuture}
\label{DogeCriminalBusiness}

It is easy to show that Dogecoin can hardly survive in the current form.


After April 2014 there will be a few more periods
in which the block reward will be halved after 69 days,
cf. Fig. \ref{DOGELTCRewardJumps}, and accordingly
the hash rate is also expected to decline twice at each moment.
Overall we expect that at the end of 2014, the hash rate of
Dogecoin will be already some 32 times smaller than what it
was in February 2014, when it was equal to that of Litecoin.
We expect that very soon Dogecoin
will become
{\bf a perfect target for criminal activity
where money can be made easily}.
Let us discuss if this is really plausible.
We restrict to the question if double-spending
attacks will be feasible.

It has already happened on April 28
that ONE SINGLE MINER had enough hash power
in order to execute a double spending attack.
The worst is however yet to come.
We claim that in the coming months it will be possible
for criminals to execute double spending attacks
with much lower investment.
Here is one possible way for an attacker to proceed:

\vskip-5pt
\vskip-5pt
\begin{itemize}
\item
The attacker needs an initial amount of say 10 times the amount of money
mined in one block, currently about 10x120 USD, he needs about 1200 USD.
\item
He sends 600 USD to some recipient and keeps 600 USD for
the cost of doing the blockchain manipulation.
\item
He executes the attack as in Fig. \ref{AltChainProfitDoubleSpendAttack} page \pageref{AltChainProfitDoubleSpendAttack}
and spends 600 USD on mining.
\item
The attack will be feasible as soon as a
certain fraction of hash power in Litecoin is available
in hosted cloud mining.
It should be at least 51 $\%$ of Dogecoin hash rate
which is going to become very easy in the coming months due to very rapid decline
in the hash rate predicted due to Table \ref{DOGELTCRewardJumps}.

There is also another even more subversive scenario
in which pools automatically provide computing power to the attacker,
without the knowledge of miners and without the knowledge of pool managers,
see Section \ref{TechnicalityFurtherManipulation}.

\item
He is then able to spend his 600 USD again as in Fig. \ref{AltChainProfitDoubleSpendAttack}.
\item
The net profit in this attack is 600 USD and it takes about 5 minutes.
\end{itemize}
\vskip-5pt

\subsection{Additional Signs of Decline}
\label{DogeDeclineSwanson}

Few days after this paper was published,
Tim Swanson from CoinDesk news service
wrote a long paper about Dogecoin
\cite{WhatDogeSurviveAfterMyPaper}
in which he has independently come to very similar
conclusions than in this paper.

\begin{figure}[!here]
\centering
\begin{center}
\hskip-1pt
\hskip-1pt
\includegraphics*[width=5.00in,height=1.7in,bb=0pt 0pt 1911pt 668pt]{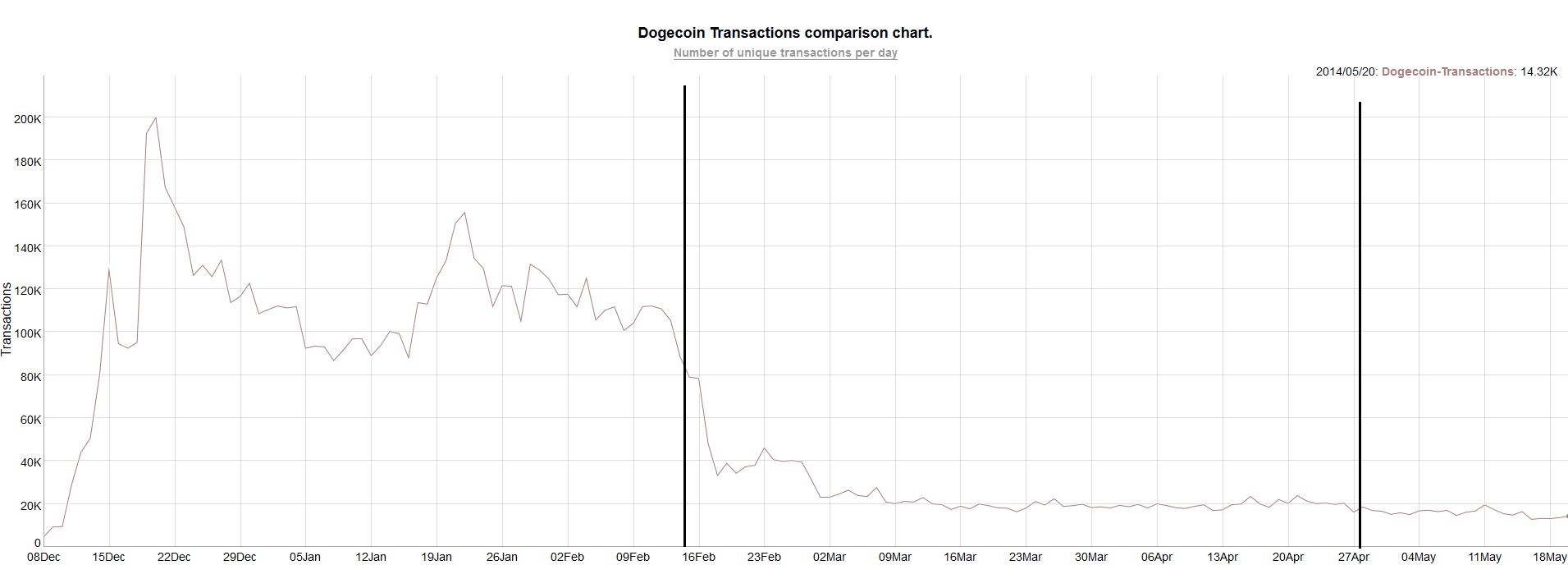}
\vskip-0pt
\vskip-0pt
\end{center}
\caption{
The decline in the number of transactions in Dogecoin observed after
successive reward halving events.
}
\label{DOGEIncredibleDeclineTxVolume}
\vskip-3pt
\vskip-3pt
\end{figure}

The paper \cite{WhatDogeSurviveAfterMyPaper}
displays a very interesting graph which
shows that the popularity of Dogecoin as a currency
has also been declining:
cf. Fig. \ref{DOGEIncredibleDeclineTxVolume}
and \cite{WhatDogeSurviveAfterMyPaper}.

%

\subsection{Better Prospects For Dogecoin in 2015?}
\label{DogeStabilizes2015}

Let us assume that Dogecoin survives until 2015,
and it is not destroyed by massive
outflows of capital, double-spending attacks
and serious for-profit blockchain manipulation or a mining cartel attack,
which will be very surprising.

The the situation is expected to stabilize in 2015.
After January 2015: there will be no more reward halving in Dogecoin.
There will be a steady production of new coins and
progressive but infinite growth of
monetary supply.

\vskip-5pt
\vskip-5pt
\begin{itemize}
\item 98 billion coins will be released by January 2015. 
\item Then some 5.2 billion more coins will be produced each year. \\
It is like a 5 $\%$ increase in the monetary supply in the first year,
slightly less in the coming years.
\end{itemize}
\vskip-5pt

Unhappily at this moment the hash rate of Dogecoin will be
maybe 50 times lower than in Litecoin,
which is what we expect from Table \ref{DOGELTCRewardJumps}.
It will be difficult for Dogecoin to compete with Litecoin.
It is expected to remain permanently weaker,
and if the specification is not changed,
it will become a permanent target for profitable criminal activity,
as shown above.
However the Dogecoin developers can apply some fixes
such as proposed in Section \ref{DoubleSpendingSolutionProposed}
and their currency will be able to function correctly
in spite of having a low hash rate.


\subsection{The Improbable Revenge of Dogecoin in the Long Run}
\label{DogeWinsAfterAll}

Ironically it is possible to see that in the long run,
like after 10, 20 or 30 years, Dogecoin hash rate should again
exceed that of Litecoin,
this is if they are still in existence at that moment
and their miner reward policies are not reformed.
This is because the monetary supply of Litecoin is fixed,
and the monetary supply of Dogecoin is unlimited.
In the long run, Litecoin will see the profitability of mining halved
many times, while it is expected to remain relatively stable in Dogecoin.
Accordingly we expect that the hash rate of Litecoin will in turn decrease
at certain moments (every 4 years, next halving expected in August 2015).
This process is expected to take a lot of time, probably many decades
because Litecoin is more popular than Dogecoin,
and some of the decreased income for miners could be compensated
by the slow appreciation of Litecoin and
higher amount of transaction fees collected in Litecoin.

\subsection{Recent Events - The Rescue Operation - August 2014}
\label{DogeMergedMining}

There is no doubt that Dogecoin can hardly survive more than a few months.
A serious reform and a hard fork of Dogecoin is needed.

This has been finally announced on 4 August 2014,
cf. \cite{SavingDogecoinByMergedMining0814}.
Josh Mohland, one of the key people
behind Dogecoin and creator of the microtransaction
service dogetipbot,
have tried to absolve the Dogecoin creators
from any responsibility in designing a faulty
financial network which exposes users to important risks.
He contended that Dogecoin
was never
"intended to function as a full-fledged
transaction network",
citing \cite{SavingDogecoinByMergedMining0814}.
He clearly agrees with us (the present paper)
that without a reform Dogecoin is in very serious trouble.
In fact he takes an even more radical view
that Dogecoin faces {\bf certain death},
well at least in the sense of double spending attacks,
cf. \cite{SavingDogecoinByMergedMining0814}.
More precisely he has stated that:

\begin{quote}
"Dogecoin was built to die quickly –
none of us expected it to grow into the absurd entity it is today.
With that said,
there's absolutely an easy way to save the coin
from its certain death
(and by death I mean 51$\%$ attacked [...])"
\end{quote}

The solution announced is to implement
merge mining with Litecoin and other similar currencies.
It is "a simple change" according to Mohland.
He also stated that

\begin{quote}
"the risk of a 51$\%$ attack
far outweighs perceived costs"
\end{quote}

Following \cite{SavingDogecoinByMergedMining0814}
the merged mining solution,
or more precisely auxiliary proof-of-work (AuxPoW),
is such that it

\begin{quote}
"enables the dogecoin block chain to receive work from other scrypt-based networks.
Current Dogecoin miners will still be able to generate blocks and receive DOGE,
but now, litecoin miners will contribute hashing power to the dogecoin network."
\end{quote}

It is important to note that this has followed many months
of intense debates in the specialist communities about what to do to save DogeCoin from destruction,
cf. also \cite{WhatDogeSurviveAfterMyPaper} and this paper.
Following \cite{SavingDogecoinByMergedMining0814}
the Litecoin creator Charlie Lee
have very generously suggested to merge the mining back in April 2014,
but it was initially not well received
by the Dogecoin community\footnote{
Or maybe simply the necessity of doing something in order to save Dogecoin from destruction
was not yet well understood.}.
Finally they have accepted this solution,
as probably (in our opinion) every other reasonable solution
would require for Dogecoin to break their monetary policy
and produce more coins, diluting the current coins.
See also Section
\ref{DeflationaryCoinsGrowthCoinsTheorySketch}
and Section
\ref{sec:CompetingCryptoCurrencies}.

\newpage
\section{Future of Bitcoin: Is Bitcoin Strong Enough to Avoid Programmed Decline?}
\label{FutureBitcoin}

Now we are going to speculate about privileged moments in time at which bitcoin
could see a decline in its hash rate.
The next block reward halving in bitcoin is predicted to happen on
{\bf 22 August 2016} according
\footnote{
However this is subject to some known irregularities
and imperfections in the automatic difficulty adjustment mechanism of bitcoin.
It is known that the bitcoin clock have been accelerating.
Some authors claim the block 420,000 and the block reward halving
will happen at up to 1 year earlier,
maybe in May 2016, maybe as early as September 2015,
see \url{https://bitcointalk.org/index.php?topic=279460.0}.
}
to \url{bitcoinclock.com}.

We predict that a major crisis of bitcoin digital currency could occur at this moment.
In fact however it does not have to be so.
We predict that bitcoin will be in trouble
{\bf only if} some preliminary conditions\footnote{
See also
Fig. \ref{BitcoinDestrTheorySummary3Things}
page \pageref{BitcoinDestrTheorySummary3Things}.
}
are also met at this date:
\vskip-7pt
\vskip-7pt
\begin{enumerate}
\item
If bitcoin mining has sufficient competition by that time,
\item
If miners are willing and able to 
reprogram their ASIC machines
to mine for other competing crypto-currencies,
\item
If overall mining market outside of bitcoin
will be large enough 
to provide a better mining income in a sustainable way:
even if there is a massive
transfer of hash power
from bitcoin to these alternative crypto currencies.
\item
If bitcoin specification is not changed
(cf. changes proposed in Section \ref{DoubleSpendingSolutionProposed}).
\end{enumerate}
\vskip-6pt

Then we predict that at this next bitcoin block reward halving
(in or before August 2016),
the hash power will massively shift to other crypto currencies.
This could possibly destroy
the reputation of bitcoin
as it might suddenly become vulnerable
to 51 $\%$-like attacks such as described
on Fig. \ref{AltChainProfitDoubleSpendAttack} page \pageref{AltChainProfitDoubleSpendAttack}.
We stress that such transition could happen nearly overnight, on some day in 2016.



\subsection{Possible Consequences}
\label{FutureBitcoinConsequences}
At a certain moment in the future we predict a rapid transition to occur
and bitcoin becoming vulnerable attacks.
We expect that such a transition can lead to a rapid decline of bitcoin
as people can switch to other competing crypto currencies very quickly
as soon as double spending suddenly becomes feasible to execute in bitcoin.
More importantly, merchants would probably
{\bf all of the sudden stop accepting
any bitcoin payments whatsoever} (the tipping point).
This would be as soon as it becomes profitable to
commit double spending attacks and therefore it will become
very risky to accept any bitcoin payments
(as they can be reversed later).

\subsection{Counter Arguments}
\label{FutureBitcoinCounterArguments}

It is very difficult to predict the future.
{\bf How can we claim} that a 50 $\%$ reduction in mining income will make
miners massively quit bitcoin mining?
This seems to be in contradiction with recent bitcoin history.
In fact the actual reward for every existing bitcoin mining machines
{\bf HAVE BEEN divided by two} countless times already.
For example it was divided by two NEARLY EVERY MONTH in the last 12 months,
see Fig. \ref{BitcoinGraph1YHashRate}.
Yet people {\bf did NOT} go to mine
for other crypto currencies
at a massive scale.
There was no important displacement of hash power, though certainly there was some
(which works in both directions,
many miners people also switched from other currencies
back to bitcoin mining,
see Fig. \ref{BitcoinUNOHashGraph}).
Overall the majority of people kept mining bitcoins as usual.

\medskip
The reason why miners did not stop mining bitcoins
is that miners had no choice so far. No plausible alternative to switch to.

\subsection{Decline or Persistent Domination?}
\label{FutureBitcoinConclusionDeclineUnlessPersistentDomination}

We observe that until now
there was not a sufficiently strong SHA256-based
bitcoin competitor to switch to (LiteCoin does not apply).
As long as bitcoin remains a {\bf dominant} monopolist crypto currency,
our predictions about decline of bitcoin simply do NOT work.

Now we anticipate that sooner or later competition to bitcoin will be there.
One or several SHA256-based crypto currencies will be able to provide higher returns for
miners contributing raw hash power.

\medskip
\medskip
{\bf Remark.}
This is more than just an opinion.
We believe that in the future
one should be able to develop a sort of economic theory
which shows that this is very likely to happen
as already explained in Section \ref{DeflationaryCoinsGrowthCoinsTheorySketch}
as a predictable 
consequence
of several contributing factors:
current monetary and reward policies which erode
the miners' income\footnote{
One argument for this (due to J. Kroll)
was that bitcoin reward policy is NOT generous enough and
does NOT reward miners well enough in the long run,
see Section \ref{KrollKeepHighIncentiveForMiners}. }
with important and sudden jumps\footnote{
Such sudden jumps have no justification whatsoever,
they can only be harmful.
They are NOT justified even if we keep
the 
premises of fixed monetary supply,
see Part 3 of \cite{MiningUnreasonable}.},
competitive markets\footnote{
When mining becomes less profitable miners are going to increase
transaction fees which is going to seriously affect
the adoption of bitcoin as a medium of exchange,
see Section \ref{IncreasingFeesArgument}.}
and other factors\footnote{
We can also argue that one of the reasons why bitcoin has attracted such a growth
was the expectation it will raise a lot, which is due to built-in unreasonable
deflationary monetary policy.
Then once bitcoin have achieves the peak of possible appreciation,
possibly already in 2014,
other crypto currencies with ``more reasonable'' policies and settings
in the sense of Section \ref{DeflationaryCoinsGrowthCoinsTheorySketch},
are likely to emerge as obvious challengers
and drive bitcoin out of business.
%
}
including precisely their
yet lower level of protection for some currencies\footnote{
Additional important shifts in hash power could occur because
several criminals might simultaneously be trying to exploit
all other SHA256-based crypto currencies
in which double spending attacks
will be easier to execute
by displacing hash power rapidly in both directions,
also possibly playing with automatic difficulty adjustments
in these currencies at the same time.}.

%

\subsection{Could Bitcoin Be In Trouble Earlier?}
\label{FutureBitcoinConclusionDeclineForReasonsOfPoorAdoption}

A predicted decline of bitcoin in the future
could due to for profit blockchain manipulation
or other reasons,
and could happen much earlier than predicted,
because of:

\vskip-5pt
\vskip-5pt
\begin{enumerate}
\item
Bad reputation:
very substantial proportion of bitcoin in circulation
are already a product of criminal activity,
cf. \cite{BitcoinExposuresList,MtGoxDoubleSpendingDecker}.

\item
A decline in bitcoin popularity as a currency for ordinary people:
Bitcoin popularity is in decline
cf. Fig. \ref{BitcoinGraph1YGoogleSearchPopularity}
page \pageref{BitcoinGraph1YGoogleSearchPopularity}
and transaction activity is in decline
cf. Fig. \ref{BitcoinGraph1YTxFeesPerDay}
page \pageref{BitcoinGraph1YTxFeesPerDay}.
There are further alarming indicators
which are not always correctly interpreted
by the news reports,
see Section \ref{IsBitcoinUsedEconomyMOE}.
Decline in popularity could somewhat inevitably follow due to centralisation of power\footnote{
For example in \cite{KaminskaRegRvasionIsntDisruptiveInnovation} we read that
"
Bitcoin's [...] use of collaborative community to police the problem of double-spending [...]
can only remain valuable for as long as it is not overtly exploited by just a few hands.
The more a few hands monetise and dominate the system, the more it threatens
to lose the users and participants
that make it have value in the first place. [...]
%
whenever you try to extract value from voluntarism,
[...] incentive to stick around will dissipate eventually."
},
cf. GHash 51$\%$ fears, centralized code development, only rich people can mine etc.

\item
There is a steady decline
in number of peer-to-peer network nodes
interested in supporting the bitcoin network.
The number of active network nodes is falling
below reasonable levels,
cf. \cite{NumberOfReachableNodesIsDroppingRecently}.

\item
Bitcoin could also be destroyed
by an unhappy network or security incident.
A well known bitcoin expert 
Antonopoulos considers that
there is a risk that "we blow it up by accident",
cf. \cite{AntonopoulosIgnorant51PcLAJan2014}.
%
%
%
%

\item
Bitcoin is also threatened by bad governance
and effective self-destruction of bitcoin
by the very people who run, develop and promote it every day.
For example due to
a promotion of mistaken ideas
and a serious lack of pro-active security engineering
in bitcoin community,
cf. Section \ref{FutureBitcoinConclusionBitcoinRegulatedCustomerFiskDevelopers}.
A well-known bitcoin core developer
Peter Todd have in June 2014 identified
"what needs to be changed" in bitcoin
in order to make it less centralized\footnote{
He postulates 3 exact points to be changed/reformed,
1. Eliminate pools 2. Make solo mining profitable 3. Eliminate ASICs.
All these 3 are about making bitcoin mining
more democratic and at the same time more resilient.
In addition it seems that goals 1. and 2. are rather realistic
cf. \cite{CornellTimeForHardForkAvoid51GHashWithholdingEtc},
while goal 3. is more problematic and might never be achieved,
cf. \cite{ToddWhyISold50PcOfMyBTC,ToddBitcoinEcosystemWillMaybeBreakDown}.}.
He fears that bitcoin community and bitcoin developers
are not up to the challenge
and are unwilling or unable to see or/and address these problems:
we hear that
"it will take a system failure to get people to agree to implementing these changes"
cf. \cite{ToddBitcoinEcosystemWillMaybeBreakDown}
and
"it might take a disaster to get the consensus to fix it"
cf. \cite{ToddWhyISold50PcOfMyBTC}.
Accordingly we read that
{\bf "Peter is preparing\footnote{
This is one of the reasons why he publicly announced
that he sold 50 $\%$ of his bitcoins
\cite{ToddWhyISold50PcOfMyBTC}
which he claims was because
"I made a promise to myself a while back that I'd sell 50$\%$ of my bitcoins
if a pool hit 50$\%$, and it's happened".
}
 for the possibility that the Bitcoin ecosystem will break down"}
\cite{ToddBitcoinEcosystemWillMaybeBreakDown,ToddWhyISold50PcOfMyBTC}.

\end{enumerate}
\vskip-5pt

\newpage

\section{The Questions of Customer Risks,
Trusting Bitcoin Developers,
How Much Can We Trust Satoshi and Academic Researchers,
and Possible Future Regulation}
\label{FutureBitcoinConclusionBitcoinRegulatedCustomerFiskDevelopers}

In this paper we have studied a lot the question of
risks related to the miner behavior.
Great majority of miners are anonymous
and bitcoin mining suffers from
dangerous centralization and insufficient network neutrality,
cf. Section \ref{DoubleSpendingSolutionCharacteristics}.
We have described endless scenarios
in which miners can influence which transactions get accepted
or miners can be abused by a sort of man-in-the middle attack
in order to take part in criminal activity
cf. Section \ref{LongestChainRuleDoubleSpendingAttacks}
and Section \ref{HiddenAttacks}.

Now there are also risks related
to the bitcoin source code development.

\subsection{The Question of Bitcoin Source Code}

Bitcoin has this ``anonymous founder'' syndrome\footnote{
Quite happily with the exception of the founder
current bitcoin developers are not anonymous,
see \cite{BitcoinMainSoftwareDistribution}}.
There were numerous security scandals
in which a lot of bitcoins have been stolen
\cite{BitcoinExposuresList,MtGoxDoubleSpendingDecker}.
Alt-coins are much more vulnerable:
the Dogecoin network is facing nearly total disintegration
see Section \ref{DogeMergedMining} and \cite{SavingDogecoinByMergedMining0814,WhatDogeSurviveAfterMyPaper},
and the same is true for Unobtanium cf. Section \ref{UnoDoubleOrDieKillSwitch}.
All this can create some uneasy feelings.

It is a common misconception to believe that
open source code is 
most probably secure.
There are several very serious questions:

\vskip-6pt
\vskip-6pt
\begin{enumerate}
\item
Why should open source code be secure
if very little or insufficient effort
is typically
made in order to make it secure?
\item
In the traditional industry
developers are paid, 
and seem to never get the security right:
we have endless security breaches and alerts.
%
Can we ever hope that bitcoin developers
will take care of security?
In fact bitcoin has always been presented
as experimental rather than mature system.

\item
In a similar way
the Dogecoin developers and promoters
do not want to admit responsibility
for their own actions and their consequences.
We hear that Dogecoin was
never
"intended to function as a full-fledged
transaction network",
citing \cite{SavingDogecoinByMergedMining0814}
and that consequently at this moment
it faces nearly "certain death",
this in fact mostly and exactly
for reasons studied in the present paper,
cf. Section \ref{DogeMergedMining}
\cite{SavingDogecoinByMergedMining0814}. 
%

\item
Actually open source software is not more secure
than closed source according to \cite{AndersonOpenClosedEquivalent}.
Moreover quite possibly, on the contrary, it will be less secure.
Malicious developers
are {\bf more} likely to work on such source code
than honest developers.
This is because rogue developers will be motivated
by profit,
while honest developers
will see no incentive to work on this code.
\end{enumerate}
\vskip-6pt

Accordingly, a recent paper
\cite{WallStreetLawyerBitcoinMoreAdvancedThanNMS}
takes the view that
customers in the area of financial services
should be 
protected
against 
security risks
through some sort of regulation
of precisely open source software systems.
We read that:

\begin{quote}
The open-source nature of the developer population
provides  opportunities  for  frivolous  or  criminal behavior
that can damage the participants in the same way that investors can be misled by promises
of  get  rich  quick  schemes.  [...]
Regulations could ensure that cybersecurity requirements are engineered into the code
[...] 
\end{quote}

This is a somewhat very surprising proposal,
see
Section
\ref{FutureBitcoinConclusionLongestChainRuleToBeRegulated}.
%

\subsection{The Question of Risk Awareness and Security Culture,
Bitcoin vs. Information Security}

In addition, there is another very serious and closely related problem.
It the quality of the public discourse about the security of bitcoin
(such as on the Internet, in the press,
in bitcoin forums, public events, specialist conferences, etc).
As a security professional
{\bf
we are under a definite impression that
insufficient attention is paid to security questions
in the bitcoin community at large}.
For example we can observe that:
\vskip-6pt
\vskip-6pt
\begin{enumerate}
\item
Extremely few professional security researchers study bitcoin.

\item
There are hundreds of conferences about bitcoin each year
but almost none of them ressembles in any way an academic
information security conference.
For example 
no community-run bitcoin conference
has a competitive open
call for papers and publishes
contributed works in the form
of proper academic papers.

\item
Furthermore
the volume of academic research on bitcoin published each is year
is astonishingly low
compared to the number of
press, media, blog and other coverage of bitcoin
in the public media space.
The effect of this is that the
many important questions
concerning security of bitcoin
are excessively simplified, badly understood,
distorted or ignored.

\item
The bitcoin foundation contains no single academic information security expert
and lacks cyber-security culture,
cf. Section \ref{CyberSecAssumeWorst} below.

\item
It is even more striking for cryptography.
Almost every day we hear sentences such as "in crypto we trust" in bitcoin community.
However this crypto currency is run by a group of people
which does not contain
a single academic cryptographer.
\end{enumerate}

The dominant discourse about bitcoin
is always excessively optimistic
and does not expand on
the security threats and risks
\cite{BitcoinExposuresList,MtGoxDoubleSpendingDecker}.
When it does,
it quite shallow, naive and superficial.
We have known this sort of situation for decades in
the area of computer software and security.
However here the situation is arguably really different.
For many people
ignoring the risks
{\bf
-- specifically in the financial sector --}
is not acceptable.
This is why the financial sector is typically regulated
in most countries,
and it is typically required that the customers
should be made aware of the risks.
This at least for assets which are likely to be used
by ordinary people to put their savings in,
which is clearly the case for bitcoin,
cf.
\cite{WallStreetLawyerBitcoinMoreAdvancedThanNMS}
and Section
\ref{FutureBitcoinConclusionLongestChainRuleToBeRegulated}.

\subsection{Optimistic vs. Pessimistic and Cybersecurity Culture}
\label{CyberSecAssumeWorst}

The golden standard in information security is
"It's always better to assume the worst"
because "when the unexpected happens,
you'll be glad you did"
this following the well-known
information security engineering
and applied cryptography guru Bruce Schneier
\cite{SchneierAssumeTheWorst}.
We don't exactly see that in bitcoin community.
Almost every day we hear certain 
bitcoin supporters and commentators
being 
very negligent
or propagating
poorly informed opinions\footnote{
Let us give just one example:
in \cite{AntonopoulosIgnorant51PcLAJan2014}
we hear
that if SHA256 is broken
"it doesn't matter at all" for bitcoin,
and further claiming that it would ONLY
somewhat affect the miners.
In fact SHA256 concerns every single bitcoin transaction which is hashed with SHA256
before signing.
If SHA256 is broken, their authenticity could no longer be guaranteed.
We have see many more similar serious 
mistakes in the bitcoin community
where serious security questions
do not get a chance to be discussed by proper security experts
who make an effort to really understand
and study these questions properly.
}
when discussing
some major risks and threats in bitcoin.
Many serious security questions we are aware of
have been already somewhat discussed
in some specialist bitcoin forums.
However we observe that:

\vskip-6pt
\vskip-6pt
\begin{enumerate}
\item
Major risks and threats
such as studied in the present paper
are far from being understood correctly.

\item
Attacks and defenses are not studied in a systematic way.

\item
The approach of bitcoin developers
is very clearly a ``risk taking'' approach
rather than avoiding the risks in order not to take chances,
cf. \cite{SchneierAssumeTheWorst,ToddWhyISold50PcOfMyBTC,ToddBitcoinEcosystemWillMaybeBreakDown}.

\item
We also see a lack of
informed expert opinions
which would warn the public about risks.
This paper alone is unlikely to solve this problem,
and we write mostly for a specialist audience. 

\item
There is an excessively large volume of text
in bitcoin forums, literally tens of thousands of pages,
and everybody is entitled to their opinion,
which makes it hard
to understand what is really going on
and how important the problems really are. 
\end{enumerate}
\vskip-4pt

Bitcoin developers and bitcoin foundation cannot be blamed for
all the security problems \cite{BitcoinExposuresList}
and cannot be blamed for not having
an army of cyber-security experts
which is there to defend bitcoin against attacks.
There might have been doing their very best efforts.
However it is a fatal mistake
for people running a financial systems used by millions of people 
not to seek help of cyber-security and cryptography professionals.
An enthusiastic optimistic promotion of bitcoin technology and software 
cannot justify an easy going approach
which dismisses 
the risks as the last thing
bitcoiners should worry about
or something reserved
for strict specialists to study,
while in fact they concern
every single user of this financial system.

\newpage
\vskip-6pt
\vskip-6pt
\subsection{More Specifically - Double Spending and 51 $\%$ Attacks}
\vskip-4pt

The question of 51 $\%$ attacks
is very frequently discussed in bitcoin community
and yet remains very poorly understood.
We have observed a 
worrying tendency to systematically
present these problems in the wrong light
and using highly misleading vocabulary,
which makes it very difficult
to see what the real problems are.
More precisely it appears that 
the majority of people
have a very restrictive
and overall simply totally incorrect view
of %
51 $\%$ attacks,
cf. Section
\ref{LongestChainRuleDoubleSpendingAttacksReasonWhySoBadlyUnderstood}
which is such that
it essentially 
ignores these important risks totally
or dismisses them under fake pretexts.
%
%
Many of these misunderstandings 
can be directly blamed on the mysterious founder
as we will see below,
however
other people need also to be blamed as will see later.

We start by recalling the point of view of Satoshi \cite{SatoshiPaper} on the 51 $\%$ attacks.

\subsection{Satoshi On 51 $\%$ Attacks}
\label{SatoshiOn51}

As already explained in Section
\ref{LongestChainRuleDoubleSpendingAttacksReasonWhySoBadlyUnderstood}
the original paper of Satoshi \cite{SatoshiPaper}
is a direct source of some very major misunderstandings in bitcoin.
In Section 6 of this paper Satoshi expands on
the monetary ``incentive'' given to miners
for mining now blocks.
It is all about how paying the miners
for mining bitcoins is expected to make them behave honestly,
and how it would be in their best interest to behave well.
Initially Satoshi writes that:

\vskip-7pt
\vskip-7pt
\begin{quote}
"The incentive may help encourage nodes to stay honest."
\end{quote}
\vskip-6pt

Until know we are inclined to agree.
It may or it may not.

\vskip5pt
At this moment the discourse becomes
much more specific about what the attacker
is expected to be like and what he is likely or/and able to do:

\vskip-5pt
\vskip-5pt
\begin{quote}
"If a greedy attacker is able to
assemble more CPU power than all the honest nodes, \\
he would have to choose between using it to defraud people
by stealing back his payments, or using it to generate new coins. \\
He ought to find it more profitable to play by the rules,
such rules that favour him with more new coins than everyone else combined,
than to undermine the system and the validity of his own wealth."
\end{quote}
\vskip-5pt

We see an image of a powerful entity which ``assembles''
a lot of CPU power under his exclusive control.
The attacker is also represented as being wealthy,
and we are inclined to believe that wealthy people
do not want to engage in fraudulent behavior of any sort.
However in most situations the attacker
does NOT need to be very powerful to run double spending attacks.
He does NOT need to be wealthy.
Miners can just be tricked to participate in an attack
without their knowledge, with man-in-the-middle approach,
and the cost of such attacks is not very large,
see Section \ref{LongestChainRuleDoubleSpendingAttacks}.

Moreover 
very clearly Satoshi
makes an important {\bf technical error} 
here.
He makes us believe that if someone commands a lot of the hash power,
he will also be capable of
"using it to generate new coins".

This is totally {\bf incorrect}
and in a great majority of cases the attacker cannot steal coins\footnote{
As an exception to the general rule,
there are known cases of attacks on pooled mining
where the attacker would be able to obtain the coins which were mined 
cf. \cite{RedirectedMiningStealingCoinsAugust2014}.
In this attack the hacker was more powerful than we generally assume in this paper:
he has hacked some major Internet service providers,
and the attack could have been prevented by standard network security techniques such as TLS.}.
The key remark is that in the mining process the miner
just needs to know the {\bf public key},
while one needs to be able to steal or modify the {\bf private key}
in order to "generate new coins" for the attacker.
There is plenty of ways for miners to operate and in most cases
the attacker will be able to make the miner work for him
without being able to ever steal his private key\footnote{
This regardless whether this private key is hold by individuals
(e.g. when  mining with Eligius) or by the pool manager (the most frequent case).}.

The founder of bitcoin can potentially
be forgiven for this enormous technical 
blunder.
After all he clearly makes another major confusion here:
he says "nodes" and he means "miners".
He clearly did not anticipate things
such as pooled mining: Satoshi has written
that in bitcoin every peer node will be
mining, cf. Section 5 of \cite{SatoshiPaper}.
Satoshi would probably be very astonished to see
that now the number of miners is now much higher than the number of peer nodes
which is reaching dangerously low levels \cite{NumberOfReachableNodesIsDroppingRecently}.

However
Satoshi is not the only person who gets it badly wrong.
For example two Cornell researchers Eyal and Sirer
\cite{CornellTimeForHardForkAvoid51GHashWithholdingEtc}
also clearly badly confuse
between miners which may "hold 49 $\%$ of the [mining] revenue",
with the control of hash power for the purpose of mining blocks,
see \cite{CornellTimeForHardForkAvoid51GHashWithholdingEtc}.
%
Similar mistake is found in \cite{CoinDeskIgnorantPaper51} 
and many other sources.
%
%
Almost every day we hear about 51 $\%$ attacks
in such a way as to
ignore the actual threats 
and somewhat obliterate
any sort of informed opinion 
about these important risks.
Additional more specific
examples will be discussed below.

\subsection{On Careful Approach To Risk}

We work in information security.
Traditionally we have the following pattern in security research.
On the one side,
the industry tends to minimize the risks and frequently
will dismiss or minimize the concerns about security problems,
cf. for example
\cite{GavinRejectsCentralizationConcernsAgain} for bitcoin.
On the other side the academics tend to play the devil's advocate \cite{SchneierAssumeTheWorst}.
Security experts have been trained to "always"
attempt to "assume the worst", see \cite{SchneierAssumeTheWorst}.
In this paper we also frequently work from this point of view\footnote{
This is not always ideal,
we also need to look at the average case, the most probable case, etc.
However assuming the worst (within limits)
remains the golden standard in security research,
this is simply security and vulnerability analysis work
which needs to be done and requires a good level of attention.}
Academic security researchers
will potentially even exaggerate the risks,
hoping to influence the industry
not to take some important risks
and to improve the security baseline.
This makes a lot of sense, what is secure
or secure enough today
will maybe no longer be secure tomorrow.

Quite interestingly, there is also plenty of examples of academics
which take careless positions,
or just not lucky and are later proven to be mistaken. 
As soon as there are systems in the design of which academics somewhat participate
or are trying to participate\footnote{
In particular open source systems like bitcoin
with an ongoing public discourse about their future improvements,
reforms or security enhancements.},
the academic discourse changes too.
We need to learn also to mistrust the academics at times.
They also tend to systematically to delude themselves
that some systems 
are very secure.
They will also on occasions claim that the risks and threats are
small to inexistent,
and may be proven badly wrong later on.

It remains that academics are typically very good at pointing out flaws
in systems designed by others 
and they dedicate a lot of time and energy to that.
For this reason the cautious and critical approach,
sometimes maybe even excessively cautious and critical,
is expected to remain dominant in the academia.

\subsubsection{More Than Careful?}
We are also going to contend that financial systems do require a slightly
{\bf more} cautious
approach than we already have
in cryptography and security.
This is because the financial sector is NOT like any other sector.
It is subject to specific stringent laws regulations,
and it is supervised and monitored by various government authorities.
A blissful lack of appreciation of dangers of technical attacks on bitcoin
is not a good idea, 
because
it may mislead the public to put their money at great risk
cf. Section \ref{FutureBitcoinConclusionLongestChainRuleToBeRegulated}
and  \cite{WallStreetLawyerBitcoinMoreAdvancedThanNMS}.
Security professionals have a moral, professional
and frequently also a legal
obligation to uphold high security standards.
Specific legal obligations
exist in the financial industry
\footnote{As an example we can cite the safeguards rule in
the US Gramm-Leach-Bliley Act [GLBA] from 1999.}
and there are here typically stronger
than in other industries.
University researchers funded by public money are also among other here
to warn and inform the public
about all the dangers of using open source systems such as bitcoin.

All these are additional reasons why we need to be very careful
when we make statements about security of bitcoin.
In addition the nature of scientific research is such that
for most questions there is no final definite answer,
and opinions vary very substantially.

In what follows we are going to delve deeper
into the questions
of what exactly bitcoin miners
which control a lot of computing power could possibly do,
as these are crucial questions in bitcoin.



\newpage

\subsection{Sirer vs. Felten Debate}
\label{PowerControversyFelterSirer}

The question of what exactly a 51 $\%$ attacker can or cannot do
is one of the most frequently discussed questions in bitcoin.
In June 2014 Felten, a well known blogger in the technology space,
has written the following words on his blog
\cite{FeltenMinersProbablyCanChangeMonPol}:

\vskip-4pt
\vskip-4pt
{\bf 
\begin{quote}
"One way to understand the potential power of a 51$\%$ attacker \\
is to consider that they can simply change the rules of Bitcoin at any time.\\
And the changes could in principle be drastic: \\
a ``pay me a 5$\%$ fee on every transaction'' rule, \\
or ``a million new Bitcoins exist and belong to me'' rule".
\end{quote}
}
\vskip-3pt

Few days later Sirer,
a well-known university researcher 
has written on his blog the following statement
\cite{CornellVotingPowerAmazingAntiEthicalClaims}
which claims exactly the contrary
in such radical terms that
it belongs to 
the far remote end
of the spectrum of possible opinions:

\vskip-4pt
\vskip-4pt
{\bf 
\begin{quote}
"the miners' hashing power has absolutely no say \\
in determining how the protocol evolves".
\end{quote}
}
\vskip-3pt

Both positions are very strong,
and represent two radically different points of view.
Moreover it is clear that it is not really possible to agree with
any of these two positions.
Inevitably the truth lies somewhere between these two statements.

Interestingly, Felten has not been so far able to
defend and justify his position very well.
When confronted with strong objections to his statement,
he disagrees with the critics,
however
even if find what he says plausible,
we do not see any convincing arguments in his blog
\cite{FeltenMinersProbablyCanChangeMonPol}.
In contrast Sirer has written a whole long blog entry
which tries very hard to justify his opinion
and presents several different arguments.
We will examine these arguments below,
and show that
in the light of the present paper
and our improved understanding of 51$\%$ attacks,
none of his arguments are really convincing.

Overall we will see that the
control of 51$\%$ of the hash rate gives
very substantial powers to miners,
but not an absolute power.
We will agree with Felten
that "a true 51 $\%$ attack is more serious than people
have generally recognized", cf. \cite{FeltenMinersProbablyCanChangeMonPol}
however we do not believe when Felten suggests that
some truly drastic changes could be imposed by miners.


\subsubsection{Can The Position of Sirer Be Justified?}
\label{PowerControversyFelterSirer2}

We present here one truly amazing citation from Sirer's blog
\cite{CornellVotingPowerAmazingAntiEthicalClaims}:

\vskip-4pt
\vskip-4pt
{\bf 
\begin{quote}
"A 51$\%$ miner does not have 51$\%$ of the vote; \\
in fact, GHash has just as much say \\
over the contents of the blockchain \\
as do I, or you, or anyone else".
\end{quote}
}
\vskip-3pt


Independently of the exact context in which this was written,
which is slightly confusing\footnote{
Maybe the answer in a long term perspective could be different
than the obvious fact that in the short run it is the miners who decide.}
we find this citation very surprising.
GHash would certainly not agree.
They have never denied that 51 $\%$ attacks are
rather a "serious threat", 
cf. \cite{GHashPressRelease}.
How is it possible to write such a statement?
Can it be ever justified?


This citation will maybe just amuse 
specialists
who will debate vigorously on its merits in 
hermetic technical papers
and dismiss 
it eventually, 
or contend 
that maybe Sirer meant something like,
well in the long run and under certain conditions
this bold statement might possibly hold.
For example Sirer introduces a notion of so called Chain Power,
which is not a well defined notion and means something like
the power to create a long term consensus about
what kind of blockchain is acceptable. 
Hypothetically the miners would only make decisions guided by
"what the buyers and sellers accept as the legitimate blockchain"
and would not do anything which a broader community would not like.

This is indeed possible,
well in theory, and in an extremely optimistic scenario.
However it is simply very
naive to count on it.
This is {\bf NOT what security engineering stands for}
which should always try to assume
that bad things are likely to happen
\cite{SchneierAssumeTheWorst}.


\subsubsection{On Voting.}
\label{PowerControversyFelterSirer3}
This is not the only very surprising statement in this blog.
It also covers the question of voting power and disputes
one of the most fundamental\footnote{
The decision by the majority of hash power was introduced by Satoshi
following the original idea by Adam Back \cite{HashCash}. 
It exists for some very good reasons:
it is believed that voting weighted by computational power 
is really the key property which allows potentially to
solve some very major difficulties in the design of secure distributed systems
such as the well-known Byzantine generals problem \cite{AntonopoulosIgnorant51PcLAJan2014}
and Sybil attacks \cite{SideChains}.
}
basic properties of bitcoin:
the fact that the voting power in bitcoin is proportional to one's hash power.
In this space Sirer claims that somewhat (presumably rather only in the long term)
in bitcoin everyone
"gets a single vote, no more, no less, on what kind of a blockchain they will accept."
We do not deny that users and bitcoin adopters could exercise some influence,
even though nobody is quite sure how\footnote{
We should ask is this ever possible at all:
users have no influence in the short run, why would they have one in the long run?}.
However in addition it is claimed that
"Miners are [..] just like every other user."
Putting an equality sign between
the powerful Ghash mining pool or
a very powerful miner who has invested hundreds of millions of dollars,
and "you, or anyone",
cf. again \cite{CornellVotingPowerAmazingAntiEthicalClaims},
potentially ordinary people involved in bitcoin transactions,
is rather irresponsible.
This citation does not only disregard major threats
which concern almost every single bitcoin transaction.
In addition ordinary users are made 
to believe that
THEY are in charge
and that they have something to say in bitcoin.
This form of wishful thinking about security
and a belief in wisdom of crowds
exists in computer science,
for example in the open source software community.
However
%
%
in information security
it is NOT a good practice to be naive and dismiss the risks.
On the contrary, cf. \cite{SchneierAssumeTheWorst}.
Researchers has historically always tried
to defend the public against threats and attacks.
Finally, again the situation becomes more serious
because and when
there is a lot of money at stake.
%

We feel obliged
to say how much this claim is
just
contrary to
not only to the common sense but also to
{\bf
almost every single
word ever written about
the bitcoin mining process
and related risks}. 
It ignores even
the most widely understood threats\footnote{
There are tens of thousands of web pages
which speak about this threat,
GHash.IO themselves have publicly stated that
they will take
"all necessary precautions to prevent
reaching 51$\%$ of all hashing power"
which is a "serious threat to the bitcoin community", cf.
\cite{GHashPressRelease}.
Some famous bitcoin developers go further
and said that they 
are preparing
"for the possibility that the Bitcoin ecosystem will break down"
and that price of bitcoins could collapse,
mainly in relation to this 51$\%$ GHash threat,
and inability or unwillingness
of the bitcoin technical architects to respond to this threat:
we hear that it could take "a system failure" or "a disaster"
for these problems to be taken seriously 
cf. \cite{ToddWhyISold50PcOfMyBTC,ToddBitcoinEcosystemWillMaybeBreakDown}.
},
not only some highly technical super subversive variants studied in this paper.
It is an example of easy-going 
excessively naive and utopian thinking
about the security of P2P financial systems
supported by
dubious
argumentation
without regard to consequences
for potential 
victims
of this intellectual
and financial security negligence.

\subsubsection{The Football Club Analogy.}
\label{PowerControversyFelterSirer4}
Sirer presents another argument.
Sirer
claims that the real power in bitcoin network
is with bitcoin adopters, or users of bitcoin wallet software,
see \cite{CornellVotingPowerAmazingAntiEthicalClaims}.
He compares miners or pool managers to
"the owner of a soccer team" who "may appear to have total control over administrative decisions".
However, he claims that in any football club,
it will be  "ultimately the fans who are fully in charge",
for example they can "routinely kick out bad management, drive away players
and override bad decisions by the seemingly powerful administration".
This argument is clearly faulty: it can be compared to claiming that
the voters who vote once every 4 years have any strong influence
on any single decision of our governments.
Moreover
owners of soccer clubs are typically not at all democratically elected,
so any claims about the very existence of the voting power of the fans
are technically void and are rather
just wishful thinking. 

\vskip-6pt
\vskip-6pt
\subsection{Bitcoin Specification and The Power of Miners}
\label{MinersHaveNoPower}
\vskip-6pt

Until now we have considered
primarily the question
of influencing the content of the blockchain
in the short or medium term.
In the long run
the objective could be much
more ambitious:
changing the bitcoin specification.
For example we recall that Felten claimed that
"a 51$\%$ attacker [...] can simply change the rules of Bitcoin at any time",
cf. \cite{FeltenMinersProbablyCanChangeMonPol},
in a rather drastic way,
like for example produce a large quantity of bitcoins.
A more moderate goal could be to
replace the hash function used in bitcoin mining.
This is less drastic.
Even though it could potentially put all ASCI miner companies out of business and
render hundreds of millions of dollars of investment obsolete,
it could be supported\footnote{
However even in this case,
we don't agree with Felten and
claim that such change is likely to fail to be accepted
(or lead to a split in bitcoin community),
see Section 6.5 and 14.1 of \cite{MiningUnreasonable} and
\cite{KaminskyPredictsEndOfSHA256}.}
by the majority of
ordinary users who are deeply concerned
by the current centralization of bitcoin mining.
Another sort of change would be to try to
make bitcoin truly anonymous,
which probably nobody would object except
the 
crime enforcement agencies.

In general the stakes are very high when
changing the bitcoin specification is considered.
In this area Sirer arrives at conclusions
which in our opinion are at odds with elementary common sense.
First it is claimed 
that
(cf. \cite{CornellVotingPowerAmazingAntiEthicalClaims}):
\vskip-9pt
\vskip-9pt
\begin{quote}
"regular users wield ultimate power in Bitcoin"
\end{quote}
\vskip-9pt

Following our discussion
in Section \ref{PowerControversyFelterSirer},
this is already something which we will find
very hard to believe,
possibly again an expression of utopian
wishful thinking.
We can however agree that this is more likely
to be true in the long-term perspective
than for the next few blocks.
We could also agree that this could hold for some truly radical
reforms, for example possibly Sirer is right
when he says that

\vskip-9pt
\vskip-9pt
\begin{quote}
"miners [...] could not,
for instance, create 10 million Bitcoins out of the thin air,
because no one would recognize those new rules",
\end{quote}
\vskip-9pt

Here many people tend to agree, because
it simply seems hard to imagine that the bitcoin monetary policy
could be changed 
as this potentially goes against interests of everybody,
including miners.
However on the other side,
probably there could be a way to convince everybody that
a different monetary policy is in everybody's interest,
and such a rule could be adopted by consensus,
cf. Section
\ref{DeflationaryCoinsGrowthCoinsTheorySketch}.

\subsubsection{Do Miners Have No Power?}
\label{MinersHaveNoPower2}
Interestingly Sirer goes again one important step further.
We recall that he claims that
"Miners are [..] just like every other user".
Moreover he further writes that:

\vskip-5pt
\vskip-5pt
\begin{quote}\bf
"Miners are subservient entities who must follow the decision of the Bitcoin community."
\end{quote}
\vskip-7pt

Here 
"the Bitcoin community" is understood as
people who run wallet software
\footnote{
This term is not sufficiently precise
in order to be able to make meaningful
statements about these questions.
No distinction is made between rather passive
``wallets'' which can be just used to store bitcoin balances
and on occasions to spend them,
without participating actively in the peer network,
and full network nodes which are the only entities which
are must be active in the real-time and are the only people who can influence
the propagation of other people's transactions into the blockchain,
and which transactions will reach miners and what moment
(those received first have more chances for being accepted).
Such active network nodes are unhappily much less numerous,
cf. \cite{NumberOfReachableNodesIsDroppingRecently}}.
Now if we make this claim more precise
and restrict it to the bare few thousands of ``full network nodes''
which propagate the transactions in the peer-to-peer network,
it may seem that there is some truth in what Sirer says.
These ``active'' bitcoin software nodes could in theory
"completely change the rules that govern the maintenance of the blockchain",
cf. \cite{CornellVotingPowerAmazingAntiEthicalClaims},
and miners will simply mine on transactions
which they will be allowed to receive.

However not every change will be well received.
In fact it is extremely difficult to claim that
miners have no power
and software nodes can change the bitcoin specification.
On the contrary.
We believe that the opposite
is going 
to happen.
Network nodes can just TRY to change the rules,
however we are not sure it they will ever succeed.

\vskip-7pt
\vskip-7pt
\subsection{Can The Bitcoin Network Implement A Reform Not Approved By A Majority Miners?}
\label{MinersHavePowerToBlock}
\vskip-7pt

Let us imagine that the Bitcoin core software client
which runs the bitcoin P2P network,
implements new rules for bitcoin
and for this purpose
they define
that when we have version=3 in the block data structure,
new rules are going to apply 
see Fig. 2 and Section 6.5 in \cite{MiningUnreasonable}. 
This is necessary to ensure a smooth transition
as bitcoin network does NOT have automatic updates and
it is totally impossible to upgrade all software nodes at once,
and it is not even possible to upgrade a large number of software nodes
at once, because nodes are run by volunteers
and they are not likely to update when
a new version is released.
All this has happened before with a very slow transition:
previously bitcoin had version=1
and for a long time both version=1 and version=2
were accepted.
Nevertheless such a transition doesn't
have to be accepted unanimously in the future.
Below we present an elaborate example.

We claim that {\bf miners can very easily maintain the status quo},
and reject all blocks with version=3.
Then peers running a new version of the software
will find themselves producing transactions which are not accepted
by all miners, only accepted by a minority of miners.
Miners who create blocks compliant with this new version=3 and include
transactions compliant with version=3 (or/and follow other specific features
or policies specified in version=3)
will risk that their blocks will be rejected by other miners.
Eventually even if a small proportion of miners produces sometimes blocks with version=3,
they risk that majority of miners will not accept these blocks.
Then the minority of miners who agree with the upgrade
will find themselves running a forked blockchain.
All this could happen ONLY because miners are too lazy to implement the upgrade,
remember that less than 10 pools control over two-thirds
of all the hash power, cf.
\cite{ToddBitcoinEcosystemWillMaybeBreakDown,MiningSubversive}.
Not because some miners are malicious or they want to disagree with some specific bitcoin reform.

Overall, we see that {\bf miners are NOT "subservient entities"}
as claimed in \cite{CornellVotingPowerAmazingAntiEthicalClaims}.
If miners and peer nodes do not agree on the new rules
which should be governing bitcoin,
or if they are just lazy or reluctant to implement them,
miners will continue to run the normal majority blockchain,
with very large hash power,
while the peers who decided to change the rules,
will have at least for some time still accept version=2
and therefore
both groups will be still sharing the same blockchain,
and blocks with version=3 will simply fail to materialize,
ever\footnote{
This sort of situation
is known in technology adoption:
for example however much IPv4 was a
"flawed protocol",
it seems that the adoption of IPv6
is just NOT happening (not at all),
see \cite{AntonopoulosIgnorant51PcLAJan2014}.
}.
After some time the peer nodes who support the reform
could decide to reject blocks with version=2,
this even though blocks with version version=3
failed to materialize.
This would be a suicide for some bitcoin software nodes who would become totally incompatible
with majority. In a rather improbable scenario,
bitcoin developers force an automatic security upgrade overnight,
on a majority of bitcoin
nodes and a minority of miners.
In this case we are sure to create a fork in bitcoin,
which fork however would be devoid of substantial hash power.
Then even if a majority of peer nodes upgrade the software,
the ``upgrade camp'' fork
will be vulnerable to double spending attacks with rapid hash displacement
from the other group,
and therefore no payment with the ``upgrade camp'' fork
will be accepted by any merchant
in a realistic time frame\footnote{
This situation was for another crypto currency qualified simply
as "certain death" of this currency,
cf. \cite{SavingDogecoinByMergedMining0814}.
} fearing possible double spend.
In contrast merchants can also safely accept all the blocks generated
by (more conservative) miners with version=2.
It appears that at any rate the upgrade will fail to be accepted if
a majority of miners are not supporting the upgrade. 

{\bf Preliminary Conclusions About The Power of Miners.}
\label{MinersHavePowerYes}
Overall we claim that
it is not correct to believe that:
"The rules are determined entirely by what the buyers and sellers accept as the legitimate blockchain.
Miners are subservient followers." as claimed in \cite{CornellVotingPowerAmazingAntiEthicalClaims}.
On the contrary, miners have the real power.
Following \cite{ToddBitcoinEcosystemWillMaybeBreakDown}
"it is the mining pool operator that chooses the software with which to mine",
cf. \cite{ToddBitcoinEcosystemWillMaybeBreakDown}.
Accordingly pool operators could
mandate changes in bitcoin software
much more easily
than any other group of people. 
Interestingly these powers of miners are waiting to be discovered.
They have been currently and temporarily given away for free.
Until now, they were de facto confiscated by large mining pools.
Yet the power of the has power is real.
In case of a disagreement,
ultimately buyers and sellers probably have no other choice than to accept the chain
with large hash power as legitimate\footnote{
Not a chain with which they would agree on some ideological technical or political grounds.}
which is exactly how bitcoin was build by Satoshi since ever,
as this provides serious protection against double spending attacks.


{\bf Related Work.}
In June 2013 a group of Princeton researchers \cite{KrollEconomicsRulesCouldBeChangedAtAnyTime}
have
published a detailed analysis of various scenarios
concerning building consensus in bitcoin.
In their conclusion they write:

\begin{quote}
[...] Bitcoin is not the fixed, rule-driven, incentive-compatible system that some advocates claim.
Although miners currently follow the original rules,
this behavior is stable only by consensus
and the rules could be changed at any time [...]
%
\end{quote}


\newpage
\subsection{Soft Power: Illusion or Reality?}
\label{UsersHaveANIllusionOfPower}


We have seen that if there is a disagreement,
miners are likely to always win over peer software
users. It remains the question of ``soft power''.
What if we are talking about not a controversial reform
but other 
changes
which maybe will not be perceived as problematic by miners.
In this case quite possibly Sirer is right.
It is quite plausible that
people or bitcoin adopters
can influence the future bitcoin spec 
and in some way "vote" for the content of the blockchain
in various indirect ways.
However we should note that:

\vskip-6pt
\vskip-6pt
\begin{enumerate}
\item
Mechanisms in place to implement such changes mandated by
a majority vote of network peers are inexistent.

\item
Options promoted by peer nodes devoided of hash power
could be just ignored by powerful miners,
unless it is supported by some authorities or the press/media.


\item
If any software changes
such as version=3 are democratically
imposed by a majority of bitcoin wallet nodes,
and even if miners accept these changes,
these changes will be very slow.
They will probably take many months,
possibly years to be implemented
with co-existence of version=2 and version=3.

This is obviously is by far too slow to prevent 51$\%$ attacks
which can be executed in the space of minutes/hours,
cf. Section \ref{LongestChainRuleDoubleSpendingAttacksMisconceptions}.



\item
Stake-holders in bitcoin are not always well informed
about certain security issues,
and for majority of them they simply don't care
because they deal with small amounts of money each time,
or maybe are resigned to wait for a long time
for their transactions to be confirmed.  

When they are informed it is maybe going to be too late,
bitcoin could suddenly be a victim of massive attacks,
and the ``soft power'' will be unable to react to them.

\item
In general there is strong asymmetry of information,
between bitcoin software architects and pool managers,
and the inert majority of bitcoin users.

\item
Users are unlikely to ever realize
that they don't want
bitcoin to function as it functions today,
and that they want bitcoin to change,
for example because they read so many misleading statements
about 51$\%$ attacks as shown
in Section \ref{LongestChainRuleDoubleSpendingAttacksMisconceptions}
and in the present Section.

\item
The current level of security awareness in the bitcoin community
is low,
bitcoin is frequently presented as almost a perfection in terms of security 
\cite{AcceptingZeroConfirmationFalselyClaimedSecure}, 
Satoshi was a genius \cite{SatoshiCryptoGenius},
and the more ``academic'' option to exercise
critical thinking and some caution with respect to future attacks and events,
and to try to improve or reform bitcoin
cf.
\cite{BitcoinFC12SecurityOverview,NumberOfReachableNodesIsDroppingRecently,FasterBitcoin,RosenfeldPoolRewardMethods,WiredKrollLMustKeepCreating} and this paper,
is 
not exercised very frequently.
\end{enumerate}
\vskip-6pt

\newpage

\vskip-7pt
\vskip-7pt
\subsection{On Power to Reform Bitcoin And Power to Block Reforms}
\label{MinersHavePowerYes2LaterCombined}
\vskip-4pt

In the previous sections we have argued
that software nodes may try impose some changes in bitcoin and it
is the miners or rather (currently) the mining pool operators
who will have the last word because
they control and can freely choose
"the software with which to mine",
cf. \cite{ToddBitcoinEcosystemWillMaybeBreakDown}.
However it seems that
it is way easier to block some upgrade
cf. Section \ref{MinersHavePowerToBlock},
rather than to impose a certain upgrade. 
%
Therefore radical changes such as creating 10 million new Bitcoins,
cf. \cite{CornellVotingPowerAmazingAntiEthicalClaims},
could be very hard to mandate,
but are NOT impossible.
They are actually claimed perfectly possible
by Princeton researchers \cite{KrollEconomicsRulesCouldBeChangedAtAnyTime}.

For sure pool operators could mandate changes in bitcoin software
more easily than anyone else but it remains an open problem
whether they could mandate some really important
reforms of bitcoin such as changing the monetary policy.
This could maybe require a slightly larger consensus than just miners.

\vskip-7pt
\vskip-7pt
\subsection{Alternative Centers of Power}
\vskip-4pt

The debate whether it is the miners or the users who have the ultimate power, or both,
or they can or cannot do is actually even more complicated than we think.
In bitcoin there exists an alternative {\bf third}
center of power. The power of holders of old coins!
For example if bitcoin code is hard forked
and people don't agree, people who own large balanced could play on one side,
like do different things with their own large balances in bitcoins on both blockchains,
potentially detrimental to some people with whom they don't agree.
More research on these questions is needed.

%

\vskip-7pt
\vskip-7pt
\subsection{Can We Agree on Some Crucial Questions In Bitcoin?}
\vskip-4pt

The reality is always more complex that our ideas about it.
However very clearly 
we find some claims of
\cite{CornellVotingPowerAmazingAntiEthicalClaims}
totally devoid of logical argumentation,
simply impossible to defend,
and in many cases
the closest thing to the reality is
just the contrary
of what is claimed in \cite{CornellVotingPowerAmazingAntiEthicalClaims}.
This is in particular the case in Sections
\ref{PowerControversyFelterSirer},
\ref{MinersHaveNoPower} through
\ref{MinersHavePowerToBlock}.

This is quite strange. 
In bitcoin we have progressively discovered
that we should not trust the miners,
that it is not exactly reasonable to trust pool managers either,
and that we should learn to be highly sceptical
of the security 
and quality of (any) open source code.
Now we also discover that
{\bf we cannot always trust university researchers}
to tell us the elementary truths
about important financial systems
used by millions of people.
All this in the context of a distributed
system which is expected
to be remove the
traditional necessity to trust
the people who build and run
our financial systems.

\newpage
\subsection{Should Blockchain Technology Be Regulated?}
\label{FutureBitcoinConclusionLongestChainRuleToBeRegulated}

This is a strange question\footnote{It is not the first time however
it has been discussed, cf. \cite{KrollEconomicsRulesCouldBeChangedAtAnyTime}}.
The question which exact US financial markets authorities
should be responsible
for regulating bitcoin in the future
is considered in a recent
article \cite{WallStreetLawyerBitcoinMoreAdvancedThanNMS}
which appears in the Wall Street Lawyer journal.

Even more surprisingly the author
suggests also that
the blockchain itself could also be regulated
and {\bf separately},
probably because it has many potential
applications outside of the world of finance.
Here are some very interesting longer
citations
from this paper:

\begin{quote}
To be clear, I am not proposing that the weightiness of bank regulation [...] be applied to tech start-ups [...] \\
I am suggesting that the codification of development standards
that good developers already use could help the network become safe\\
The open-source nature of the developer population
provides  opportunities  for  frivolous  or  criminal behavior
that can damage the participants in the same way that investors can be misled by promises
of  get  rich  quick  schemes.  [...]\\
a self- regulatory organization (SRO) [...] 
could be created
to  oversee  and  examine  [...] 
the engineers who create the code [...]\\
SRO  could qualify  and  register  developers  and  participants
in the Bitcoin ecosystem [...]\\
Regulations could ensure that cybersecurity requirements are engineered into the  code
and  could  ensure  that  the network would recover from a failure by building
in redundancy. \\
One of the biggest risks that we face as a society in the digital age [...]
is the quality of the code that will be used to run our lives.
\end{quote}


\newpage

\vskip-5pt
\vskip-5pt
\section{Summary and Conclusion}
\label{sec:conclusion}
\vskip-6pt

Bitcoin has a number of features and properties
which are sometimes presented as
very interesting and positive.
In fact they are
closer to engineering mistakes.
These features have been blindly
copied by other currencies, so called alt-coins.
Naive customers
(cf. Section \ref{FutureBitcoinConclusionBitcoinRegulatedCustomerFiskDevelopers})
are presented with software systems which are
claimed to be payment systems and currencies
which creates expectations that they
will be relatively stable
and that they are protected against attacks.
In reality serious problems are programmed
right there in the DNA of these currencies.
Sudden jumps and rapid phase transitions
{\bf are programmed at fixed dates in time}
and are likely to ruin the life of these currencies.
In this paper
we show that most crypto currencies
simply {\bf do NOT have
a good 
protection} against double spending:
the current protection
is flawed or/and ineffective.
Bitcoin and other crypto currencies which has
copied the same mechanisms 
make such attacks 
too easy.
We have been brainwashed with ideas about static 51 $\%$ attacks
while dynamic redirection attacks which just temporarily displace 100$\%$ or more of hash power
are perfectly feasible,
cf. Sections \ref{LongestChainRuleDoubleSpendingAttacks},
\ref{HiddenAttacks},
\ref{FutureBitcoinConclusionBitcoinRegulatedCustomerFiskDevelopers}
and Fig. \ref{DOGEIncredible5xIncreaseDogecoin}.
%

\vskip-8pt
\vskip-8pt
\subsection{What's Wrong?}
\label{sec:conclusionLongestChain}
\vskip-6pt

We discovered that
{\bf neither Satoshi
nor bitcoin developers
have mandated any sort of transaction timestamp}
in bitcoin software.
This can be seen as an expression of some sort of strange
{\bf ideology}:
giving an 
impression that maybe the {\em Longest Chain Rule}
does solve the problems in an appropriate way.
However clearly this rule is inadequate, 
it has definite perverse effects and
it is in fact simply dangerous.
Double spending events
are not only facilitated
by this exact rule
as we show in this paper
but they are not even recorded in the current bitcoin 
network, cf.
\cite{MtGoxDoubleSpendingDecker}.

The {\em Longest Chain Rule}
is probably OK for deciding for
which blocks miners will obtain a monetary
reward
(though more stable
mechanisms could be proposed).
However there is no reason why {\bf the same
exact slow and unstable mechanism would also
be used to decide which transactions are valid}.
{\bf This is NOT a feature, it is a bug, }
An engineering 
mistake on behalf of Satoshi Nakamoto,
the founder of bitcoin.
It affects not only the security of bitcoin 
but also its usability:
it makes transactions unnecessarily slow,
especially for larger transactions which require more confirmations,
cf. also \cite{FTACourtoisVideoHorseCarriage}.

\vskip-7pt
\vskip-7pt
\subsection{A Vulnerability Which is Programmed To Get Worse}
\label{sec:conclusionDestructionTheory}
\vskip-6pt

In this paper we initiate something which could be called a 
{\em Theory of Programmed Self-Destruction of Crypto Currencies}.
We look at built-in properties in crypto currencies
and we point out the combined effect of several factors.
We observe that vulnerability to double spending attacks
is very closely affected by build-in 
deflationary miner reward policies and the fact that these policies
mandate abrupt and sudden jumps.
These moments are likely to coincide with dates on which
the hash power is going to dramatically fall,
most probably in August 2016 for bitcoin,
and much sooner,
at several moments during 2014 for Dogecoin, Unobtanium
and many other existing coins.
At one moment the protection cushion
which is provided 
by the high 
hash rate disappears.
It becomes easier 
to execute double spending attacks.
More importantly,
we show that such attacks can be executed WITHOUT the knowledge
of miners which participate in the attack,
see Section \ref{HiddenAttack}.
In Section \ref{TechnicalityFurtherManipulation}
we describe a further realistic
attack scenario in which
this is done without the knowledge of pool managers.

\vskip-8pt
\vskip-8pt
\begin{figure}[!here]
\centering
\begin{center}
\vskip1pt
\vskip1pt
\includegraphics*[width=5.0in,height=2.0in,bb=0pt 0pt 730pt 270pt]{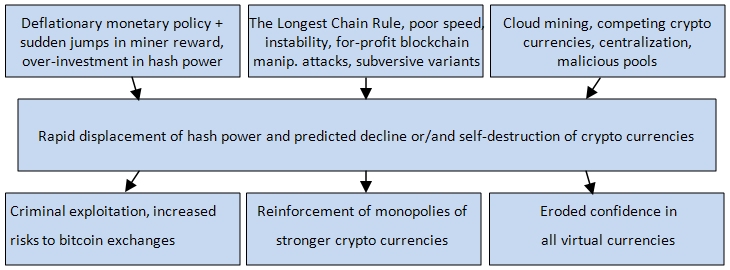}
\vskip-0pt
\vskip-0pt
\end{center}
\vskip-9pt
\vskip-9pt
\caption{
The built-in risks and dangers in current digital currencies. 
}
\label{BitcoinDestrSummary}
\vskip-3pt
\vskip-3pt
\end{figure}
\vskip-6pt
\vskip-4pt

In this paper
we have identified the DNA responsible for the epidemics
of programmed self-destruction
which is already affecting more than one crypto coin quite badly
with rapid outflow of hash power within days/hours:
cf. Section
\ref{UNODestructionSection}
and
\ref{DogeCoinandLitecoin}.
We conjecture that for small coins,
the Longest Chain Rule alone is sufficient to kill them.
For large coins which dominate the market,
it is still most probably fatal
in the long run
in combination with
deflationary monetary policies,
and in a competitive 
environment
plagued by numerous moral hazards.

\vskip-8pt
\vskip-8pt
\subsection{How To Fix It}
\label{sec:conclusionReformPath}
\vskip-6pt

There is no doubt that the virtual currency technology
could be improved or fixed.
At present a majority of existing crypto currencies
have copied this problematic 
{\em Longest Chain Rule} of bitcoin
and made things substantially worse by mandating
substantially faster transitions
in monetary policy and reward rules. 

Our main claim is that bitcoin software MUST change and implement
additional lower latency mechanisms 
in order to prevent and police
double spending attacks better than with blockchain alone.
It is urgent 
to modify the process of
deciding which transactions are valid
in a crypto currency.
Our main claims are 
that 1) the order and timing of transactions SHOULD 
be used in order to decide which transactions are accepted,
and that 2)
in order to facilitate fast zero confirmation transactions
double spending attacks should be increasingly difficult as time passes by
and 3)
bitcoin needs to create new incentives
for more peers to support the network 
cf. \cite{NumberOfReachableNodesIsDroppingRecently}.
The exact details remain an open problem.
As a quick fix,
in Section \ref{DoubleSpendingSolutionProposed}
we discuss possible solutions using timestamps
and peer confirmations.
Overall we expect to improve the security against double spending 
and also dramatically improve the speed of transactions
in bitcoin and all other crypto currencies.
Our solutions also promote better {\bf network neutrality}: 
timing information makes decisions of the network less arbitrary
and miners have less discretionary powers
which could help the attackers.
%

\vskip-9pt
\vskip-9pt
\subsection{Discussion}
\label{sec:conclusionMonopolySatoshiStupid}
\vskip-5pt

We should think twice before saying
that what Satoshi did was wrong or
mistaken. 
In Section \ref{FutureBitcoin}
we show that current bitcoin specification
makes that bitcoin currency 
has a privileged position. 
Smaller bitcoin competitors which use the same hash function
are rather unable to survive,
cf. Sections
\ref{UNODestructionSection}, \ref{DogeCoinandLitecoin}
and \cite{SavingDogecoinByMergedMining0814,WhatDogeSurviveAfterMyPaper}.
Bitcoin tends to remain in a monopoly situation
while smaller alt-coins are in trouble,
even if they copy its mechanisms exactly.
Satoshi and other early adopters may then hope that nobody will
challenge bitcoin and they will be able to earn
hundreds of millions of dollars selling their coins,
cf. Section \ref{InvestorEconomicsAB2Billion} and
\ref{FutureBitcoinConclusionDeclineUnlessPersistentDomination}.

{\bf Remark:} Litecoin which uses a different hash function escapes this rule
and creates a dominating position in its own space
\cite{LiteCoinPoolNear51Percent}.
Here it has been recently challenged by Dogecoin
which has achieved a comparable hash rate in February 2014.
Unhappily as we show in this paper,
the hash rate of Dogecoin is now bound 
to substantially deflate. 
It has already become highly vulnerable to double spending attacks,
which can be executed by one single miner,
cf. Section \ref{DogeCriminalBusiness}.

\vskip-5pt
\vskip-5pt
\subsection{Investors and Alt-Coin Designers in Trouble}
\label{sec:conclusionMonopolyInvestorsAltCoins}
\vskip-5pt

In this paper we have studied how
hundreds of millions of dollars were invested
in bitcoin. On one side it is a bubble,
on the other side it is an investment.
An investment in building secure
distributed hashing infrastructure
which has costed hundreds of
millions dollars
and consumes tens of
megawatts in electricity.
In this paper we show that this investment
does {\bf NOT} do the job correctly.
We claim that large hash power is {\bf neither necessary nor sufficient}
in order to run a digital currency system.
We contend that this expensive electronic notary infrastructure
is {\bf not} needed for bitcoin to function correctly.
It is {\bf not} justified by security against double spending.
Now it may appear necessary,
because bitcoin and other digital currencies
have not really tried to protect
themselves against double spending attacks.
Current digital currencies simply
do allow blockchain manipulation
to affect transactions too easily
(cf.
Fig. \ref{AltChainProfitDoubleSpendAttack} page \pageref{AltChainProfitDoubleSpendAttack}).

The current monopoly rent situation for bitcoin (if there is one)
is more accidental than deserved.
It is rather due to the fact that competitors of bitcoin
have not done enough in order to design reasonable crypto currencies
(cf. Section \ref{DoubleSpendingSolutionProposed}).
In fact it is possible to believe that they have been
excessively naive 
and they have fallen
into a specific sort of deadly trap.
They have copied
{\bf those exact mechanisms in bitcoin which mandate
programmed destruction of all (weaker) crypto currencies
which implement them}.
Moreover many alt-coins have accelerated this processus
greatly by
programming many consecutive very fast transitions
to occur within months.
Current alt-coin crypto currencies are also ideal candidates
for ``pump and dump'' investment strategies
in which some form of decline,
possibly a "certain death" \cite{SavingDogecoinByMergedMining0814}
is bound to happen
at exact predicted moments in time.

\medskip
{\bf Acknowledgments:}
We thank
Xavier Alexandre, George Danezis, Gerald Davis,
Pinar Emirdag, Michael Folkson, Cl\'{e}ment Francomme,
Pawel Krawczyk, Emin G\"un Sirer,
Guangyan Song, Tim Swanson
and John Shawe-Taylor
for their extremely helpful suggestions, observations and comments.



\vfill
\pagebreak




\end{document}